\documentclass[a4paper, english, 11pt]{article}

\usepackage[utf8]{inputenc}
\usepackage[T1]{fontenc}
\usepackage[english]{babel}
\usepackage{csquotes}
\usepackage{textcomp}
\usepackage{varioref}
\usepackage{amsmath}
\usepackage{amssymb}
\usepackage{amsthm}
\usepackage{graphicx}
\usepackage{color}
\usepackage{titlesec}
\usepackage{algpseudocode}
\usepackage{algorithmicx}
\usepackage{algorithm}
\usepackage{hyperref}
\usepackage{enumitem}
\usepackage{caption}
\usepackage{subcaption}
\usepackage{bookmark}
\usepackage{multirow}
\usepackage{booktabs}
\usepackage{setspace}
\usepackage{tikz}
\usetikzlibrary{arrows.meta}
\makeatletter
\newtheorem*{rep@theorem}{\rep@title}
\newcommand{\newreptheorem}[2]{%
\newenvironment{rep#1}[1]{%
 \def\rep@title{#2 \ref{##1}}%
 \begin{rep@theorem}}%
 {\end{rep@theorem}}}
\makeatother

\theoremstyle{plain}
\newtheorem{mytheorem}{Theorem}
\newtheorem{myprop}[mytheorem]{Proposition}
\newreptheorem{myprop}{Proposition}

\theoremstyle{definition}

\newcommand{\comp}{\mathsf{c}}

\newcommand{\bA}{\mathbf{A}}
\newcommand{\bb}{\mathbf{b}}

\newcommand{\bI}{\mathbf{I}}
\newcommand{\bJ}{\mathbf{J}}
\newcommand{\bK}{\mathbf{K}}
\newcommand{\bQ}{\mathbf{Q}}
\newcommand{\bS}{\mathbf{S}}
\newcommand{\bu}{\mathbf{u}}

\newcommand{\bv}{\mathbf{v}}
\newcommand{\bV}{\mathbf{V}}

\newcommand{\bW}{\mathbf{W}}
\newcommand{\bx}{\mathbf{x}}
\newcommand{\bX}{\mathbf{X}}

\newcommand{\bone}{\mathbf{1}}

\newcommand{\bmu}{\boldsymbol{\mu}}

\newcommand{\bSigma}{\boldsymbol{\Sigma}}

\newcommand{\hatbmu}{\hat{\boldsymbol{\mu}}}

\newcommand{\hatbSigma}{\hat{\boldsymbol{\Sigma}}}

\newcommand{\hatbQ}{\hat{\mathbf{Q}}}

\newcommand{\barbx}{\bar{\mathbf{x}}}

\newcommand{\diag}{\text{\normalfont{diag}}}

\newcommand{\argmax}{\text{\normalfont{argmax}}}

\newcommand\numberthis{\addtocounter{equation}{1}\tag{\theequation}}

\makeatletter
\newcommand*{\tp}{%
{\mathpalette\@tp{}}%
}
\newcommand*{\@tp}[2]{%
\raisebox{\depth}{$\m@th#1\intercal$}%
}
\makeatother

\algnewcommand\algorithmicinput{\textbf{Input:}}
\algnewcommand\INPUT{\item[\algorithmicinput]}

\algnewcommand\algorithmicmyreturn{\textbf{Return:}}
\algnewcommand\myRETURN{\item[\algorithmicmyreturn]}

\def\checkmark{\tikz\fill[scale=0.4](0,.35) -- (.25,0) -- (1,.7) -- (.25,.15) -- cycle;}
\usepackage{natbib}
\setcitestyle{authoryear, semicolon, aysep = {}, yysep = {,}}
\usepackage[margin=1in]{geometry}

\newcommand{\mvcapacor}{CAPA-CC}
\newcommand{\mvcptcor}{CPT-CC}

\begin{document}

\title{Scalable changepoint and anomaly detection in cross-correlated data with an application to condition monitoring}
\author{Martin Tveten \thanks{Department of Mathematics, University of Oslo, Oslo, Norway.}
\and Idris A. Eckley \thanks{Mathematics and Statistics, Lancaster University, Lancaster, LA1 4YF, UK.}
\and Paul Fearnhead \footnotemark[2]}
\maketitle

\begin{abstract}
Motivated by a condition monitoring application arising from subsea engineering we derive a novel, scalable approach to detecting anomalous mean structure in a subset of correlated multivariate time series. Given the need to analyse such series efficiently we explore a computationally efficient approximation of the maximum likelihood solution to the resulting modelling framework, and develop a new dynamic programming algorithm for solving the resulting Binary Quadratic Programme when the precision matrix of the time series at any given time-point is banded. Through a comprehensive simulation study, we show that the resulting methods perform favourably compared to competing methods both in the anomaly and change detection settings, even when the sparsity structure of the precision matrix estimate is misspecified. We also demonstrate its ability to correctly detect faulty time-periods of a pump within the motivating application.
\end{abstract}


\newpage

\section{Introduction} \label{sec:intro}
Modern machinery can be perplexingly complicated and interlinked.
The interruption of one machine may cause downtime of a whole operation, in addition to a repair being both costly, time consuming and arduous.
This has spawned an enormous interest in (remote) condition monitoring of industrial equipment to detect deviations from its normal operation, such that optimal uptime can be achieved and impending faults discovered before they occur.
Overviews of condition monitoring techniques for different equipment exist for pump-turbines \citep{egusquiza_condition_2015}, wind turbines \citep{tchakoua_wind_2014}, and audio and vibration signals \citep{henriquez_review_2014}, among others.
A common theme is the decision problem of when the machinery is running abnormally---a problem that lends itself well to statistical changepoint analysis.

The current work is motivated by a problem of detecting time-intervals (segments) of suboptimal operation of an industrial process pump.
We will refer to these segments as "anomalies" or "anomalous segments", because they correspond to deviations from some predefined baseline pump behaviour.
The pump is equipped with sensors that measure temperatures and pressures over time at various locations.
Other operational variables such as the flow rate and volume fractions for the different fluids being pumped are also recorded.
If present, the aim is to estimate the start- and end-point of anomalies, as well as indicate which variables are anomalous.
This is useful information to the operators of the pump to pin-point the source of historical problems and learn from them.
Another reason for performing such an analysis is to create a clean reference data set that can be used  to train a model of the equipment's baseline behaviour on, before deploying the method for online condition monitoring.
The particular data set we consider contains four anomalies that have been manually labelled by engineers familiar with this data,  based on retrospectively looking for signs in the data of degrading performance.

\begin{figure}[t]
  \centering
  \includegraphics[width=0.85\textwidth] {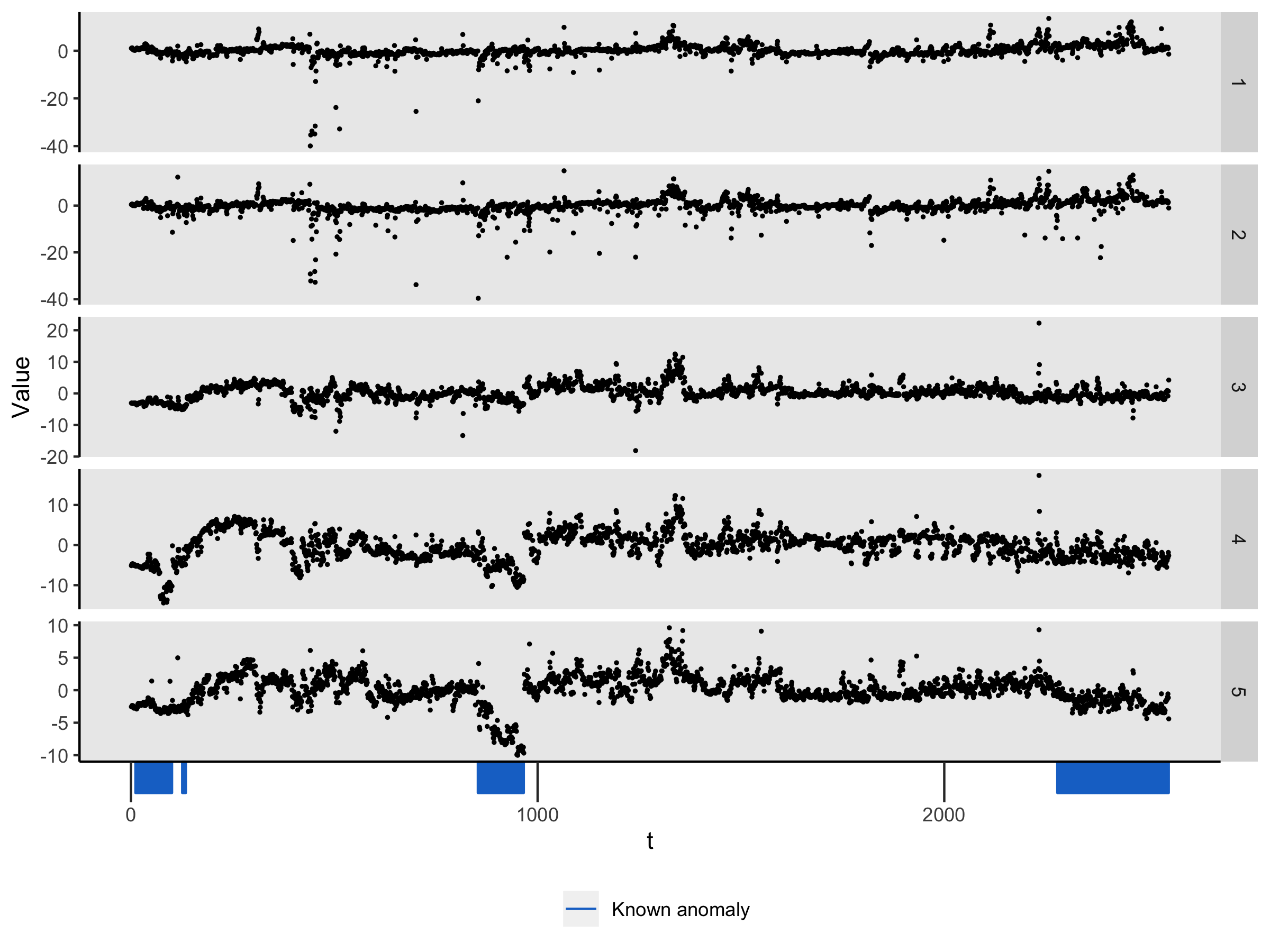}
  \caption{Pump data after preprocessing with four known segments of suboptimal operation marked by blue bars. The correlation between variables 1 and 2 is 0.89 and the pairwise correlations between variables 3, 4 and 5 are all above $0.6$.}
  \label{fig:raw_data}
\end{figure}

The starting-point of our methodology is to assume that during normal operation of the pump, the data follows a baseline stationary distribution, and during suboptimal operation, the mean of the distribution changes abruptly for some period of time before it reverts back to the baseline mean.
This is known as an \textit{epidemic changepoint} model in the literature \citep{kirch_detection_2015}, but in the presence of our application, we will refer to it as the \textit{anomaly} model.
A challenge with the pump data is that the mean changes as a consequence of what is being pumped and other operating conditions in addition to suboptimal operation.
To decrease the dependence on the operating conditions and thus increase the signal from changes due to suboptimal operation, we divide the variables into sets of \textit{state} variables and \textit{monitoring} variables, and regress the monitoring variables onto the state variables (similar to \citet{klanderman_fault_2020}).
The remaining five-variate time series of monitoring residuals are shown in Figure \ref{fig:raw_data}, where the known anomalies are marked on the time axis.
Observe that the strength of the known anomalies vary as well as which variables seem to be affected.
It is also apparent that the mean changes outside of the known anomalous segments.
Detecting and estimating these segments is also important as they may correspond to previously unknown anomalies or constitute data for which the current model between state and monitoring variables fit poorly, and hence point to how it should be improved.

\begin{figure}[t]
  \centering
  \includegraphics[width=0.85\textwidth] {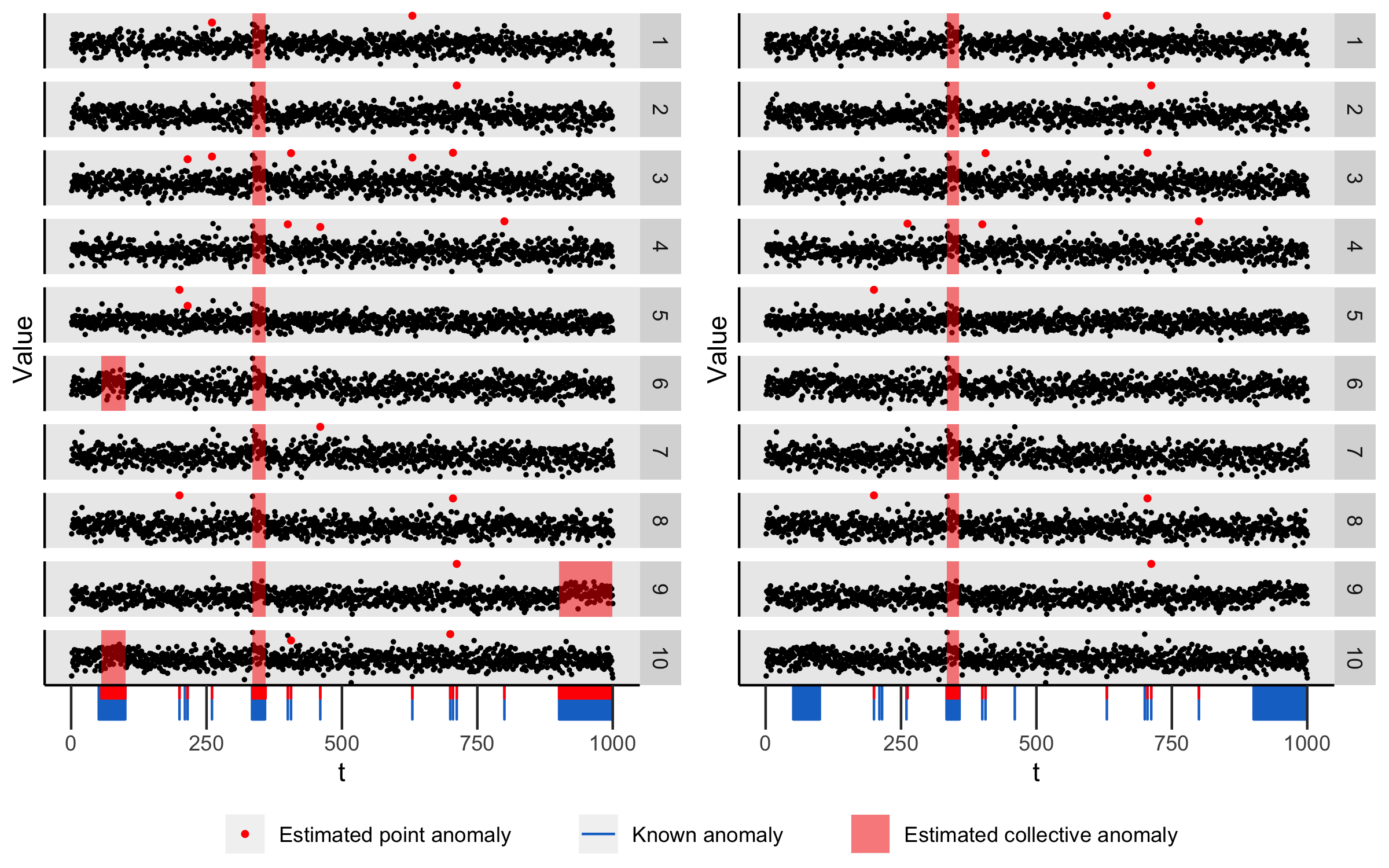}
  \caption{Modelling cross-correlation increases detection power for a fixed Type I error probability, especially for sparse changes.
  Both plots show the same set of 1000 simulated observations from a 10-variate Gaussian distribution with a global constant correlation of $0.5$, containing three collective anomalies at $t \in (50, 100], (333, 358], (900, 1000]$, affecting the means of variables $\{ 6, 10 \}$, $\{ 1, \ldots, 10 \}$ and $\{ 9 \}$, respectively, and 12 point anomalies affecting two random variables each.
  The left plot displays the estimates of collective and point anomalies of our method, which incorporates cross-correlations, while the right plot shows estimates when the method ignores cross-correlations.
  As both methods were tuned to achieve $0.05$ probability of a false positive under the global correlation null model, the two sparse anomalies are not detected in the right plot as a trade-off with error control.}
  \label{fig:simple_example}
\end{figure}

The pump data after preprocessing also exhibit strong cross-correlation due to the proximity of the sensors to each other, with the correlation of variables 1 and 2 being 0.89 and the pairwise correlations between variables 3, 4 and 5 all being above 0.6.
Most existing methods for detecting a change or anomaly in a subset of variables ignore cross-correlation \cite[though see][]{wang_high_2018}.
If not accounted for, however, cross-correlation will hamper the detection of more subtle anomalies as illustrated by the simulated example in Figure \ref{fig:simple_example}.
The benefit of undertaking multivariate changepoint detection is to borrow strength between variables to detect smaller changes than would be possible if each variable were considered separately, and including cross-correlation if sufficiently strong will increase the power of detection.
This is particularly true for sparse changes, an observation also made by \citet{liu_minimax_2019}.

Our main methodological contribution is to develop a novel test statistic based on a penalised cost approach for detecting multiple anomalies/epidemic changes in a subset of means of cross-correlated time series.
The test is designed to be powerful for both sparse and dense alternatives, as well as being computationally fast and scalable.
This is crucial for our method to also be useful for anomaly detection problems of higher dimensionality than our process pump example.
Anomalies are then detected by using the test within a PELT-type algorithm \citep{killick_optimal_2012} to optimise exactly over all possible start- and end-points of anomalies.

Through the work on making the method scalable, we derive an algorithm which may be of independent interest within combinatorial optimisation.
Our test statistic is an approximation to the maximum likelihood solution of our problem, formulated as what is known as an unconstrained Binary Quadratic Program (BQP).
We show that such optimisation problems can be solved exactly by a dynamic programming  algorithm scaling linearly in the number of variables, $p$, if the matrix in the quadratic part of the objective function is sparse in a banded fashion.
In the anomaly detection problem, this corresponds to having a banded precision matrix. We present a simple pre-processing step for obtaining a banded estimate of the precision matrix of our data, and show empirically that detecting the anomalies using such an estimate leads to gains in power over methods that ignore cross-correlation even when the banded assumption is incorrect.

A further challenge in many applications, such as the pump data of Figure \ref{fig:raw_data}, is the presence of outliers.
If left unattended, it is well-known that they will interfere with the detection of changes \citep{fearnhead_changepoint_2019}.
To handle outliers, we incorporate the distinction between point and collective anomalies, introduced in the CAPA (Collective And Point Anomalies) and MVCAPA (MultiVariate CAPA) methods of \citet{fisch_linear_2019, fisch_subset_2019}.
A point anomaly is defined as an anomalous segment of length one---a single anomalous observation---while a collective anomaly is an anomalous segment of length two or longer.
This distinction enables the method to classify sporadic outliers as point anomalies rather than confusing them with a collective anomaly.
We call our anomaly detection algorithm \mvcapacor\, short for Collective And Point Anomalies in Cross-Correlated data.

To the best of our knowledge, there are no other methods designed specifically for the multiple point and collective anomaly detection problem in multivariate, cross-correlated data with both sparse and dense anomalies. Current approaches to detect collective anomalies assume independence across series \citep{fisch_subset_2019,jeng_simultaneous_2013}. Alternatively, methods like \citet{kirch_detection_2015} model correlated series, but focus on detecting changes in the cross-correlation.

For the general changepoint problem of a sparse or dense change in the mean, the literature is mostly concentrated on methods that either allow for sparse changes but assume cross-independence \citep{xie_sequential_2013, jirak_uniform_2015, cho_multiple-change-point_2015, cho_change-point_2016, bardwell_most_2019}, or allow cross-dependence but assume changes are dense \citep{horvath_change-point_2012, li_change_2019, bhattacharjee_change_2019, westerlund_common_2019}.
The inspect method of \citet{wang_high_2018} is a notable exception to this rule, as it is designed to estimate sparse changes in the mean of potentially cross-correlated data. Whilst general changepoint methods can also be used for the anomaly detection problem, some power is expected to be lost as there is no assumption of a shared baseline parameter.

The paper is organised as follows:
We first describe the anomaly detection problem in detail in Section \ref{sec:problem}, before considering our solution in Section \ref{sec:anomalies}.
Particular focus is put on the single collective anomaly case and our BQP solving algorithm for approximating the maximum likelihood solution.
We then briefly describe how the same ideas can be applied to the general changepoint detection problem in Section \ref{sec:changepoints}.
In Section \ref{sec:implementation}, we cover a useful strategy for robustly estimating the precision matrix with a given sparsity structure, and we suggest strategies for tuning our method.
Section \ref{sec:simulations} contains an extensive simulation study for assessing the performance of our method.
We conclude by presenting the analysis of the pump data in Section \ref{sec:real_data}.

\section{Problem description} \label{sec:problem}
Suppose we have $n$ observations $\{\bx_t\}_{t = 1}^n$ of $p$ variables $\bx_t = (x_t^{(1)}, \ldots x_t^{(p)})$, where each $\bx_t$ has mean $\bmu_t$ and a common precision matrix $\bQ$ encoding the conditional dependence structure between the variables.
Our interest is in detecting collective anomalies that are characterised by a change in the mean of the data.

In our anomaly detection problem, segments of the data will be considered anomalous if the mean $\bmu_t$ is different from a baseline mean $\bmu_0$.
Let $K$ be the number of collective anomalies, 
where the $k$th anomaly, for $k = 1, \ldots, K$, starts at observation $s_{k} + 1$, ends at observations $e_k$, and affects the components in a subset $\bJ_k \subseteq [p]$. So, the model assumes that the mean vectors $\bmu_t$ are given by
\begin{equation}
  \mu_t^{(i)} = \begin{cases}
    \mu_1^{(i)} & \text{if } s_1 < t \leq e_1 \text{ and } i \in \bJ_1, \\
                & \vdots \\
    \mu_K^{(i)} & \text{if } s_K < t \leq e_K \text{ and } i \in \bJ_{K}, \\
    \mu_0^{(i)} & \text{otherwise},
  \end{cases}
  \label{eq:mean_model}
\end{equation}
where $e_k \leq s_{k + 1}$, such that no overlapping anomalous segments are allowed. To distinguish collective anomalies from point anomalies, which we will consider later, we make the assumption that collective anomalies are of length at least 2, i.e.  $e_k - s_k \geq 2$. The rationale is that point anomalies, that is, anomalies that affect data at isolated time points, are likely to be caused by different factors than collective anomalies. In our application, point anomalies may be due to sensor errors, whereas collective anomalies indicate underlying issues with the machinery.
In some cases, one may also be given information about the minimum and maximum segment length of a collective anomaly, $l \geq 2$ and $ l < M \leq n$, respectively, such that $l \leq e_k - s_k \leq M$ for all $k$.

Our aim is to infer the number of collective anomalies $K$, as well as their locations within the data $(s_k, e_k, \bJ_k)_{k = 1}^{K}$ together with the anomalous means $\bmu^{(i)}_k$, for $i \in \bJ_k$, in a computationally efficient manner. 

During method development, we assume that the baseline parameter $\bmu_0$ and the precision matrix $\bQ$ is known.
In practice, these will be estimated from the data using robust statistical methods described in Section \ref{sec:precision_estimation}.
Later, to enable quick computation, we will also assume that $\bQ$ or an estimate of $\bQ$ is sparse in a banded fashion.
A sparse precision matrix corresponds to cases where only a few of the variables are conditionally dependent.

\section{Detecting anomalies} \label{sec:anomalies}
\subsection{A single collective anomaly} \label{sec:single_anomaly}
In this section, we consider the anomaly detection problem described in Section \ref{sec:problem} for $K \leq 1$. Our approach is to model the data as being realisations of multivariate Gaussian random variables, independent over time, and to use a penalised likelihood approach to detect an anomaly.

We will use the following notation: For a $p$-vector $\bx$ and set $\bJ \subseteq [p]$, $\bx^{(\bJ)} := (x^{(i)})_{i \in \bJ}$ and $\bx(\bJ) := \big(x^{(i)}I\{ i \in \bJ \}\big)_{i = 1}^p$, where $I\{i \in \bJ\})$ is the indicator function. For a matrix $\bX$, $\bX_{\bJ, \bK}$ denotes the sub-matrix of rows $\bJ$ and columns $\bK$.
Both $-\bJ$ and $\bJ^\comp$ refer to the complement of a set $\bJ$.
The $k$-subscripts enumerating the anomalies will be skipped when the referenced anomaly is clear from the context.

Define the cost of introducing an anomaly from time-point $s + 1$ to $e$ in variables $\bJ$ as twice the negative log-likelihood of multivariate Gaussian data
\begin{align*}
  C\left(\bx_{(s + 1):e}, \bmu(\bJ)\right) &= 
\sum_{t = s + 1}^e \left(\bx_t - \bmu(\bJ)\right)^\tp \bQ \left(\bx_t - \bmu(\bJ)\right), \numberthis \label{eq:cost}
\end{align*}
where for simplicity we have dropped added constants.
Now, for ease of presentation, without loss of generality we assume $\bmu_0 = \mathbf{0}$.  Then the log-likelihood ratio statistic of the observations $\bx^{(\bJ)}_{(s + 1):e}$ being anomalous is given by
\begin{equation}
  S(s, e, \bJ) = C(\bx_{(s + 1):e}, \mathbf{0}) - \underset{\bmu(\bJ)}{\min}\; C\left(\bx_{(s + 1):e}, \bmu(\bJ)\right).
  \label{eq:savings}
\end{equation}
We refer to $S(s, e, \bJ)$ as the \textit{saving} realised by allowing the observations $\bx^{(\bJ)}_{(s + 1):e}$ to have a different mean from $\mathbf{0}$.
In a maximum likelihood spirit, the aim is to maximise the savings $S(s, e, \bJ)$ over start-points $s$, end-points $e$, and subset $\bJ$, and infer the anomalous segment thereof.
However, as we vary $\bJ$ we are optimising over differing numbers of means in the anomalous segment -- and the savings will always increase as we optimise over more parameters.
One way of dealing with this is to introduce a penalty that is a function of the number of anomalous variables, $P(|\bJ|)$, and maximise the penalised savings instead.
This gives us the following anomaly detection statistic:
\begin{equation}
    S := \underset{l \leq s - e \leq M}{\max}\; S(s, e)
      := \underset{l \leq s - e \leq M}{\max}\; \underset{\bJ}{\max}\; \big[S(s, e, \bJ) - P(|\bJ|)\big].
  \label{eq:max_penalied_savings}
\end{equation}
Recall that $l$ and $M$ are the minimum and maximum segment length, respectively.
An anomaly is declared if \eqref{eq:max_penalied_savings} is positive, and the maximising $(s, e, \bJ)$ is a point-estimate of the anomaly's position in the data.

Throughout this article, we use a piecewise linear penalty function of the form
\begin{equation}
  P(|\bJ|) = \min(\alpha_\text{sparse} + \beta |\bJ|, \alpha_\text{dense})
  = \begin{cases}
    \alpha_\text{sparse} + \beta|\bJ|,& |\bJ| < k^* \\
    \alpha_\text{dense},& |\bJ| \geq k^*
  \end{cases},
  \label{eq:penalty}
\end{equation}
where $k^* = (\alpha_\text{dense} - \alpha_\text{sparse}) / \beta$.
We will refer to $|\bJ| < k^*$ as being in the \textit{sparse regime} and ${|\bJ| \geq k^*}$ as being in the \textit{dense regime}.
Such a penalty function ensures that our method can be powerful against both sparse and dense alternatives.
In addition, we can apply the results from \citet{fisch_subset_2019} where it is shown that, if our modelling assumptions are correct, setting ${\alpha_\text{dense} = p + 2\sqrt{p\psi} + 2\psi}$, ${\alpha_\text{sparse} = 2 \psi}$ and ${\beta = 2\log(p)}$, for $\psi = \log(n)$, results in a false positive rate that tends to 0 as $n$ grows.
Furthermore, \citet{fisch_subset_2019} show that scaling the penalty function \eqref{eq:penalty} by a factor $b$ is appropriate in many situations where the modelling assumptions do not hold, such as when there is dependence over time.

Note that $[p]$ is always the maximiser in the dense regime, and that $\beta$ is the additional penalty for adding an extra variable to the anomalous subset in the sparse regime. We will exploit these properties when deriving an efficient optimisation algorithm in Section \ref{sec:anomaly_approx}.

To compute the anomaly detection statistic $S$, we need the maximum likelihood estimator (MLE) $\hatbmu(\bJ)$ of $\bmu(\bJ)$, where the means of variables $j \in \bJ$ are allowed to vary freely while the others are restricted to $0$.
Optimising the multivariate Gaussian likelihood \eqref{eq:cost} with respect to such a subset restricted mean results in the following MLE for the mean components in $\bJ$:
\begin{equation}
  \hat{\bmu}^{(\bJ)}_{(s + 1):e} = \barbx_{(s + 1):e}^{(\bJ)} + \bQ_{\bJ, \bJ}^{-1} \bQ_{\bJ, -\bJ}\barbx_{(s + 1):e}^{(-\bJ)}.
  \label{eq:MLE}
\end{equation}
The corresponding $p$-vector $\hat{\bmu}(\bJ)$ is constructed by placing $\hat{\bmu}^{(\bJ)}$ at indices $\bJ$ and zeroes elsewhere.
Finally, putting the MLE back into the expression for the saving, and suppressing the subscripts $(s + 1):e$ to not clutter the display, gives us that
\begin{equation}
  S(s, e, \bJ) = (e - s)(2\barbx - \hat{\bmu}(\bJ))^\tp\bQ\hat{\bmu}(\bJ).
  \label{eq:MLE_savings}
\end{equation}

Unfortunately, the complicated form of the MLE \eqref{eq:MLE} means that the number of operations required for finding the exact maximum penalised saving over subsets $\bJ$ is $O(2^p)$.
The optimisation problem is not only combinatorial, but also nonlinear, and as far as we know, there is no reformulation of the saving \eqref{eq:MLE_savings} that would make the problem notably more tractable.
We thus opt for an approximation to the saving \eqref{eq:MLE_savings} to achieve scalability.

\subsection{Approximate savings for anomaly detection} \label{sec:anomaly_approx}
Our idea for a computationally efficient approximation of the subset-maximised penalised savings $S(s, e)$, is to replace the MLE in \eqref{eq:MLE_savings} with the subset-truncated sample mean,
\begin{equation}
  \barbx(\bJ) = \barbx \circ \bu,
  \label{eq:subset_sample_mean}
\end{equation}
where $\bu = (I\{i \in \bJ\})_{i = 1}^p$ and $\circ$ is the element-wise (Hadamard) product.
That is, under the sparse regime, we aim to maximise the approximate penalised saving;
\begin{equation}
  \tilde{S}(s, e)
  := \max_\bJ \big[\tilde{S}(s, e, \bJ) - P(|\bJ|)\big]
  = \max_\bJ \left[(e - s)(2\barbx - \barbx(\bJ))^\tp\bQ\barbx(\bJ) - \beta|\bJ|\right] - \alpha_\text{sparse}.
  \label{eq:aMLE_savings}
\end{equation}
Under the dense regime, the exact maximum is given by $S(s, e, [p]) - \alpha_\text{dense}$.

An important motivation for using $\barbx(\bJ)$ is that finding $\tilde{S}(s, e)$ corresponds to what is known as a \textit{binary quadratic program} (BQP).
The unconstrained version of such optimisation problems are of the form
\begin{equation}
  \underset{\bu \in \{ 0, 1 \}^p}{\max}\; \bu^\tp\bA\bu + \bu^\tp\bb + c,
  \label{eq:BQP}
\end{equation}
where $\bA$ is a real, symmetric, $(p \times p)$-dimensional matrix, $\bb$ is a real, $p$-dimensional vector and $c$ is a real scalar.
BQPs are NP-hard in general \citep{garey_computers_1979}, even if $\bA$ is positive or negative definite.
If $\bA$ is $r$-banded, however, we show that BQPs can be solved with $O(p2^r)$ operations.
Proposition \ref{prop:BQP} confirms that $\max_\bJ [\tilde{S}(s, e, \bJ) - P(|\bJ|)]$ is indeed a BQP.
\begin{myprop} \label{prop:BQP}
  Let $\alpha, \beta \geq 0$, $\barbx \in \mathbb{R}^p$ and $\barbx(\bJ) = \bu \circ \barbx$, where $\bu$ is a binary vector with $1$ at positions $\bJ$ and $0$ elsewhere.
  Then solving
  \begin{equation}
    \max_\bJ \left[(e - s)(2\barbx - \barbx(\bJ))^\tp\bQ\barbx(\bJ) - \beta|\bJ|\right] - \alpha
    \label{eq:approx_optimal}
  \end{equation}
  corresponds to a BQP with $\bA = -(e - s)\barbx \barbx^\tp \circ \bQ$, $\bb = 2(e - s)(\barbx \circ \bQ\barbx) - \beta$ and $c = -\alpha$.
\end{myprop}

To explain the dynamic program (Algorithm \ref{alg:BQP_solver}) for solving the BQP when the precision matrix $\bQ$, and hence $\bA$, is $r$-banded, it is illustrative to consider the case of $r = 1$. The key idea is that if we cycle through the variables in turn, then the choice of which of the variables $d,\ldots,p$ are anomalous will depend on the variables $1,\ldots,d - 1$ only through whether variable $d - 1$ is anomalous or not. Thus we can obtain a recursion by considering these two possibilities separately.

In the case of $r = 1$, the BQP for $\max_\bJ[\tilde{S}(s, e, \bJ) - P(|\bJ|)]$ is given by
\begin{equation}
  \underset{\bu \in \{ 0, 1 \}^p}{\max}\; \sum_{d = 1}^p (b_d + A_{d, d}) u_d + 2\sum_{d = 2}^p A_{d, d - 1} u_d u_{d - 1} + c,
  \label{eq:1bandedBQP}
\end{equation}
where $A_{d, i} = -(e - s)Q_{d, i}\bar{x}_d\bar{x}_i$ for $i = d, d - 1$, $b_d = 2(e - s)\bar{x}_d\sum_{i = d - 1}^{d + 1} Q_{d, i} \bar{x}_i - \beta$, and $c = -\alpha$.
Let $\tilde{S}_{1}(d)$ and $\tilde{S}_{0}(d)$ be the maximal approximate penalised savings of variables $1, \ldots, d \leq p$ conditional on variable $d$ being anomalous ($u_d = 1$) or not ($u_d = 0$) for a fixed $s$ and $e$.
Moreover, we write $\tilde{S}_{(0, u)}(d)$ and $\tilde{S}_{(1, u)}(d)$ for $u = 0, 1$ when additionally conditioning on variable $d - 1$ being $0$ or $1$.
Then, by initialising from $\tilde{S}(0) := c$, $\tilde{S}_0(1) = \tilde{S}(0)$ and $\tilde{S}_1(1) = \tilde{S}(0) + b_1 + A_{1, 1}$, the following two-stage recursion holds for $d = 2, \ldots, p$:
\begin{equation}
  \begin{split}
      \tilde{S}_{(0, u)}(d) &= \tilde{S}_{u}(d - 1), \\
      \tilde{S}_{(1, u)}(d) &= \tilde{S}_u(d - 1) + b_d + A_{d, d} + 2u A_{d, d - 1},
  \end{split}
  \label{eq:band1_grow_children}
\end{equation}
for $u = 0, 1$, and
\begin{equation}
  \tilde{S}_{u}(d) = \max\big( \tilde{S}_{(u, 0)}(d), \tilde{S}_{(u, 1)}(d) \big),
  \label{eq:band1_select_parents}
\end{equation}
such that $\max(\tilde{S}_0(p), \tilde{S}_1(p)) = \max_{\bJ}[\tilde{S}(s, e, \bJ) - P(|\bJ|)]$ when $r = 1$.
Note that the computational complexity of finding the optimum in this case is only $O(p)$.

\begin{figure}[t]
\centering
\tikzstyle{fillwhite}=[rectangle,draw,fill=white!10]
\tikzstyle{fillgray}=[rectangle,draw,fill=black!10]
\tikzstyle{fillred}=[rectangle,draw,fill=red!10]
\tikzstyle{filldarkred}=[rectangle,draw,fill=red!40]
\tikzstyle{fillblue}=[rectangle,draw,fill=blue!10]
\tikzstyle{filldarkblue}=[rectangle,draw,fill=blue!40]

\begin{tikzpicture}
\begin{scope}[every node/.style={rectangle}]
    \node (A) at (2,4) [fillgray] {$\tilde{S}(0)$};
    \node (B) at (0,3) [filldarkred] {$\tilde{S}_{0}(1)$};
    \node (C) at (4,3) [filldarkblue] {$\tilde{S}_{1}(1)$};
    \node (D) at (-1,2) [fillred] {$\tilde{S}_{00}(2)$};
    \node (E) at (1,2) [fillblue] {$\tilde{S}_{10}(2)$};
    \node (F) at (3,2) [filldarkred] {$\tilde{S}_{01}(2)$} ;
    \node (G) at (5,2) [filldarkblue] {$\tilde{S}_{11}(2)$} ;
    \node (H) at (-1,0.5) [fillred] {$\tilde{S}_{00}(3)$};
    \node (I) at (1,0.5) [filldarkblue] {$\tilde{S}_{10}(3)$};
    \node (J) at (3,0.5) [filldarkred] {$\tilde{S}_{01}(3)$} ;
    \node (K) at (5,0.5) [fillblue] {$\tilde{S}_{11}(3)$} ;
    \node (L) at (-1,-1) [filldarkred] {$\tilde{S}_{00}(4)$};
    \node (M) at (1,-1) [fillblue] {$\tilde{S}_{10}(4)$};
    \node (N) at (3,-1) [fillred] {$\tilde{S}_{01}(4)$} ;
    \node (O) at (5,-1) [filldarkblue] {$\tilde{S}_{11}(4)$} ;
    \node (P) at (-1,-2.5) [filldarkred] {$\tilde{S}_{00}(p)$};
    \node (Q) at (1,-2.5) [fillblue] {$\tilde{S}_{10}(p)$};
    \node (R) at (3,-2.5) [fillred] {$\tilde{S}_{01}(p)$} ;
    \node (S) at (5,-2.5) [fillblue] {$\tilde{S}_{11}(p)$} ;
\end{scope}

\begin{scope}[>={Stealth[black]},
              every edge/.style={draw=black,thick}]
    \path [->] (A) edge node {} (B);
    \path [->] (A) edge node {} (C);
    \path [->] (B) edge node {} (D);
    \path [->] (B) edge node {} (E);
    \path [->] (C) edge node {} (F);
    \path [->] (C) edge node {} (G);
    \path [->] (F) edge node {} (H);
    \path [->] (F) edge node {} (I);
    \path [->] (G) edge node {} (J);
    \path [->] (G) edge node {} (K);
    \path [->] (J) edge node {} (L);
    \path [->] (J) edge node {} (M);
    \path [->] (I) edge node {} (N);
    \path [->] (I) edge node {} (O);
    \path [->] (O) edge [draw=black,dotted] node {} (R);
    \path [->] (O) edge [draw=black,dotted] node {} (S);
    \path [->] (L) edge [draw=black,dotted] node {} (P);
    \path [->] (L) edge [draw=black,dotted] node {} (Q);
\end{scope}
\end{tikzpicture}
\caption{The unbalanced binary tree structure of the dynamic program for solving \eqref{eq:approx_optimal} for 1-banded $\bQ$ and fictitious data. The blue and red nodes refer to conditioning on whether the current variable/level $d$ is anomalous ($u_d = 1$) or not anomalous ($u_d = 0$), respectively. At each level, the darker coloured nodes are the selected parent within every colour group, while the edges correspond to the step of growing children nodes. Observe that the maximum value to the BQP in this example is $\tilde{S}_{00}(p)$, with "position" $\bu = (1, 1, 0, 0, \ldots, 0)$.}
\label{fig:simple_tree}
\end{figure}
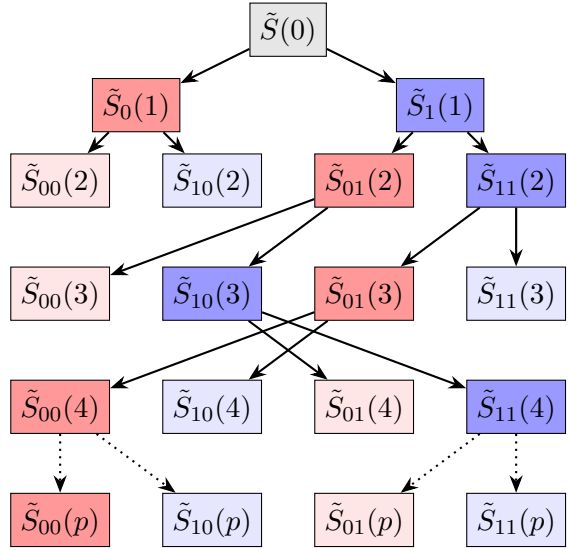

To extend the recursion to more general precision matrices, observe that the dynamic program given by \eqref{eq:band1_grow_children} and \eqref{eq:band1_select_parents} can be described by an unbalanced binary tree (Figure \ref{fig:simple_tree}).
Initialisation occurs at levels 0 and 1 of the tree.
Thereafter, two selected nodes at level $d - 1$ grow children nodes according to \eqref{eq:band1_grow_children}, before two of the four nodes at level $d$ are selected as parents for the next level by the $\max$ operation in \eqref{eq:band1_select_parents}.
The path from the maximum node at the final level back to the root encodes the optimal $\bu$.
In the following, we will refer to the vector of $0$'s and $1$'s along the path from a certain node back up to the root as the "position" of a node.

\begin{figure}[t]
  \centering
  \includegraphics[scale = 0.04] {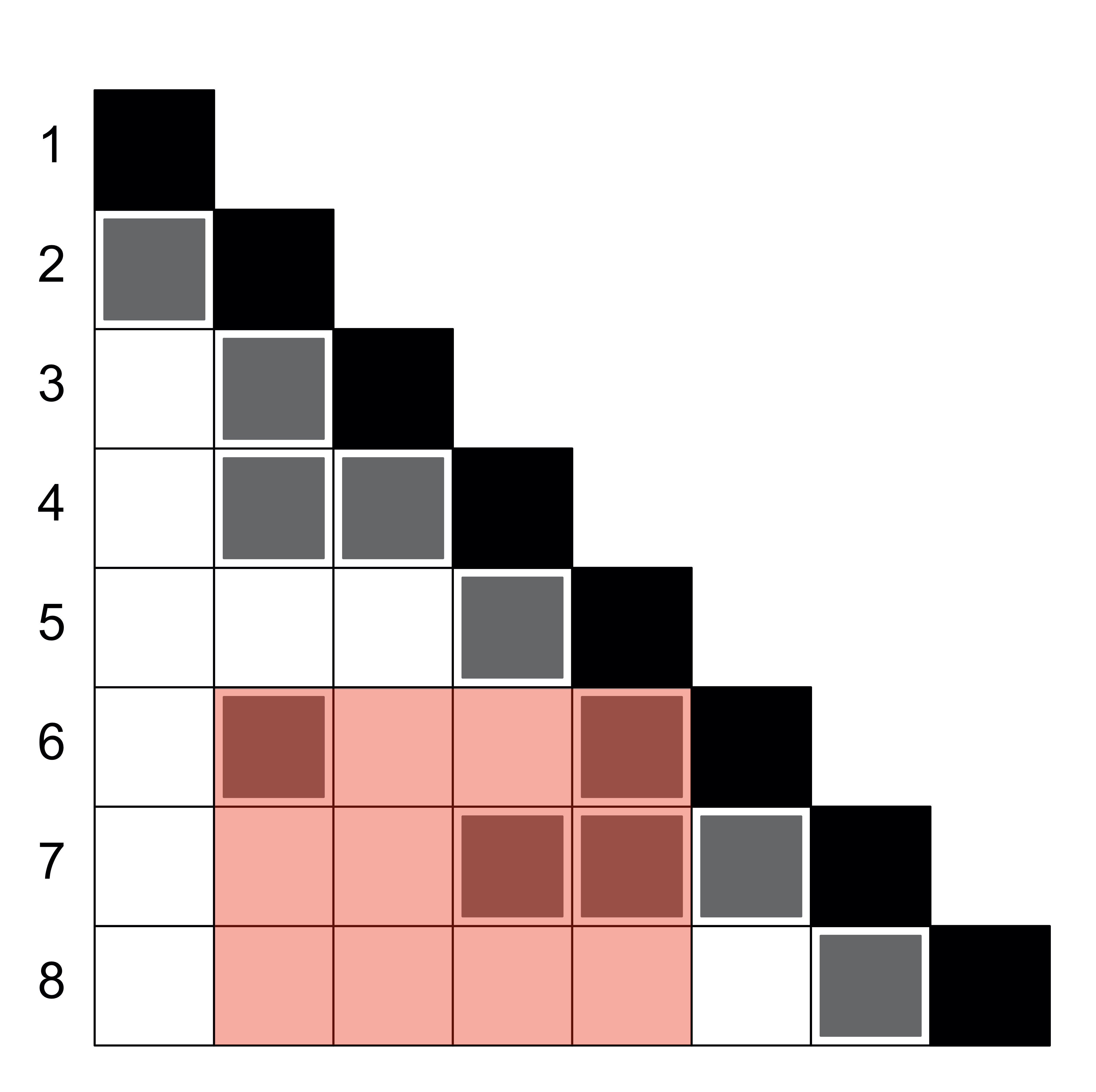}
  \caption{An example $4$-banded $\bA$ matrix where the diagonal is black, other non-zero elements are gray, and zero-elements are white.
  The red region illustrates how the extended neighbours of $d = 6$ are found;
  the column indices of the red region correspond to $P_6^< = \{2, 3, 4, 5\}$, but variable $3$ can be excluded as it is not in any of the coming neighbourhoods, making ${M_6 = \{ 2, 4, 5 \}}$.
  The other extended neighbourhoods in this example are $M_1 = \emptyset$, $M_2 = \{ 1 \}$, $M_3 = \{ 2 \}$, $M_4 = \{ 2, 3 \}$, $M_5 = \{ 2, 4 \}$, $M_7 = \{ 4, 5, 6 \}$ and $M_8 = \{7\}$.}
  \label{fig:extended_nbs}
\end{figure}

By using the tree description, it is easier to generalise the algorithm to any neighbourhood structure of each variable $d$.
When $r = 1$, we only have to consider the two options of variable $d - 1$ being $0$ or $1$ at every step $d$, whereas for a general band, we have to consider all combinations of variables $d - r, \ldots, d - 1$ being $0$ or $1$.
A further adaptation to the precision matrix at hand can be made by excluding those variables among $d - r, \ldots, d - 1$ that will never be visited again, at each step $d$.
To be precise, let us define the neighbours of variable $d$ by $N_d := \{i : A_{d, i} \not= 0 \}$,
and the potential lower neighbours of $d$ by $P_d^< := \{ \max(1, d - r), \ldots, d - 1 \}$ for $d \geq 2$ and $P_1^< := \emptyset$.
At each step $d$, we have to condition on all $0$-$1$-combinations of the variables in
\begin{equation}
M_d := P_d^< \setminus \big(\cup_{i = d}^{d + r} N_i\big)^\comp = P_d^< \cap \big(\cup_{i = d}^{d + r} N_i\big).
\label{eq:extended_nb}
\end{equation}
We call the variables in $M_d$ the \textit{extended neighbours} of $d$.
See Figure \ref{fig:extended_nbs} for an example of how the $M_d$'s are constructed.

To accomodate more complicated neighbourhood structures,
we have to extend the scalar indicators $u$ needed when $r = 1$, to vector indicators $\bu_d \in \{ 0, 1 \}^{|M_d|}$ that give us the position of a node in the tree relative to $M_d$.
I.e., $\bu_d$ tells us which extended neighbours of $d$ are on ($1$) or off ($0$).
At each level $d$, all $2^{|M_d|}$ possible on-off-combinations must be conditioned on, resulting in $2^{|M_d| + 1}$ recursive updates, given by
\begin{equation}
  \begin{split}
    \tilde{S}_{(0, \bu_d)}(d) &= \tilde{S}_{\bu_d}(d - 1) \\
    \tilde{S}_{(1, \bu_d)}(d) &= \tilde{S}_{\bu_d}(d - 1) + b_d + A_{d, d} + 2 \bu_d^\tp \bA_{d, M_d},
  \end{split}
  \label{eq:grow_children}
\end{equation}
where $(0, \bu_d)$ and $(1, \bu_d)$ indicates the positions of the 0-child and 1-child nodes relative to $M_d$.
All these children nodes constitute the nodes at level $d$, and we will refer to them as $\{ \tilde{S}(d) \}$.

The parent-selecting step in the general case also becomes more complex since the extended neighbourhoods can evolve in many different ways.
To explain this step in detail, we use the notation $\text{position}(\tilde{S}(d))$ to refer to the $0$-$1$-vector that gives the position of a given node in our binary tree representation of the algorithm.
For example, $\text{position}(\tilde{S}_{10}(4)) = (1, 1, 0, 1)$ in Figure \ref{fig:simple_tree}.
Now the parent for each $\bu_d$ are determined by maximising over the variables that will never be visited again;
\begin{equation}
  \tilde{S}_{\bu_d}(d - 1) = \max_{\bv \in \bV}\; \tilde{S}_\bv(d - 1),
  \label{eq:select_parents}
\end{equation}
where
$\bV = \{ \bv \in \text{positions}(\{ \tilde{S}(d - 1) \}) : \bv^{(M_d)} = \bu_d\}$ is the set of positions at level $d - 1$ that match the on-off pattern indicated by $\bu_d$ relative to $M_d$.

The final algorithm is summarised in Algorithm \ref{alg:BQP_solver} and \ref{alg:approx_saving}.
Note that we also keep track of the minimum number of anomalous variables at each level $d$ through the term $\underline{k}$.
In this way, the recursions can be stopped as soon as the anomaly is guaranteed to lie in the dense regime.
For an $r$-banded matrix the computational complexity is bounded by $O\big(\sum_{d = 1}^p2^{|M_d|}\big) \leq O(p2^r)$, and if the anomaly is estimated as dense, the number of operations may be substantially less.

\begin{algorithm}
  \caption{Dynamic programming BQP solver for banded matrices} \label{alg:BQP_solver}
  \begin{algorithmic}[1]
    \INPUT{$\bA$, $\bb$, $c$, $\{ M_d \}_{d = 1}^p$, $k^*$}
    \State $d = 1$, $\underline{k} = 0$, $\tilde{S}(0) = c$.
    \While{$d \leq p$ and $\underline{k} \leq k^*$}
      \For{$\bu_d \in \{0, 1\}^{|M_d|}$}
          \State $\bV = \{ \bv \in \text{positions}(\{ \tilde{S}(d - 1) \}) : \bv^{(M_d)} = \bu_d\}$.
          \State $\tilde{S}_{\bu_d}(d - 1) = \max_{\bv \in \bV}\; \tilde{S}_\bv(d - 1)$.
        \State $\tilde{S}_{(0, \bu_d)}(d) = \tilde{S}_{\bu_d}(d - 1)$.
        \State $\tilde{S}_{(1, \bu_d)}(d) = \tilde{S}_{\bu_d}(d - 1) + b_d + A_{d, d} + 2 \bu_d^\tp \bA_{d, M_d}$.
      \EndFor
      \State $\underline{k} = \min_{\bv \in \text{positions}\{ \tilde{S}(d) \}} \bv^\tp \mathbf{1}$.
      \State $d = d + 1$.
    \EndWhile
    \State $\tilde{\bJ} = \argmax \{ \tilde{S}(p) \}$.
    \State $\tilde{S} = \max \{\tilde{S}(p)\}$.
    \State \textbf{return:} $\tilde{S}$, $\tilde{\bJ}$.
  \end{algorithmic}
\end{algorithm}

\begin{algorithm}
  \caption{The approximate penalised saving for anomaly detection used in \mvcapacor}
  \label{alg:approx_saving}
  \begin{algorithmic}[1]
    \INPUT{$\barbx$, $\bQ$, $\{ M_d \}_{d = 1}^p$, $\beta$, $\alpha_\text{sparse}$, $\alpha_\text{dense}$, $k^*$, $e$, $s$.}
    \State $\bA = -(e - s)\barbx \barbx^\tp \circ \bQ$.
    \State $\bb = 2(e - s)(\barbx \circ \bQ\barbx) - \beta$.
    \State $c = -\alpha_\text{sparse}$
    \State $\tilde{S}$, $\tilde{\bJ}$ from Algorithm \ref{alg:BQP_solver} with input ($\bA$, $\bb$, $c$, $\{ M_d \}_{d = 1}^p$, $k^*$)
    \State $S = S(s, e, [p]) - \alpha_\text{dense}$.
    \State \textbf{if} $\tilde{S} \geq S$ \textbf{return:} $\tilde{S}, \tilde{\bJ}$.
    \State \textbf{else} \textbf{return:} $S$, $[p]$.
  \end{algorithmic}
\end{algorithm}

\subsection{Properties of the approximation} \label{sec:theory_anomaly_approx}
Our main evaluation of the approximation's performance is done through simulations, where in Section \ref{sec:Bapprox_performance} of the Supplementary Material we demonstrate that the approximation and the MLE give almost equal results for low $p$.
Some properties regarding how $\tilde{S}(s, e)$ compares to $S(s, e)$, however, can be derived theoretically.

Firstly, under the dense penalty regime, the approximate MLE is equal to the MLE because the optimal $\bJ$ is $[p]$ in both cases, making $\hatbmu(\bJ) = \barbx$.
Thus, we are only approximating the savings under the sparse penalty regime.

Secondly, $\tilde{S}(s, e) \leq S(s, e)$ for all start- and end-points $s$ and $e$.
This follows by definition of the MLE, which is present in $S(s, e)$; $\hatbmu(\bJ)$ is the minimiser in \eqref{eq:savings}, and consequently, no other estimator can make the saving larger.
Using the approximation will therefore not increase the probability of falsely detecting anomalies.
The only effect it may have is a reduction in power.

In addition to the lower bound of $0$ on the approximation error, Proposition \ref{prop:aMLE_bound} gives an upper bound which is useful for distilling what drives a potential decrease in performance. The proof is given in Section \ref{sec:aMLE_bound_proof} in the Supplementary Material.

\begin{myprop} \label{prop:aMLE_bound}
  Let $\bW(\bJ)$ be the matrix where $\bW(\bJ)_{\bJ, -\bJ} = \bQ_{\bJ, \bJ}^{-1} \bQ_{\bJ, -\bJ}$ and is $0$ elsewhere, and ${\hat{\bJ} = \argmax_\bJ \big[S(s, e, \bJ) - P(|\bJ|)\big]}$. Then the following bound on the approximation error holds for all $s < e$:
  \begin{equation}
    0 \leq S(s, e) - \tilde{S}(s, e) \leq (e - s)\lambda_{\max} \big(\bQ\bW(\hat{\bJ})\big) \|\barbx_{(s + 1):e}(\hat{\bJ}^\comp)\|^2.
    \label{eq:aMLE_bound}
  \end{equation}
\end{myprop}

The right-hand side of \eqref{eq:aMLE_bound} suggests that the relative approximation error will be largest for sparse anomalies in strongly correlated data---as this is the situation when $\|\barbx_{(s + 1):e}(\hat{\bJ}^\comp)\|^2$ is largest (see Section \ref{sec:aMLE_bound_proof} in the Supplementary Material).
The simulation results in Section \ref{sec:Bapprox_performance} of the Supplementary Material support this conclusion that the greatest difference in performance occurs when there is a sparse anomaly in strongly correlated data, although the difference is small in the tested settings.

\subsection{Multiple point and collective anomalies} \label{sec:multiple_anomalies}
We can extend the described method for detecting a single collective anomaly to detecting multiple collective anomalies, and also to allow for point anomalies within the baseline segments. To incorporate point anomalies, we follow the approach of \citet{fisch_linear_2019, fisch_subset_2019} by defining point anomalies as collective anomalies of length $1$.
Thus, the optimal approximate saving of a point anomaly at time $t$ can be defined as
\begin{equation}
  \tilde{S}'(t) = \max_\bJ \big[ \tilde{S}(t, t, \bJ) - \beta'|\bJ| \big].
\end{equation}
In accordance with \citet{fisch_subset_2019}, we set $\beta' = 2\log p + 2\psi$, where $\psi = \log n$ as in Section \ref{sec:single_anomaly}.
As for the collective anomaly penalty function, $\beta'$ can be scaled by a constant factor $b'$ to achieve appropriate error control.

We can now extend our penalised likelihood framework. The estimates for the collective anomalies, $\tilde{K}$ and $(\tilde{s}_k, \tilde{e}_k, \tilde{\bJ}_k)$ for $k = 1, \ldots, \tilde{K}$, and point anomalies, $\tilde{O}$ and $\tilde{\bJ}_t$ for $t \in \tilde{O}$, can then be obtained by minimising the penalised cost
\begin{equation}
  \max_{K \in [\lfloor n/l \rfloor], s_k, e_k} \sum_{k = 1}^K \tilde{S}(s_k, e_k) + \max_{O \subseteq [n]} \sum_{t \in O} \tilde{S}'(t),
  \label{eq:point_and_collective_anomalies_objective}
\end{equation}
subject to $\tilde{e}_k - \tilde{s}_k \geq l \geq 2$, $\tilde{e}_k \leq \tilde{s}_{k + 1}$ and $(\cup_k[\tilde{s}_k + 1, \tilde{e}_k])\cap \tilde{O} = \emptyset$.

The optimisation problem \eqref{eq:point_and_collective_anomalies_objective} can be solved exactly by a pruned dynamic program, using ideas from the PELT algorithm of \citet{killick_optimal_2012}.
Defining $C(m)$ as the maximal penalised approximate savings for observations $\bx_{1:m}$, the basis for our PELT algorithm is the following recursive relationship:
\begin{equation}
  C(m) = \max\left( C(m - 1), \underset{0 \leq t \leq m - l}{\max} \Big[C(t) + \tilde{S}(t, m) \Big], C(m - 1) + \tilde{S}'(t) \right),
  \label{eq:OP_anomalies}
\end{equation}
for $C(0) = 0$.
The first term in the outer maximum corresponds to no anomaly at $m$, the second term to a collective anomaly ending at $m$, and the third term to a point anomaly at $m$.

The computationally costly part of $\eqref{eq:OP_anomalies}$ is the maximisation over all possible starting-points $t$ in the term for collective anomalies.
Due to this term, the runtime of this dynamic program scales quadratically in $n$.
If one specifies a maximum segment length $M$, however, the runtime is reduced to $O(Mn)$ at the risk of missing collective anomalies that are longer than $M$.
The PELT algorithm is able to prune those $t$'s in the term for the collective anomalies that can never be the maximisers, thus reducing computational cost whilst maintaining exactness.
Proposition \ref{prop:pruning} gives a condition for when $t$ can be pruned.
The proof is given in the Supplementary Material.

\begin{myprop} \label{prop:pruning}
  If there exists an $m \geq t - l$ such that
  \begin{equation}
    C(t) + \tilde{S}(t, m) + \alpha_\text{dense} \leq C(m)
    \label{eq:pruning}
  \end{equation}
  then, for all $m' \geq m + l$, $C(m') \geq C(t) + \tilde{S}(t, m')$.
\end{myprop}

Proposition \ref{prop:pruning} states that if $\eqref{eq:pruning}$ is true for some $m \geq t - l$, $t$ can never be the optimal  start-point of an anomaly for future times $m' \geq m + l$, and can therefore be skipped in the dynamic program.
\citet{killick_optimal_2012} show that if the number of changepoints increases linearly in $n$, then such a pruned dynamic program can scale linearly.
In the worst case of no changepoints, however, the scaling is still quadratic in $n$.

Calculating $C(n)$ in \eqref{eq:OP_anomalies} by PELT with savings computed from Algorithm \ref{alg:approx_saving} constitutes our \mvcapacor\ algorithm.



\section{Relation to general changepoint detection} \label{sec:changepoints}
So far, we have considered the anomaly detection problem, which is a special case of the changepoint detection problem.
In the changepoint model, the changing parameter can change freely at every changepoint.
The anomaly model restricts the changepoint model by assuming there is a (known) baseline parameter the data reverts to at every other changepoint.
In terms of the anomaly model \eqref{eq:mean_model}, the changepoint model is given by setting $e_k = s_{k + 1}$ and $e_K = n$ and assuming all mean vectors to be unknown, including $\bmu_0$.
In this section, we highlight the benefits of making a distinction between changes and anomalies in light of our application, and briefly describe how our method can be adapted to changepoint detection in general.

In our application of condition monitoring a process pump, as well as in other anomaly detection applications, the aim is to classify observations as either conforming to some baseline behaviour or being anomalous.
Moreover, the majority of observations belong to the baseline group (or else the anomalies would not really be anomalies), and the remaining, anomalous observations may have any location and grouping within the data set (see Figure \ref{fig:raw_data}).
The anomaly model adapts the changepoint model to this setting by assuming that the baseline behaviour is characterised by a common stationary distribution, and that each anomaly is characterised by two (unknown) changepoints---one change from the baseline distribution to some other distribution, and another change back to the baseline.
In this way, the model clearly distinguishes between which segments are in line with the baseline and those that are anomalous.
In a general changepoint model with $K$ changepoints, on the other hand, observations are classified into $K + 1$ distinct segments.
These segments would subsequently have to be labelled as either baseline or anomalous by some additional rule if anomaly detection was the aim.
In addition, the anomaly model enables borrowing strength across the entire data set for estimating the baseline distribution, rather than separately estimating each parameter between each anomaly, increasing the power of detecting anomalies. 

%
%


If, however, a classical changepoint analysis is of interest, the methodology described in Section \ref{sec:anomalies} can be adapted.
The overarching strategy in the corresponding changepoint problem is to embed a test statistic for a single changepoint within binary segmentation or a related algorithm, such as wild binary segmentation \citep{fryzlewicz_wild_2014} or seeded binary segmentation \citep{kovacs_seeded_2020}.
The test statistic for a single changepoint can be derived in a similar fashion as the test for a single anomaly given in \eqref{eq:aMLE_savings}.
For a detailed derivation, algorithm and simulation results for the changepoint detection problem, see the Supplementary Material, Sections \ref{sec:Achangepoints} and \ref{sec:Bsim_cpt}.

\section{Implementational details} \label{sec:implementation}
\subsection{Robustly estimating the mean and precision matrix} \label{sec:precision_estimation}
To detect anomalies in practice, we need an estimate of $\bQ$ and $\bmu_0$, as they are very rarely known \textit{a priori}.
We will use the median of each series $\bx_{1:n}^{(i)}$ to estimate $\bmu_0^{(i)}$.
To estimate $\bQ$ we use a robust version of the GLASSO algorithm \citep{friedman_sparse_2008}.
This algorithm takes as input an estimate of the covariance matrix, $\hatbSigma$, and an adjacency matrix $\bW$.
An estimate $\hatbQ(\bW)$ of $\bQ$ is then computed by maximising the penalised log-likelihood
\begin{equation}
  \log \det \Theta - \text{tr}(\hatbSigma \Theta) - \| \Gamma \circ \Theta \|_1
  \label{eq:GLASSO}
\end{equation}
over non-negative definite matrices $\Theta$, where we define the entries of $\Gamma$ to be $\gamma_{ij} = 0$ if $w_{ij} = 1$ or $i = j$ and $\gamma_{ij} = \infty$ otherwise.
This can be seen as producing the closest estimate of $\bQ$ based on $\hatbSigma^{-1}$ subject to the sparsity pattern imposed by $\bW$.
To compute $\hatbQ$ efficiently, we use the R package \texttt{glassoFast} \citep{sustik_glassofast_2012}. 

As input for $\hatbSigma$ we use an estimate, $\bS$, of the covariance in the raw data that is robust to the presence of anomalies.
Our robust estimator is constructed from the Gaussian rank correlation and the median absolute deviation estimator of the standard deviation, as suggested by \citet{ollerer_robust_2015}.
To be precise, let $\text{mad}(\bx^{(i)})$ be the median absolute deviation of all measurements of variable $i$, and
\begin{equation}
  r_\text{Gauss}(\bx^{(i)}, \bx^{(j)}) := r\left(
  \Phi^{-1}\left(R\big(\bx^{(i)}\big)/(n + 1)\right),
  \Phi^{-1}\left(R\big(\bx^{(j)}\big)/(n + 1)\right)
  \right)
\end{equation}
be the Gaussian rank correlation between variables $i$ and $j$, where $r$ is the sample Pearson correlation, and $R\big(\bx\big)$ is a vector of the ranks of each $x_t$ within $\bx$.
Then the robust pairwise covariances are estimated by
\begin{equation}
  s_{ij} = \text{mad}\big(\bx^{(i)}_{1:n}\big)\text{mad}\big(\bx^{(j)}_{1:n}\big) r_\text{Gauss}\big(\bx^{(i)}_{1:n}, \bx^{(j)}_{1:n}\big).
  \label{eq:robust_cov_mat}
\end{equation}

\subsection{Tuning} \label{sec:tuning}
There are two primary tuning parameters in \mvcapacor: The adjacency matrix $\bW$ in the precision matrix estimator of Section \ref{sec:precision_estimation}, and the scaling factor $b$ in the penalty function for collective anomalies.
This section contains guidelines for tuning them after some notes on the remaining tuning parameters.

The scaling factor for the point anomaly penalty, $b'$, can be tuned separately from $b$ if the application dictates it, but a reasonable default is to let $b' = b$ and tune both penalties simultaneously.
The minimum and maximum segment lengths of collective anomalies, $l$ and $M$, are tuning parameters solely for the convenience of speeding up computation if there is knowledge of such limits in an application.
Otherwise, they default to $l = 2$ and $M = n$.

A number of different considerations can go into choosing $\bW$.
From a modelling perspective, selecting $\bW$ corresponds to deciding on a model for the conditional independence structure; $w_{ij} = 0$ means variables are assumed to be conditionally independent, while $w_{ij} = 1$ means variables are conditionally dependent.
For spatial data, for example, the choice of $\bW$ is the same as choosing the neighbourhood structure in a conditional autoregressive model, where $w_{ij} = 1$ if and only if spatial region $i$ is a neighbour of spatial region $j$.
In our process pump example this would mean specifying which sensors are neighbours.

Computational considerations can also guide the choice of $\bW$, however.
As we have seen, \mvcapacor\ scale exponentially in the band of $\bQ$.
Hence, the band of $\bW$ governs the run-time of our algorithms to a large extent.
A reasonable default choice of $\bW$ is therefore a low value of $r$ in the $r$-banded adjacency matrix $\bW(r)$, defined by
\begin{equation}
w_{ij} =
\begin{cases}
  1 & \text{if } 0 <|i - j| \leq r, \\
  0 & \text{otherwise}.
\end{cases}
\label{eq:banded_adjmat}
\end{equation}
In the simulations of the next section, we illustrate that good performance can be achieved even when specifying $\bW$ to have a much narrower band than the true $\bQ$.

In cases where the precision matrix is sparse but not banded, bandwidth reduction algorithms such as the Cuthill-McKee algorithm \citep{cuthill_reducing_1969} and the Gibbs-Poole-Stockmeyer algorithm \citep{lewis_algorithm_1982} can be a useful pre-processing step before running \mvcapacor.

Several strategies can also be employed for tuning the penalty scaling factor $b$.
The first strategy requires a training set containing only baseline observations.
This training set can either be used to estimate a model (e.g. Gaussian) of the baseline behaviour of the data, or constitute the empirical distribution of the baseline data.
Anomaly-free data sets can then be sampled parametrically or non-parametrically from the baseline model to obtain bootstrap estimates $\hat{\alpha}$ of $\alpha = P(\hat{K} > 0 | K = 0)$ for a fixed $b$.
A practitioner can thus select a target probability of false positives $\alpha$, and find $b$ that meets this criterion within a selected interval of error, $\alpha \pm \delta$, and level of confidence governed by the number of repetitions used to calculate $\hat{\alpha}$ per $b$.

A second criterion is to find the smallest $b$ such that a user-selected tolerable number of false alarms is raised in the training set.
This strategy is much less computationally intensive as it avoids the bootstrap sampling, but the error control hinges more strongly on how generalisable the training set is.

If there is no training set available, a third tuning strategy is to adjust $b$ until a desired number of anomalies are output by \mvcapacor.
As $b$ is increased, the ordering of the anomalies in terms of significance will gradually be revealed.
We explore the pump data set by this tuning strategy in Section \ref{sec:real_data}.


\section{Simulation study} \label{sec:simulations}
We next turn to examine the power and estimation accuracy of \mvcapacor\ in a range of data settings.
In almost all cases, we test the robustness of the method against an incorrectly specified adjacency matrix in the precision matrix estimate.
We concentrate on the single anomaly setting first, before comparing several state-of-the-art methods in the multiple anomaly setting.

We have chosen a widely used one-parameter version of the \textit{conditional autoregressive} (CAR) model, called the \textit{row-standardised} CAR model, as our primary testbed (see for instance \citet{ver_hoef_relationship_2018} for a concise introduction).
This CAR model is given by
\begin{equation}
    \bQ_\text{CAR}(\rho, \bW) := \diag(\bW \bone) - \rho\bW,
    \label{eq:CAR}
\end{equation}
where $\bW$ is an adjacency matrix as before.
$\bQ_\text{CAR}$ is then standardised so that $\bQ^{-1}$ becomes a correlation matrix, and we let $\bmu_0 = \mathbf{0}$ throughout.
Conveniently, the sparsity structure of $\bQ_\text{CAR}$ follows directly from the design of $\bW$.
In our simulations, we consider data with precision matrices corresponding to the $r$-banded neighbourhood structures given in \eqref{eq:banded_adjmat} and regular lattice neighbourhood structures.
To define the $m \times m$ lattice adjacency matrix, let $(u, v)$ denote the coordinate of a node in the lattice, for $0 \leq u, v \leq m$.
The neighbourhood of $(u, v)$ is considered to be $\{(u - 1, v), (u + 1, v), (u, v - 1), (u, v + 1) \}$.
Coordinates are then enumerated by $i = (u - 1)m + v$, such that the square lattice adjacency matrix $\bW_\text{lat}$ can be defined by $w_{ij} = 1$ if $i$ and $j$ are neighbours and $0$ otherwise.
For the sake of brevity, we also define $\bQ_\text{lat}(\rho) := \bQ_\text{CAR}(\rho, \bW_\text{lat})$ and $\bQ(\rho, r):= \bQ_\text{CAR}(\rho, \bW(r))$.
In addition to the CAR models, we also test performance under the constant correlation model, given by
\begin{equation}
  \bQ_\text{con}(\rho) := (\rho\bone\bone^\tp + (1 - \rho)\bI)^{-1}.
  \label{eq:constant_cor_model}
\end{equation}
Note that we use $\bW^*$ to refer to the true adjacency matrix of the data.

If more than one series changes, the power of different methods may depend on how similarly each series change.
To investigate this, we consider the following ways of simulating anomalous means, $\bmu_k$, $k = 1, \ldots, K$:
$\bmu_k^{(\bJ_k)} \sim N(\mathbf{0}, \bSigma_{\bJ_k, \bJ_k})$, where $\bSigma$ is the data covariance matrix,
and $\bmu_k^{(\bJ_k)} \sim N(\mathbf{0}, (\bQ_\text{con}(\rho))^{-1})$.
We refer to anomalies being drawn from the former and latter classes, respectively, by $\bmu_{(\bSigma)}$ and $\bmu_{(\rho)}$. Note that $\rho = 0$ and $\rho = 1$ correspond to the special cases of the means being independent and equal for the changing variables, respectively.
After sampling a mean vector, it is scaled by a constant to achieve a specific signal strength $\vartheta_k := \| \bmu_k - \bmu_{0} \|_2 = \| \bmu_k \|_2$.
Moreover, unless stated otherwise, we let $\bJ_k = \{ 1, 2, \ldots, J_k \}$, where $J_k \in [p]$ denotes the number of changing variables.

In all simulations, the penalty functions or detection thresholds are tuned to achieve $\alpha = 0.05 \pm 0.02$ probability of false positives in data simulated from the appropriate true null distribution (see Section \ref{sec:tuning}).
1000 and 500 bootstrap repetitions were used for each $b$ to obtain $\hat{\alpha}$ for $p = 10$ and $p = 100$, respectively.

\subsection{Single anomaly detection} \label{sec:sim_anomaly}
To the best of our knowledge, there are no other statistical methods tailored for jointly detecting sparse and dense anomalies in correlated multivariate data.
A comparison between methods for independent multivariate data was performed by \citet{fisch_subset_2019}, where their MVCAPA method was shown to generally outperform other competitors.
Hence, we focus on comparing MVCAPA against a range of \mvcapacor scenarios, including various incorrectly specified versions, exploring the trade-offs between the two methods. We evaluate methods in terms of power to detect an anomaly of increasing signal strength, and also assess the correctness of the estimated subset of anomalous variables, $\bJ$.

In the following, \enquote{Whiten + MVCAPA} means that the input to MVCAPA are the whitened observations $\bS^{-1/2}\bx_t$, where $\bS$ is the robust covariance matrix estimate \eqref{eq:robust_cov_mat}, whereas a plain \enquote{MVCAPA} takes the raw data $\bx_t$.
Note that Whiten + MVCAPA scrambles the sparsity structure of an anomalous mean such that the recovery of $\bJ$ is lost.
It is, however, still interesting to include in the comparisons of detection power as no sparsity structure has to be imposed on the covariance or precision matrix, in contrast to \mvcapacor.
We therefore expect Whiten + MVCAPA to perform well when the precision matrix as well as the change is dense.

\subsubsection{Independence vs. dependence} \label{sec:single_anom_sim}
As the performance of the anomaly detection methods we consider ultimately hinges on the performance of a test statistic at each pair $(s, e)$, we compare performance assuming that the location of the collective anomaly is known \textit{a priori}.
That is, we fix the collective anomaly at $(s, e) = (n/2, n/2 + 10)$,
and compare the power of $\tilde{S}(s, e)$ with the corresponding test statistic assuming cross-independence used within MVCAPA.
In \mvcapacor, we test using the true precision matrix $\bQ$, an estimate based on the true adjacency structure $\hat{\bQ}(\bW^*)$, as well as misspecified banded adjacency structures with $r = 1, 2, 4$.
The power at each point along the power curve is estimated from $1000$ ($p = 10$) or $500$ ($p = 100$) simulated datasets, and the same datasets were used for all methods.
The full set of tested scenarios include all combinations of $\{ (n, p), \bQ, \rho, J, \bmu_{(\cdot)} \}$ for $(n, p) = (100, 10), (200, 100)$, $\bQ = \bQ(2), \bQ_\text{lat}, \bQ_\text{con}$, $\rho = 0.3, 0.5, 0.7, 0.9, 0.99$, $J = 1, \lfloor\sqrt{p}\rfloor, p$, and change classes $\bmu_{(\bSigma)}$, $\bmu_{(0)}$, $\bmu_{(0.8)}$, $\bmu_{(0.9)}$ and $\bmu_{(1)}$.
In addition, we have also varied which series are anomalous for selected scenarios.
Note that \mvcapacor($\bQ$) represents the performance of an oracle method.
For larger $n$ relative to $p$, however, the difference between \mvcapacor($\bQ$) and \mvcapacor($\hat{\bQ}(\bW^*)$) will decrease.

\begin{figure}[t]
  \centering
  \includegraphics[width=0.85\textwidth] {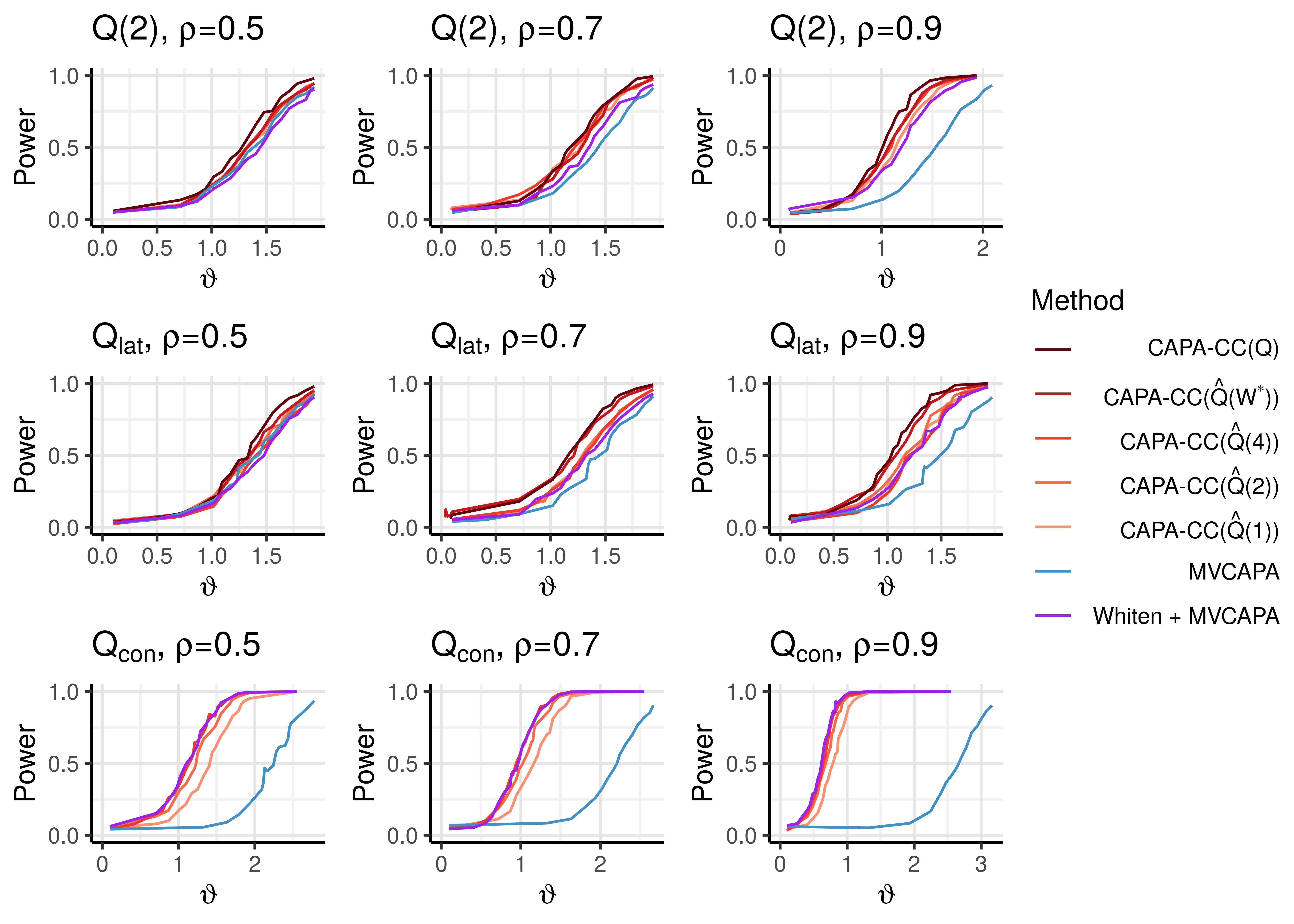}
  \caption{Power curves for correct and misspecified versions of \mvcapacor\ for a single known anomaly at $(s, e) = (100, 110)$ when $J = 1$ and $p = 100$. The existing MVCAPA method for iid variables is marked in blue, the MVCAPA method after whitening the data is marked in purple, and the red colours correspond to versions \mvcapacor.
  A lighter red colour roughly means increasing misspecification of the precision matrix's structure. Results for 2-banded, lattice and globally constant correlation precision matrices are shown from top to bottom, with increasing $\rho$ from left to right. Other parameters: $n = 200$, $\alpha = 0.05$, and 500 repetitions were used during tuning and for each point along the power curves.}
  \label{fig:known_anom_J1}
\end{figure}

A first main finding, illustrated in Figure \ref{fig:known_anom_J1}, is that for detecting a single anomalous variable, incorporating correlations leads to higher power, even when misspecifying the structure of the precision matrix estimate.
The stronger the correlation, the higher the gain in power.
For a collection of densely correlated variables, even using a 1-banded estimate of the precision matrix leads to a big improvement in power for sparse anomalies compared to MVCAPA (the bottom row of plots).
It is somewhat surprising that Whiten + MVCAPA performs comparably to CAPA-CC in this setting of a very sparse change.

\begin{figure}

    \begin{subfigure}{\textwidth}
        \centering
        \includegraphics[width=0.85\linewidth]{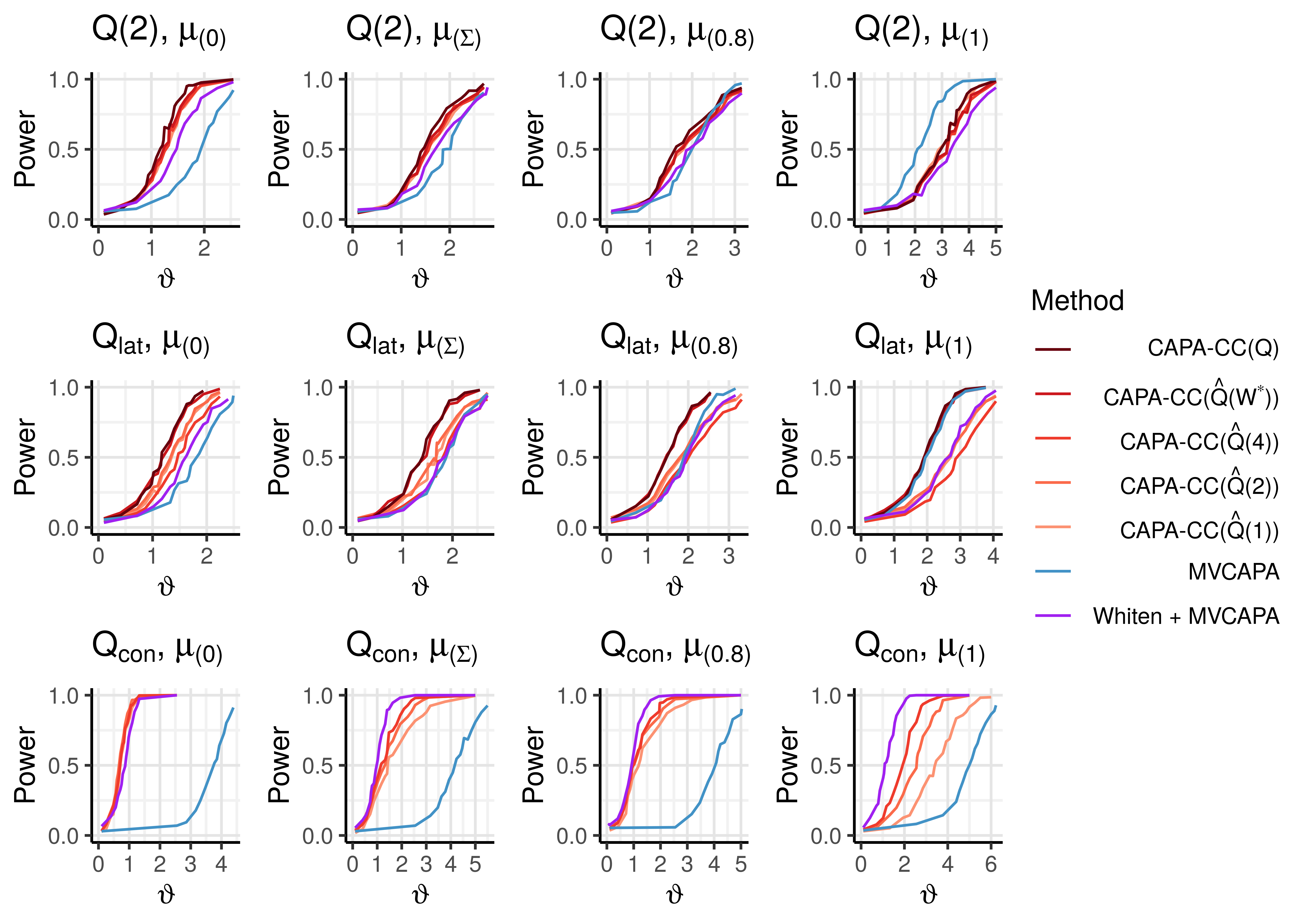}
        \caption{$J = 10$.}
        \label{fig:known_anom_J10}
    \end{subfigure}

    \vskip+2ex

    \begin{subfigure}{\textwidth}
        \centering
        \includegraphics[width=0.85\linewidth]{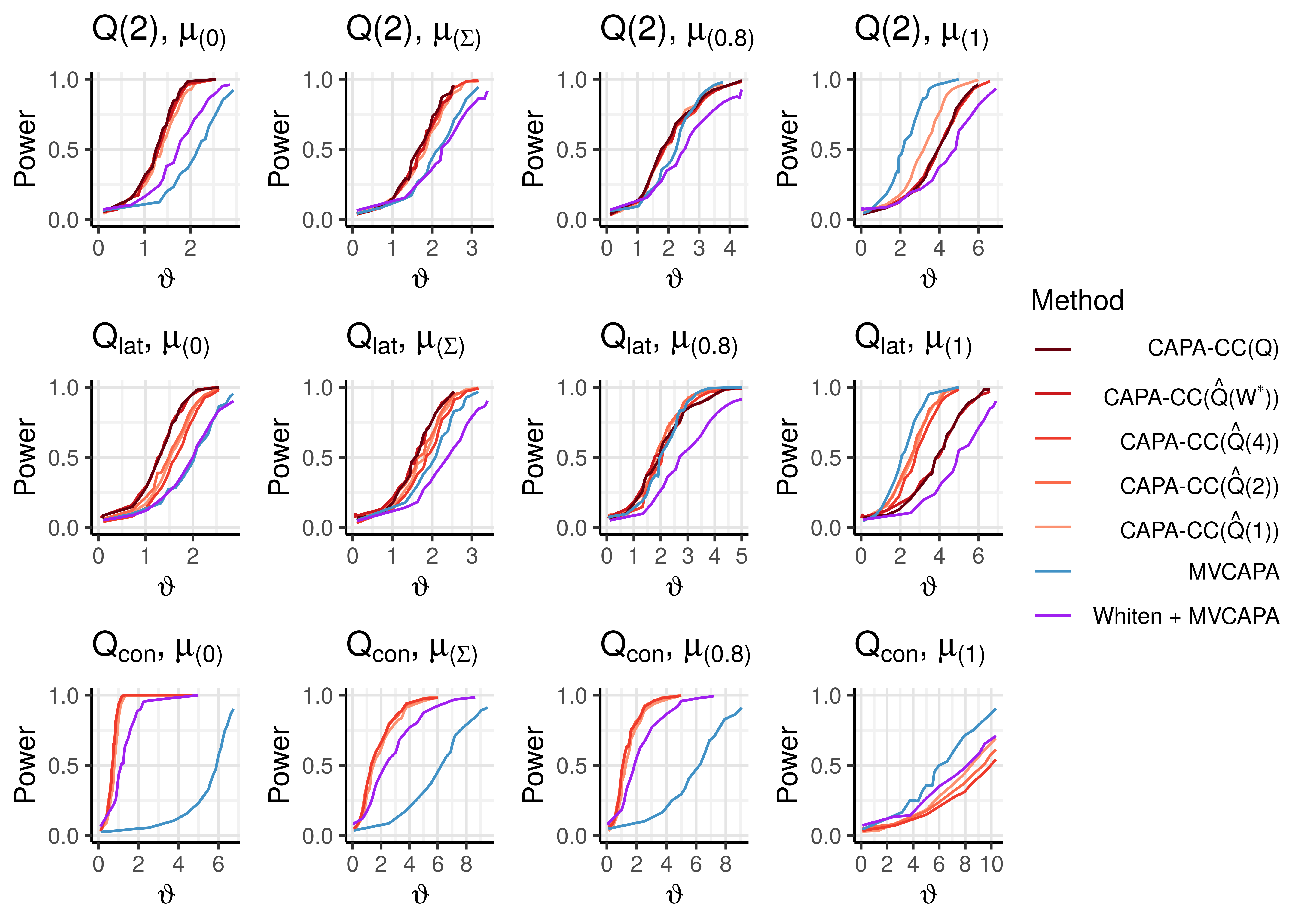}
        \caption{$J = 100$.}
        \label{fig:known_anom_J100}
    \end{subfigure}

    \caption{Power curves for a single known anomaly at $(s, e) = (100, 110)$ and (a) $J = 10$ and (b) $J = 100$, when $p = 100$ and $\rho = 0.9$. The methods are the same as in Figure \ref{fig:known_anom_J1}.
    From left to right, the columns of plots show results for the anomalous means being sampled from $N(\mathbf{0}, \bI)$, $N(\mathbf{0}, \bSigma)$, $N(\mathbf{0}, \bQ_\text{con}^{-1}(0.8))$, and $\mu^{(i)} = \mu$ for all $i \in \bJ$ in the right-most column.
    From top to bottom are results for 2-banded, lattice and global constant correlation data precision matrices.
    Other parameters: $n = 200$, $\alpha = 0.05$, and 500 repetitions per point along the power curves.}
    \label{fig:known_anom_J10_J100}
\end{figure}


The picture for more than one anomalous variable is more complex.
Figure \ref{fig:known_anom_J10_J100} displays the results for different precision matrices and classes of changes for $p = 100$ and $\rho = 0.9$ when (a) $J = 10$ and (b) $J = 100$.
Observe that for all precision matrices and $J$'s (entire first column), \mvcapacor\ is superior for anomalous means sampled from the independent normal distribution ($\bmu_{(0)}$).
This is also the case when the anomalous means are sampled from a normal distribution with the data correlation matrix ($\bmu_{(\bSigma)}$) (entire second column), with the exception of $J = 10$ and global constant correlation.
The power of \mvcapacor\ decreases, however, when the anomalous means have very similar or equal values, as in the case of means being sampled from to $\bmu_{(0.8)}$ and $\bmu_{(1)}$.
Surprisingly, for the special case of equally sized anomalous means and a banded or lattice precision matrix, MVCAPA is more powerful than using the true model for the precision in \mvcapacor($\bQ$).
For $J = 100$, this is also the case for equal changes in the global constant correlation model.
The same phenomenon can be observed for other methods as well (see Section \ref{sec:Achangepoints} in the Supplementary Material), and we discuss it further in Section \ref{sec:discussion}.
As expected, Whiten + MVCAPA performs well for $\bQ_\text{con}$ precision matrices, but the misspecified versions of CAPA-CC outperforms it when $J = 100$.
For low values of $\rho$, we observe almost no difference between the different methods, which is why we focus on $\rho \geq 0.5$.
For higher values of $\rho$ than $0.9$, the gain from incorporating correlations in the method increases.
For $p = 10$, the corresponding results look qualitatively similar.
See Section \ref{sec:Banomaly_detection} of the Supplementary Material for more details.

\subsubsection{Variable selection} \label{sec:variable_selection}
Although \mvcapacor\ is not designed to estimate $\bJ$ consistently, it is worth investigating the behaviour of $\hat{\bJ}$ so that it is interpreted with sufficient caution.
Note that we now use $\hat{\bJ}$ to refer to the output estimate of $\bJ$ for all algorithms.
Also recall that we let $J := |\bJ|$ and $\hat{J} := |\hat{\bJ}|$.

For $p = 10$ and $100$, the precision and recall of $\hat{\bJ}$ from MVCAPA as well as both true and misspecified versions of \mvcapacor\ were compared in the single known anomaly setting, described in Section \ref{sec:single_anom_sim}.
We also included the exact ML method for $p = 10$.
Whiten + MVCAPA is excluded from these simulations since the decorrelation transform breaks up the sparsity structure of the anomalies.

\begin{figure}[t]
  \centering
  \includegraphics[width=0.85\linewidth] {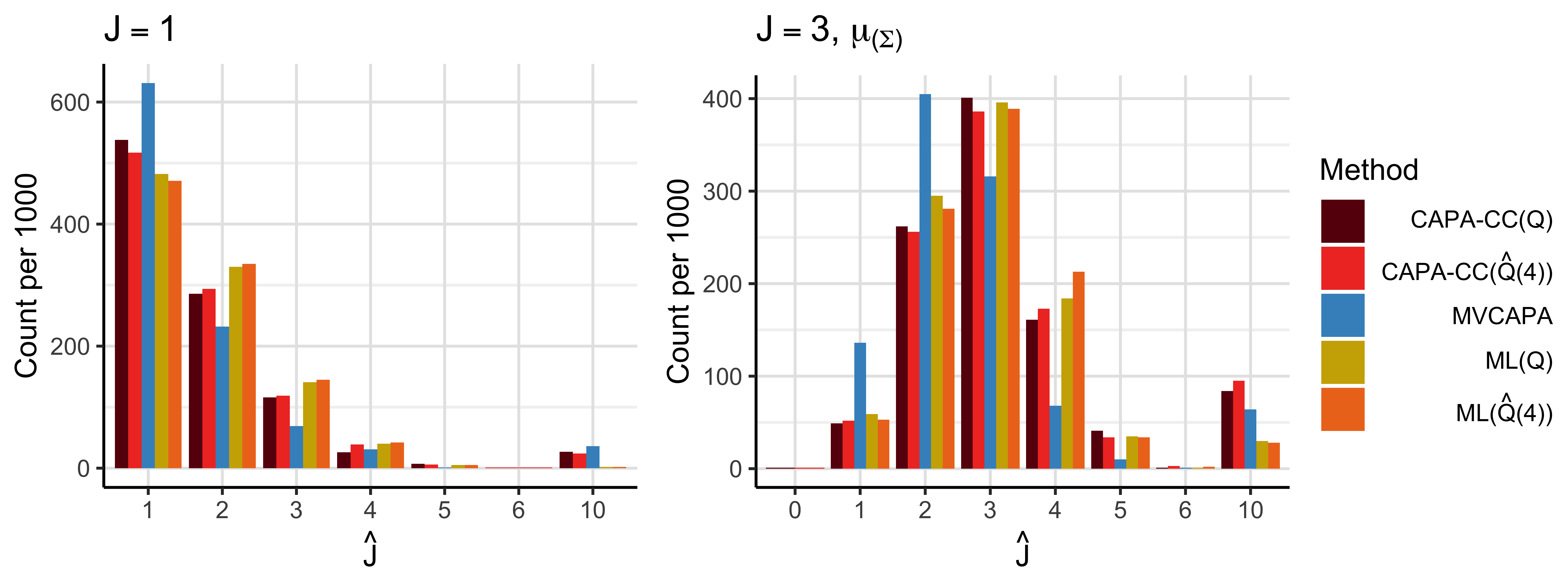}
  \caption{Estimated sizes of $\bJ$ for $\bJ = \{1\}$ (left) and $\bJ = \{1, 2, 3\}$ (right) when $p = 10$ and the location of the anomaly is assumed known.
  Other parameters: $n = 100$, $\bQ = \bQ(2, 0.9)$, $s = 10$, $e = 20$, $\vartheta = 2$, $\bmu_{(\bSigma)}$ , $\alpha = 0.005$.}
  \label{fig:variable_selection_p10}
\end{figure}

Under a 2-banded precision matrix model, we see from Tables \ref{tab:subset_est_p10} and \ref{tab:subset_est_p100}  in the Supplementary Material that both \mvcapacor\ and the exact ML method tend to have higher recall, but slightly lower precision, than MVCAPA.
The reason for this is illustrated in Figure \ref{fig:variable_selection_p10}, where it can be observed that all the methods that incorporate cross-correlations overestimate $J$ more frequently than MVCAPA.
In particular, \mvcapacor\ more often estimates anomalies as dense.
This effect is seen more clearly for $p = 100$ (Figure \ref{fig:variable_selection_p100} in the Supplementary Material), where estimating $J$ becomes increasingly hard as $J$ grows closer to the boundary $k^*$ between sparse and dense changes.
Moreover, we found that the estimated subset is quite sensitive to the scaling of the penalties relative to the signal strength $\vartheta$.
If a more accurate estimate of $\bJ$ is desired, we thus recommend running a post-processing step by optimising the penalised saving for each anomalous segment using only the sparse penalty regime.

\subsection{Multiple anomaly detection} \label{sec:multiple_anom_sim}
The simulation study is concluded by comparing the following methods in a multiple anomaly setting with and without point anomalies:
\begin{itemize}
    \item \mvcapacor\ with a misspecified precision matrix $\hat{\bQ}(4)$.
    \item MVCAPA and Whiten + MVCAPA.
    \item The inspect method of \citet{wang_high_2018}. We test both the version assuming cross-independence implemented in the R-package \texttt{InspectChangepoint} as well as the version including cross-correlations discussed in their paper.
    To distinguish the two versions, we refer to the former as inspect($\bI$) and the latter as inspect($\hat{\bQ}$), where $\hat{\bQ}$ is the inverse of the robust covariance matrix estimator \eqref{eq:robust_cov_mat}.
    \item The group fused LARS method of \citet{bleakley_group_2011}, implemented in the R-package \texttt{jointseg}.
\end{itemize}
In addition, we tested the methods of \citet{wang_localizing_2020} and \citet{safikhani_joint_2020} for detecting changes in vector autoregressive models, but they were excluded due to poor computational scaling in $p$ or $n$.
For example, the method of \citet{wang_localizing_2020} with a maximum segment length of $100$ takes around 13 minutes to complete on a single $p = 10$, $n = 1000$ data set on a typical computer, and the method of \citet{safikhani_joint_2020} scales exponentially in $\hat{K}$. The included methods are all tuned to a specific false positive probability $\alpha$ on data sets of size $\min(n, 200)$, except the group fused LARS, which uses the default model selection procedure of \texttt{jointseg} proposed in \citet{bleakley_group_2011}.
To speed up computation we set the maximum segment length of \mvcapacor\ and MVCAPA to $M = 100$

Performance is measured by the \textit{adjusted rand index} \citep[ARI;][]{hubert_comparing_1985} of classifying observations as either anomalous (point or collective) or baseline.
The ARI measures the accuracy of the classification, but adjusts for the sizes of the classes.
It is therefore suitable in an unbalanced classification problem such as ours.

As inspect and the group fused LARS method are not made specifically for the anomaly setting, as opposed to MVCAPA and \mvcapacor, we do not expect them to be competitive.
However, since they could be used for the purpose, we include them to measure the gain of using a dedicated anomaly detection method rather than a generic changepoint detection method.
Our heuristic for turning the changepoint detection methods into an anomaly classifier is as follows:
If the sample mean of an estimated segment has $L_2$ norm greater than $1$, the observations within the segment are classified as anomalous, and if the $L_2$ norm is smaller than or equal to $1$, they are classified as baseline.
Adjacent segments where both are classified as collective anomalies by this rule are also merged to a single collective anomaly if the sign of $\sum_{j = 1}^p \bar{x}_{(s + 1):e}^{(j)}$ in each of the two segments is the same.

Also note that we use a misspecified precision matrix in \mvcapacor\ since this is most realistic, but improved performance on the order of what can be seen in Figures \ref{fig:known_anom_J1} and \ref{fig:known_anom_J10_J100} could be achieved by selecting the correct model.

\begin{table}[htb]
\centering
\begin{tabular}{ccccccc}
\toprule
$\mathbf{Q}$ & $\rho$ & Pt. anoms & CAPA-CC($\hat{\mathbf{Q}}(4)$) & W + MVCAPA & MVCAPA & inspect($\hat{\mathbf{Q}}$) \\
\midrule
$\mathbf{Q}(2)$ & 0.5 & -- & $\mathbf{0.23}$ & $0.09$ & $0.20$ & $0.05$ \\
$\mathbf{Q}(2)$ & 0.5 & \checkmark & $\mathbf{0.40}$ & $0.25$ & $0.37$ & $0.01$ \\
$\mathbf{Q}(2)$ & 0.7 & -- & $\mathbf{0.34}$ & $0.19$ & $0.12$ & $0.06$ \\
$\mathbf{Q}(2)$ & 0.7 & \checkmark & $\mathbf{0.43}$ & $0.30$ & $0.31$ & $0.00$ \\
$\mathbf{Q}(2)$ & 0.9 & -- & $\mathbf{0.53}$ & $0.43$ & $0.05$ & $0.13$ \\
$\mathbf{Q}(2)$ & 0.9 & \checkmark & $\mathbf{0.61}$ & $0.46$ & $0.26$ & $0.03$ \\
$\mathbf{Q}_\text{lat}$ & 0.5 & -- & $\mathbf{0.21}$ & $0.08$ & $0.12$ & $0.05$ \\
$\mathbf{Q}_\text{lat}$ & 0.5 & \checkmark & $\mathbf{0.29}$ & $0.26$ & $0.25$ & $0.08$ \\
$\mathbf{Q}_\text{lat}$ & 0.7 & -- & $\mathbf{0.27}$ & $0.21$ & $0.13$ & $0.05$ \\
$\mathbf{Q}_\text{lat}$ & 0.7 & \checkmark & $\mathbf{0.35}$ & $0.31$ & $0.25$ & $0.10$ \\
$\mathbf{Q}_\text{lat}$ & 0.9 & -- & $\mathbf{0.34}$ & $0.28$ & $0.09$ & $0.08$ \\
$\mathbf{Q}_\text{lat}$ & 0.9 & \checkmark & $0.33$ & $\mathbf{0.42}$ & $0.18$ & $0.14$ \\
$\mathbf{Q}_\text{con}$ & 0.5 & -- & $0.44$ & $\mathbf{0.52}$ & $0.00$ & $0.06$ \\
$\mathbf{Q}_\text{con}$ & 0.5 & \checkmark & $\mathbf{0.50}$ & $0.49$ & $0.11$ & $0.03$ \\
$\mathbf{Q}_\text{con}$ & 0.7 & -- & $0.60$ & $\mathbf{0.65}$ & $0.00$ & $0.08$ \\
$\mathbf{Q}_\text{con}$ & 0.7 & \checkmark & $\mathbf{0.66}$ & $0.64$ & $0.10$ & $0.04$ \\
$\mathbf{Q}_\text{con}$ & 0.9 & -- & $0.66$ & $\mathbf{0.82}$ & $0.00$ & $0.26$ \\
$\mathbf{Q}_\text{con}$ & 0.9 & \checkmark & $0.71$ & $\mathbf{0.82}$ & $0.09$ & $0.10$ \\
\bottomrule
\end{tabular}
\caption{ARI of classifying baseline and anomalous observations when $p = 100$, $n = 1000$, $(\vartheta_k)_{k = 1}^3 = (1, 2, 3)$, the change class is $\bmu_{(\bSigma)}$, $\{(s_k, e_k)\}_{k = 1}^3 = \{ (300, 330), (600, 620), (900, 910) \}$ and $\bJ_1 = \{1\}$, $\bJ_2 = \{ 1, \ldots, 10 \}$, $\bJ_3 = \{1, \ldots, 10, 46, \ldots, 55, 91, \ldots, 100 \}$, based on 100 repetitions.
Point anomalies are placed at 10 fixed locations, each randomly affecting a single variable with size sampled from $N(0, 4\log p)$.  The largest value for each data setting is given in bold.
Note that the results for inspect($\bI$) and the group fused LARS methods are excluded from the table since their ARIs are approximately 0 in all the tested scenarios.}
\label{tab:ari_p100_vartheta1_shape6}
\end{table}

Table \ref{tab:ari_p100_vartheta1_shape6} displays the results for $p = 100$, $n = 1000$ with three evenly spaced collective anomalies of lengths (30, 20, 10), different affected subsets, affected means sampled from $\bmu_{(\bSigma)}$ and $\vartheta = 1$ in signal strengths of sizes $\vartheta (1, 2, 3)$.
The results are again generally favourable for \mvcapacor($\hat{\bQ}(4)$), in particular for the banded and lattice precision matrices, while Whiten + MVCAPA is slightly better for the global constant correlation matrix when point anomalies are absent.
The group fused LARS and inspect($\bI$) methods achieved approximately $0$ ARI on all the tested scenarios, including the different signal strength parameters of $\vartheta = 1, 1.5, 2$.

The full set of multiple anomaly simulation results, covering anomalous means sampled from $\bmu_{(0)}$ and $\bmu_{(0.8)}$ in addition to $\bmu_{(\bSigma)}$, and $\vartheta = 1.5, 2$ in addition to $\vartheta = 1$, can be found in Section \ref{sec:Bmultiple_anom_sim} of the Supplementary Material.
The results for $\bmu_{(0.8)}$ are very similar to the results for $\bmu_{(\bSigma)}$, and the results for $\bmu_{(0)}$ are slightly more favourable for CAPA-CC compared to the other methods.
As $\vartheta$ increases, the ARI of all methods increase, and the differences in performance decrease.
In the scenarios with point anomalies when $\vartheta = 1.5$ and $\vartheta = 2$, a lot is gained by using \mvcapacor\ or (Whiten +) MVCAPA rather than inspect.

\section{Pump data analysis} \label{sec:real_data}
\begin{figure}[htbp]
  \centering
  \includegraphics[scale = 0.025]{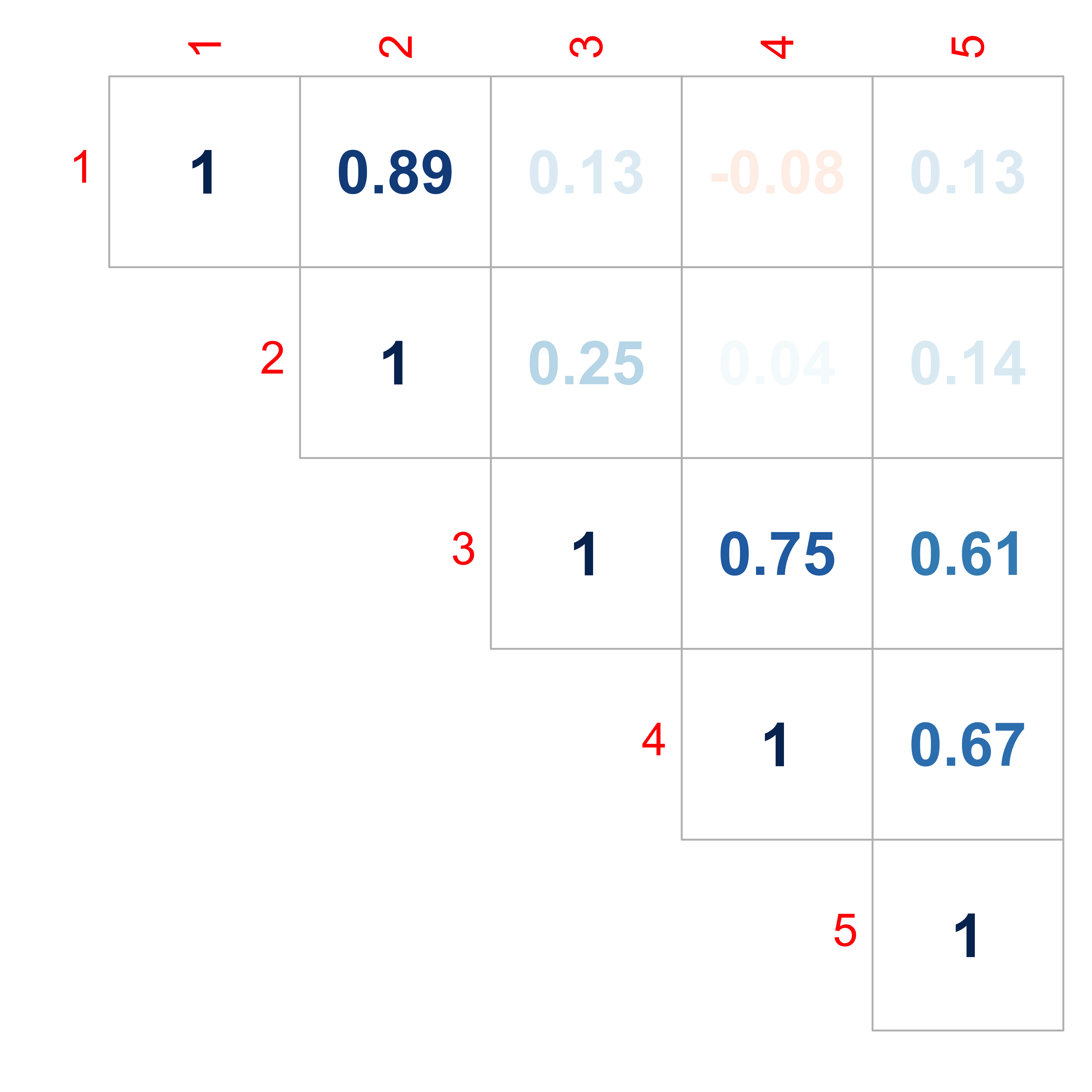}
  \caption{The robustly estimated correlations (see \eqref{eq:robust_cov_mat}) of the pump data after preprocessing.}
  \label{fig:data_cor}
\end{figure}

We now return to the problem of inferring anomalous segments and variables in the pump data described in the introduction.
Recall that the data was preprocessed by regressing a set of monitoring variables onto a set of state variables, such that we are left with five series of residuals to detect anomalies in (Figure \ref{fig:raw_data}).
Some of the residuals are strongly correlated (Figure \ref{fig:data_cor}), suggesting that incorporating cross-correlations when modelling them is advantageous based on our simulation study.

Before running \mvcapacor\ on the pump data, the penalties must be tuned and input parameters selected.
The tuning of the penalties accounts for all features in the data that we have not modelled, e.g. auto-correlation, a non-stationary correlation matrix and trends in the data's mean not associated with segments of suboptimal operation.
As we do not have training data guaranteed to only contain baseline observations, we instead tune the penalties such that a chosen number of the most significant anomalies are output, as discussed in Section \ref{sec:tuning}.
To test performance, we tune $b$ such that the correct number of collective anomalies---four---are output to see how they align with the known ones.
Since there are many outliers in the data set, we want to retain the default level of outlier-robustness, and therefore keep the point anomaly scaling at $1$, while adjusting $b$.
This tuning procedure resulted in a scaling factor of $b = 11$.
For the remaining inputs, we set $\bQ$ to the inverse of the correlation matrix in Figure \ref{fig:data_cor}, a minimum segment length $l = 5$, and use no maximum segment length.

\begin{figure}[t]
  \centering
  \includegraphics[width=0.85\textwidth]{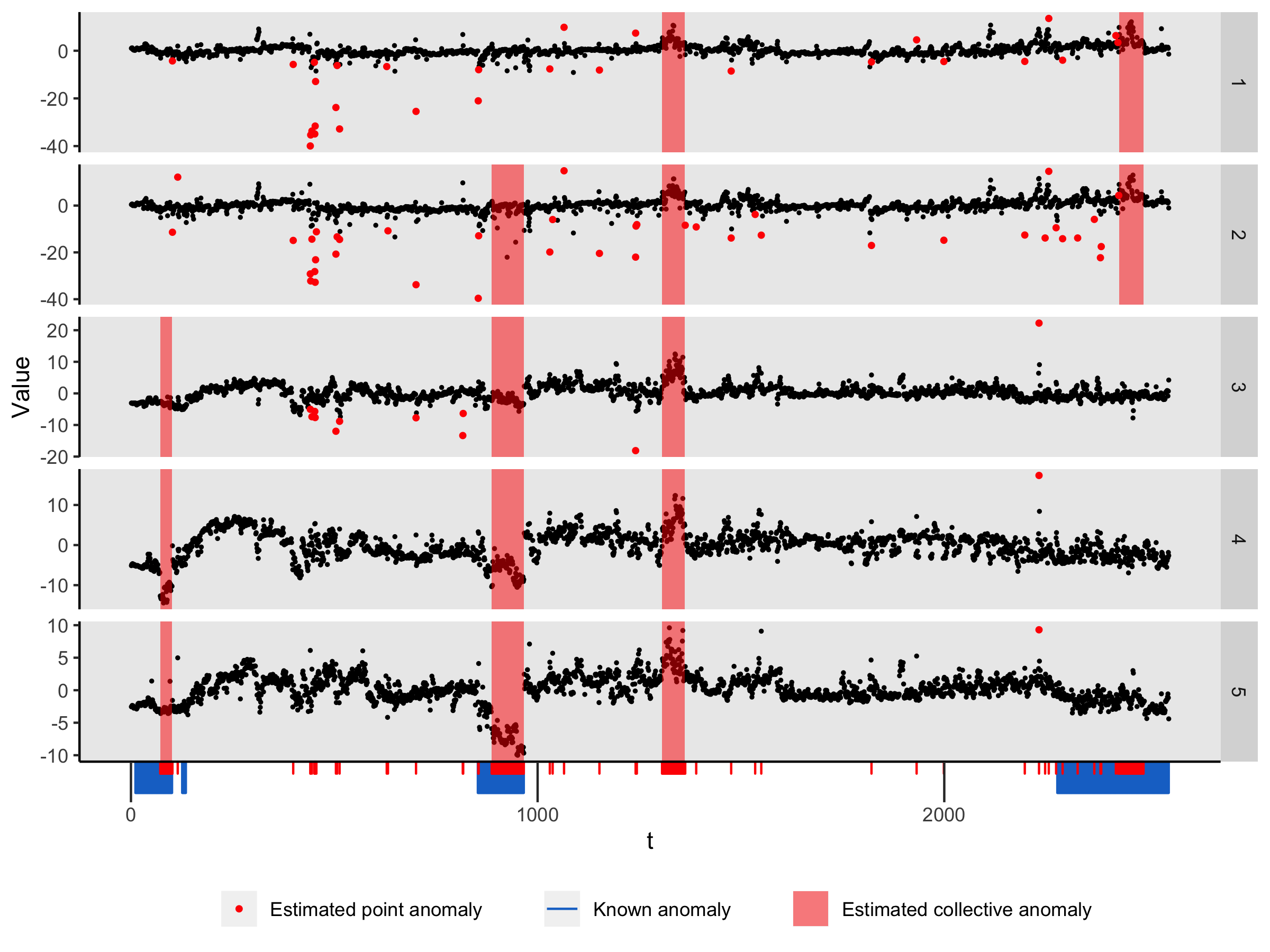}
    \caption{The four most significant estimated collective anomalies in the five residual times series derived from the pump data. Tuning parameters: $b = 11$, $b_\text{point} = 1$, $l = 5$ and $M = n$.}
  \label{fig:data_anoms}
\end{figure}

The final result is shown in Figure \ref{fig:data_anoms}.
Before interpreting the output, it is important to know that the start points of the known anomalies are more uncertain than the end points; the end point is the time where the pump was brought back to normal operation, whereas the start point has been set based on a retrospective analysis by the engineers.
With this in mind, we observe that three out of four estimated collective anomalies are within three separate known anomalous segments, with the estimated end points being more accurate than the estimated start points.
The short known anomaly from $t = 125$ to $t = 135$ is missed as there is virtually no signal of it in the data.
The estimated anomaly from $t = 1306$ to $t = 1362$, however, does not overlap with a known anomaly, but it clearly looks anomalous by eye.
This segment is also of interest to detect since it may correspond to an unknown segment of suboptimal operation.
If not, this segment points to a part of the data that fits our linear regression model poorly, indicating that a more sophisticated model might be in order if fewer false alarms are required.
In general we expect that a better model for linking the state variables with the monitoring variables would improve the results even further because more of the trend in the mean not associated with the known anomalies would be absorbed by the model rather than leaking into the residuals.

In addition, notice the importance of including point anomalies in the analysis for this application.
Rerunning \mvcapacor\ on the data without inferring point anomalies resulted in four additional false collective anomalies being inferred for $b = 11$.

\section{Conclusions} \label{sec:discussion}
In this article, we have proposed computationally efficient penalised cost-based methods for detecting multiple sparse and dense anomalies or changes in the mean of cross-correlated data.
In addition to estimating the locations of the anomalies/changepoints, the methods indicate which components are affected by a change.
This is important to understand why and how changes or anomalies have occured.
At the computational core of these methods lies a novel dynamic programming algorithm for solving banded unconstrained binary quadratic programs which approximate the Gaussian likelihood ratio test for a subset mean change.

The motivation of our methodological development comes from condition monitoring of an industrial process pump, where strong cross-correlations between spatially adjacent sensor measurements could be observed. Although several modelling assumptions were violated, three out of four known anomalies could be detected, with only one potential false alarm, when analysing the data with \mvcapacor. Even better results can be expected by using a more accurate model to remove trends not associated with anomalies. Also of interest for this application is being able to detect collective anomalies in real-time. The CAPA framework we have adopted has been shown to be able to be applied in online settings \citep{fisch_real_2020}, and similar ideas could be used to produce a sequential version of \mvcapacor.

When assessing the method's performance empirically, special attention was paid to how incorporating cross-correlations in the model affected the results compared to ignoring it as most existing methods do.
We found that for low to medium levels of dependence there was almost no difference in power or estimation accuracy; e.g. for $\rho < 0.5$ in the 2-banded and lattice precision matrices, and $\rho < 0.2$ for the constant correlation matrix, in the case of $p = 100$ variables.
For increasingly stronger dependence above these levels, either in the form of a denser precision matrix or higher correlation parameter, the benefit of including cross-correlation in the model of the data grows in almost all tested cases.

The exception to this rule is connected to the somewhat surprising finding that the shape of the change in mean across variables influences the magnitude of the advantage of including cross-correlations quite strongly.
In positively correlated data, changes that affect many series and are of very similar, or the same, size for each series can be harder to detect when including cross-correlations in the model.
For example, in a model with strong positive correlations, it is much harder to detect if a moderately large amount of variables changes by the same amount in the same direction, than if these variables changes by varying amounts in wildly varying directions.
The intuition behind this is that in the former case, the change mimics the expected behaviour of the data given the variables' strong positive dependence, while in the latter, the change strongly violates the model's expectation.
The model assuming independence, on the other hand, is completely agnostic to the shape of the changed mean vector.
As a result, the benefits of including correlations in the model is small, or perhaps even negative, if variables in the data is strongly dependent, and interest lies on detecting moderately sparse to dense and similarly changing variables.

\section*{Acknowledgements}
We are grateful to OneSubsea for sharing their data with us, and to Alex Fisch and Daniel Grose for helpful discussions.
This work is partially funded by the Norwegian Research Council project 237718 (Big Insight) and EPSRC grant EP/N031938/1 (STATSCALE).

\section*{Supplementary Material}
\begin{description}
 \item[Supplementary material]
 Proofs of the propositions, additional comments to Proposition \ref{prop:aMLE_bound}, derivation of the related changepoint test, and detailed results from the simulation study, for both anomaly and changepoint detection.
 \item[Code]
 Efficient implementations of the \mvcapacor\ and \mvcptcor\ algorithms as well as the code for reproducing the simulation study is available in the R package \texttt{capacc}, downloadable at \url{https://github.com/Tveten/capacc}. \mvcapacor\ will be included in a future version of the R package \texttt{anomaly} on CRAN, which contains the CAPA family of methods.
\end{description}

\renewcommand{\refname}{\large References}
\bibliography{capacc_library}

\newpage
\setcounter{section}{0}
\setcounter{page}{1}
\renewcommand{\thesection}{\Alph{section}}
{\centering{\LARGE\bf{Supplementary Material}}}

\section{Proofs and additional comments} \label{sec:appendixA}
\subsection{Proof of Proposition \ref{prop:BQP}} \label{proof:prop1}
First rewrite the optimisation problem in terms of the binary vector $\bu$:
\begin{align*}
  \tilde{S}(s, e) &= \underset{\bu}{\max}\;(e - s)\left[2\barbx^\tp\bQ(\barbx \circ \bu) - (\barbx \circ \bu)^\tp\bQ(\barbx \circ \bu) \right] - \beta\mathbf{1}^\tp\bu.
\end{align*}
The proof is completed by using properties of the Hadamard product and its relations to the regular matrix product to reexpress the optimal savings as
\begin{align*}
  \tilde{S}(s, e) &= \underset{\bu}{\max}\;(e - s)\left[2\barbx^\tp\bQ\diag(\barbx)\bu - \bu^\tp(\barbx \barbx^\tp \circ \bQ)\bu \right] - \beta\mathbf{1}^\tp\bu \\
  &= \underset{\bu}{\max}\;\left[2(e - s)\barbx^\tp\bQ\diag(\barbx) -\beta \right]\bu + \bu^\tp\left[- (e - s)\barbx \barbx^\tp \circ \bQ\right]\bu. \\
  &= \underset{\bu}{\max}\;\bu^\tp\left[2(e - s)\barbx \circ \bQ \barbx -\beta \right] + \bu^\tp\left[- (e - s)\barbx \barbx^\tp \circ \bQ\right]\bu.
\end{align*}

\subsection{Proof of Proposition \ref{prop:aMLE_bound} with comments} \label{sec:aMLE_bound_proof}
Let $\tilde{\bJ} := \argmax_\bJ \big[ \tilde{S}(s, e, \bJ) - P(|\bJ|) \big]$ and $\hat{\bJ} := \argmax_\bJ \big[ S(s, e, \bJ) - P(|\bJ|) \big]$.
In the following, we omit $s$ and $e$ in the notation of $S(s, e, \bJ)$ and $\tilde{S}(s, e, \bJ)$, such that $S(\hat{\bJ}) = S(s, e)$ and $\tilde{S}(\tilde{\bJ}) = \tilde{S}(s, e)$ in Proposition \ref{prop:aMLE_bound}.

The lower bound $S(\hat{\bJ}) - \tilde{S}(\tilde{\bJ}) \geq 0$ follows from $S$ using the exact MLE and $\tilde{S}$ using an alternative estimator.
I.e., $S(\bJ) \geq \tilde{S}(\bJ)$ for all subsets $\bJ$.
Hence, $S(\hat{\bJ}) \geq S(\tilde{\bJ}) \geq \tilde{S}(\tilde{\bJ})$.

To obtain the upper bound, let $\Delta S(\bJ) := S(\bJ) - \tilde{S}(\bJ)$.
Now, as $\tilde{\bJ}$ is the maximiser of $\tilde{S}$, we get that
\begin{equation*}
  S(\hat{\bJ}) - \tilde{S}(\tilde{\bJ}) = \tilde{S}(\hat{\bJ}) + \Delta S(\hat{\bJ}) - \tilde{S}(\tilde{\bJ})
  \leq \Delta S(\hat{\bJ}).
\end{equation*}
The final result immediately follows;
\begin{align*}
  \Delta S(\hat{\bJ})
  &= (e - s)(2\barbx(\hat{\bJ}^\comp) - \bW(\hat{\bJ})\barbx(\hat{\bJ}^\comp))^\tp \bQ\bW(\hat{\bJ})\barbx(\hat{\bJ}^\comp) \\
  &\leq 2(e - s)\barbx(\hat{\bJ}^\comp)^\tp \bQ\bW(\hat{\bJ})\barbx(\hat{\bJ}^\comp) \\
  &\leq 2(e - s)\lambda_{\max}\big(\bQ\bW(\hat{\bJ})\big) \| \barbx(\hat{\bJ}^\comp) \|^2.
\end{align*}
The first inequality is due to the positive semi-definiteness of the quadratic form in the second term, while the second inequality is a standard result on quadratic forms. This concludes the proof.

The following arguments suggests that the worst-case scenario for the approximation is sparse changes in strongly correlated data, as is also observed in the simulations of Section \ref{sec:Bapprox_performance}.
First observe that $\|\barbx_{(s + 1):e}(\hat{\bJ}^\comp)\|^2$ grows as $|\hat{\bJ}|$ becomes smaller.
Moreover, under the cross-correlated multivariate Gaussian model with means equal to $0$ and variances equal to $1$,
\begin{equation}
  (e - s) \big\| \barbx_{(s + 1):e}\big(\hat{\bJ}^\comp\big) \big\|^2
  = \sum_{j \in \hat{\bJ}^\comp} (e - s) \big( \bar{x}_{(s + 1):e}^{(j)} \big)^2,
  \label{eq:aMLE_bound_part}
\end{equation}
where $(e - s)\big(\bar{x}_{(s + 1):e}^{(j)}\big)^2$ for $j \in \hat{\bJ}^\comp$ are dependent $\chi^2_1$ random variables.
By standard rules of expectation and variance, we get that the expected value of \eqref{eq:aMLE_bound_part} is
$|\hat{\bJ}^\comp|$
and the variance is $\big(|\hat{\bJ}^\comp| + \sum_{i \not= j \in \hat{\bJ}^\comp} \omega_{i, j}\big)$,
where $\omega_{i, j}$ is the pairwise covariance between $\big(\bar{x}_{(s + 1):e}^{(i)}\big)^2$ and $\big(\bar{x}_{(s + 1):e}^{(j)}\big)^2$.
If we assume the zero-mean model to hold for the variables in the estimated non-anomalous variables $\hat{\bJ}^\comp$, we thus see that approximation may get worse as the change becomes sparser and the strength of the correlation increases in positive direction.

Note that this analysis only contains half of the picture as $\lambda_{\max} \big(\bQ\bW(\hat{\bJ})\big)$ seems intractable to study theoretically even for simple examples of $\bQ$.
Our numerical experimentation, however, suggests that $\lambda_{\max} \big(\bQ\bW(\hat{\bJ})\big)$ also grows as the correlations increase.
In addition, the simulation results in Section \ref{sec:Bapprox_performance} in the Supplementary Material agree with the conclusion that the greatest difference in performance occurs when there is a sparse change in strongly correlated data, although the difference is small in the tested low $p$ settings.

\subsection{Proof of Proposition \ref{prop:pruning}}
This proof follows the lines of the proof of Theorem 3.1 in \citet{killick_optimal_2012}.
First, recall the expression for the approximate savings,
\begin{equation*}
  \tilde{S}(s, e, \bJ) = (e - s)\left[ 2\barbx^\tp\bQ\barbx(\bJ) - \barbx(\bJ)^\tp\bQ\barbx(\bJ)\right],
\end{equation*}
and that we write $\tilde{S}(s, e) = \max_\bJ [\tilde{S}(s, e, \bJ) - P(\bJ)]$ for the optimal penalised approximate savings.
Next, observe that
\begin{align*}
  \max_\bJ \big[\tilde{S}(t, m, \bJ) - P(\bJ)\big] + \max_\bJ \tilde{S}(m, m', \bJ) &\geq \max_\bJ \big[\tilde{S}(t, m', \bJ) - P(\bJ) \big] \\
  \max_\bJ \big[\tilde{S}(t, m, \bJ) - P(\bJ)\big] + \max_\bJ \big[\tilde{S}(m, m', \bJ) - P(\bJ)\big] + \max_\bJ P(\bJ) &\geq \max_\bJ \big[\tilde{S}(t, m', \bJ) - P(\bJ) \big] \\
  \tilde{S}(t, m) + \tilde{S}(m, m') + \max_\bJ P(\bJ) &\geq \tilde{S}(t, m') \numberthis \label{eq:approx_savings_ineq}
\end{align*}
The inequality follows because of the basic fact that we are maximising over more parameters on the left-hand side than on the right-hand side, while adding the maximum penalty in the left-hand side guarantees that the additional penalty term is canceled out.
As a consequence of \eqref{eq:approx_savings_ineq}, and assuming that
\[
  C(t) + \tilde{S}(t, m) + \max_\bJ P(\bJ) \leq C(m)
\]
holds, we see that for all future times $m' \geq m + l$,
\begin{align*}
  C(t) + \tilde{S}(t, m) + \tilde{S}(m, m') + \max_\bJ P(\bJ) &\leq C(m) +  \tilde{S}(m, m') \\
  C(t) + \tilde{S}(t, m') &\leq C(m').
\end{align*}
The proof is concluded by noting that for the penalty given in \eqref{eq:penalty}, $\max_\bJ P(\bJ) = \alpha_\text{dense}$.

\subsection{Proof of Proposition \ref{prop:BQP_cpt}}
The proof follows the same steps as in the proof of Proposition 1 in Section \ref{proof:prop1}.

\section{Changepoint detection} \label{sec:Achangepoints}
In this section, we derive a test statistic for the single changepoint detection problem that utilises the approximation used for anomaly detection.
Multiple changepoints can be detected by embedding the test for a single changepoint within binary segmentation or a related algorithm, such as wild binary segmentation \citep{fryzlewicz_wild_2014} or seeded binary segmentation \citep{kovacs_seeded_2020}.
We call this changepoint detection method \mvcptcor.

Recall that the single changepoint problem is like the anomaly detection problem, with the exception that $e = n$ and all means are unknown.
In addition, we use $\tau$ to denote the changepoint.
Without loss of generality assume the sample mean for each series is 0.
To be able to use the same approximation as in the anomaly detection case we will base our cost on the log-likelihood under the assumption that the mean of the data is 0 for each series if there is no change.
The resulting changepoint saving for a fixed $\tau$ and $\bJ$ is given by
\begin{align*}
    S(\tau, \bJ) :&= C\left(\bx_{1:n}, \mathbf{0}\right) - \underset{\bmu(\bJ)}{\min}\; C\left(\bx_{1:\tau}, \bmu(\bJ)\right) - \underset{\bmu(\bJ)}{\min}\; C\left(\bx_{(\tau + 1):n}, \bmu(\bJ)\right) \numberthis \label{eq:cpt_saving} \\
    &= S(1, \tau, \bJ) + S(\tau + 1, n, \bJ),
\end{align*}
where $C$ is defined in \eqref{eq:cost} and $S(s, e, \bJ)$ in \eqref{eq:savings}.
Note that $\bJ$ is the same both before and after a changepoint to restrict the change vector $\bmu_1 - \bmu_0$ to be nonzero only in  $\bJ$.
Mirroring the anomaly detection case, we obtain our changepoint test statistic by subtracting the penalty function and maximising over $\tau$ and $\bJ$;
\begin{equation}
  \underset{l \leq \tau \leq n - l}{\max}\; S(\tau) := \underset{l \leq \tau \leq n - l}{\max} \left[\underset{\bJ}{\max}\; S(\tau, \bJ) - P(|\bJ|)\right],
  \label{eq:max_penalied_savings_cpt}
\end{equation}
where $l \geq 1$ is the minimum segment length as before.

Next, we once again replace the MLE of $\bmu(\bJ)$ with the subset-truncated sample mean, $\barbx(\bJ)$, defined in \eqref{eq:subset_sample_mean}.
The optimal approximate penalised savings for the single changepoint problem is thus given by
\begin{equation}
\tilde{S}(\tau) =
\underset{\bJ}{\max}\; \left[
\tau(2\barbx_1 - \barbx_1(\bJ))^\tp\bQ\barbx_1(\bJ) +
(n - \tau)(2\barbx_2 - \barbx_2(\bJ))^\tp\bQ\barbx_2(\bJ)
- \beta|\bJ| - \alpha \right],
\label{eq:approx_optimal_cpt}
\end{equation}
where $\barbx_1 := \barbx_{1:\tau}$ and $\barbx_2 := \barbx_{(\tau + 1):n}$.
A changepoint is detected when $\max_\tau \tilde{S}(\tau) > 0$.
Proposition \ref{prop:BQP_cpt} confirms that \eqref{eq:approx_optimal_cpt} is also a BQP, such that Algorithm \ref{alg:BQP_solver} can be used to find the optimum efficiently for a banded precision matrix $\bQ$. Algorithm \ref{alg:approx_saving_cpt} summarises the method.

\begin{myprop} \label{prop:BQP_cpt}
  Let $\alpha, \beta \geq 0$, $\barbx \in \mathbb{R}^p$ and $\barbx(\bJ) = \bu \circ \barbx$, where $\bu$ is $1$ at $\bJ$ and $0$ elsewhere.
  Then solving \eqref{eq:approx_optimal_cpt}
  corresponds to a BQP with $c = -\alpha$ and
  \begin{align*}
    \bA &= -\tau(\barbx_1 \barbx_1^\tp \circ \bQ)
          - (n - \tau)(\barbx_2 \barbx_2^\tp \circ \bQ)\\
    \bb &= 2\tau(\barbx_1 \circ \bQ\barbx_1)
          + 2(n - \tau)(\barbx_2 \circ \bQ\barbx_2) - \beta.
  \end{align*}
\end{myprop}

\begin{algorithm}
  \caption{The approximate penalised saving for changepoint detection used in \mvcptcor}
  \label{alg:approx_saving_cpt}
  \begin{algorithmic}[1]
    \INPUT{$\barbx$, $\bQ$, $\{ M_d \}_{d = 1}^p$, $\beta$, $\alpha_\text{sparse}$, $\alpha_\text{dense}$, $k^*$, $e$, $s$.}
    \State $\bA = -\tau(\barbx_1\barbx_1^\tp \circ \bQ) - (n - \tau)(\barbx_2 \barbx_2^\tp \circ \bQ)$
    \State $\bb = 2\tau(\barbx_1 \circ \bQ\barbx_1) + 2(n - \tau)(\barbx_2 \circ \bQ\barbx_2) - \beta$.
    \State $c = -\alpha_\text{sparse}$
    \State $\tilde{S}$, $\tilde{\bJ}$ from Algorithm \ref{alg:BQP_solver} with input ($\bA$, $\bb$, $c$, $\{ M_d \}_{d = 1}^p$, $k^*$)
    \State $S = S(s, e, [p]) - \alpha_\text{dense}$.
    \State \textbf{if} $\tilde{S} \geq S$ \textbf{return:} $\tilde{S}, \tilde{\bJ}$.
    \State \textbf{else} \textbf{return:} $S$, $[p]$.
  \end{algorithmic}
\end{algorithm}


\subsection{Simulation experiments for a single changepoint} \label{sec:sim_cpt}
We now look at changepoint detection and estimation in a single changepoint scenario, where we also compare our method to the inspect method of \citet{wang_high_2018}.
We focus on using \mvcptcor\ with a $\hat{\bQ}(4)$ precision matrix.
I.e., we assume that the precision matrix model is misspecified since this is most realistic, but note that improved performance on the order of what can be seen in Figures \ref{fig:known_anom_J1} and \ref{fig:known_anom_J10_J100} in the main text could be achieved by selecting a more correct model.
The version of inspect that assumes independence is available in the R package \texttt{InspectChangepoint}, and we refer to it by inspect($\bI$).
\citet{wang_high_2018} also discuss how inspect can be extended to include cross-correlations, and we have implemented this version into inspect($\hat{\bQ}$).
The inspect method does not require $\bQ$ to be sparse, and thus we estimate it using the same robust estimator \eqref{eq:robust_cov_mat} that we plug into the GLASSO method for estimating $\bQ$ in \mvcptcor.

\begin{figure}[t]
  \centering
  \includegraphics[width=0.85\textwidth] {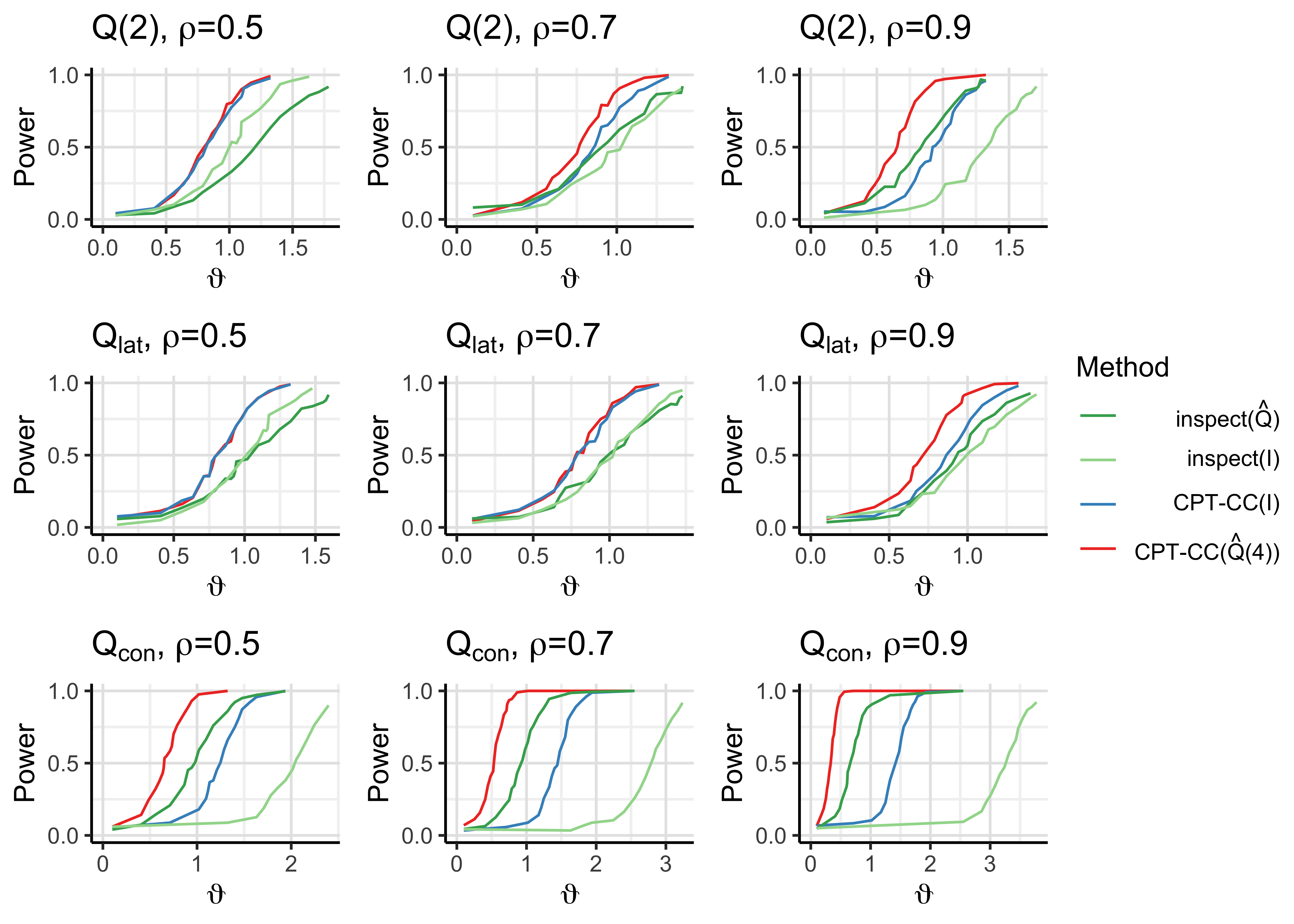}
  \caption{Power curves for a single known changepoint at $\tau = 170$ when $J = 1$ and $p = 100$.
  Results for 2-banded, lattice and globally constant correlation precision matrices are shown from top to bottom, with increasing $\rho$ from left to right. Other parameters: $n = 200$, $\alpha = 0.05$, and 1000 simulated data sets were used during tuning and power estimation.}
  \label{fig:known_cpt_J1}
\end{figure}

\begin{figure}[t]
  \centering
  \includegraphics[width=0.85\textwidth] {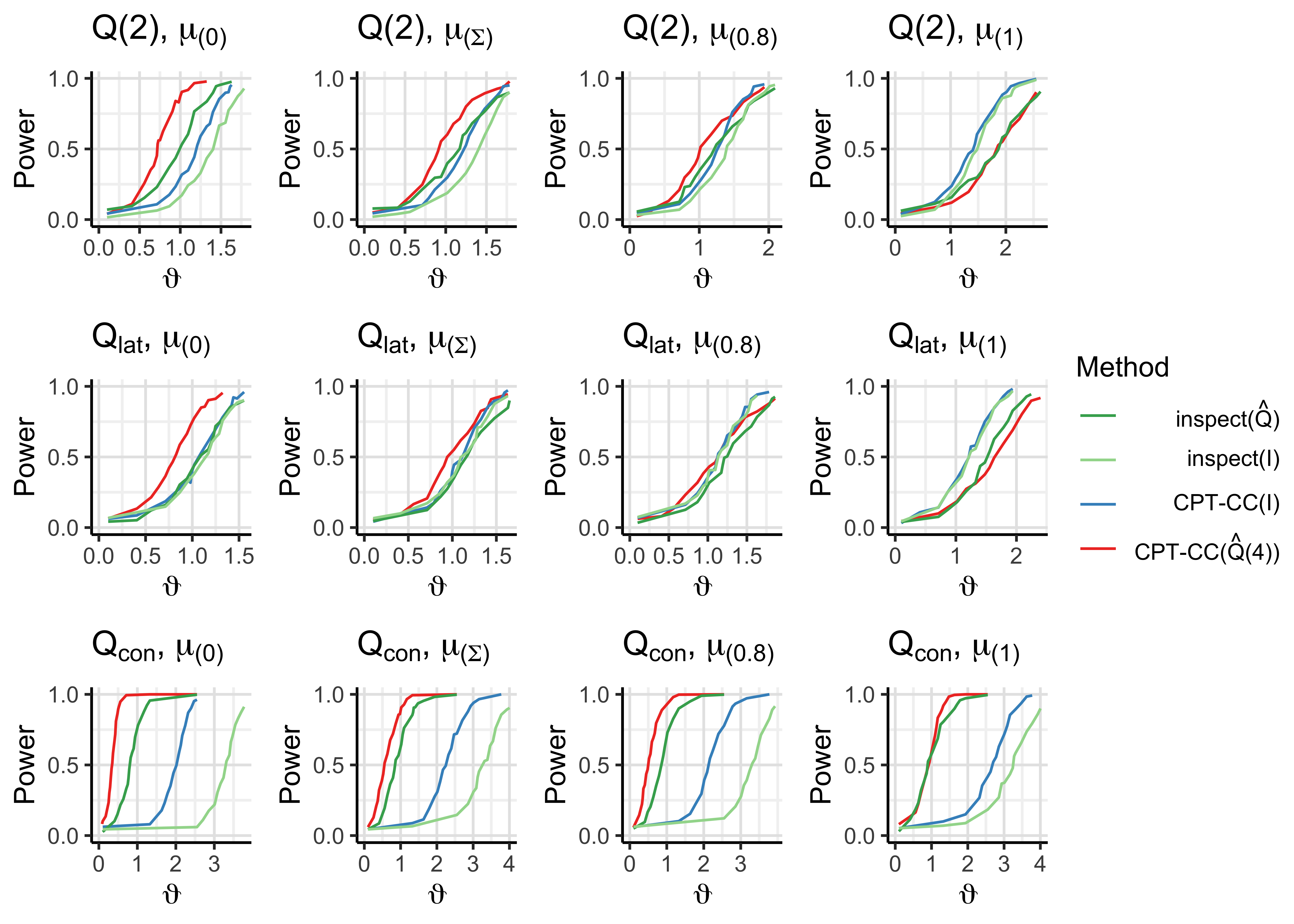}
  \caption{Power curves for a single known changepoint at $\tau = 170$ when $J = 10$ and $p = 100$.
  Results for 2-banded, lattice and globally constant correlation precision matrices are shown from top to bottom, with increasingly similar entries in the changed mean from left to right. Other parameters: $n = 200$, $\rho = 0.9$, $\alpha = 0.05$, and 1000 simulated data sets were used during tuning and power estimation.}
  \label{fig:known_cpt_J10}
\end{figure}

For comparing the method's power, we assume that the changepoint $\tau$ is known \text{a priori}, like in the anomaly setting in Section \ref{sec:sim_anomaly} in the main text.
We let $\tau = n - 30$ within the same scope of data scenarios as in the anomaly setting of Section \ref{sec:single_anom_sim}.
A brief summary of the results is given by Figures \ref{fig:known_cpt_J1} and \ref{fig:known_cpt_J10}.
The main conclusion is that \mvcapacor($\hat{\bQ}(4)$) generally is the most powerful method for the models we consider.
An exception from this rule can be observed for changes where more than $J/p \gtrapprox 0.1$ adjacent variables change in a very similar ($\bmu_{(\rho)}$ with $\rho \geq 0.9$) or equal way, where MVCAPA is better.
Interestingly, observe that the same pattern of whether it is best to include correlations or not also holds when comparing the two versions of inspect.
For more details, see Section \ref{sec:Bsim_cpt}.

\begin{table}
\centering
\begin{tabular}{ccccccc}
\toprule
$\bQ$ & $\rho$ & $J$ & CPT-CC($\hat{\bQ}(4)$) & CPT-CC($\bI$) & inspect($\hat{\bQ}$) & inspect($\bI$) \\
\midrule
$\mathbf{Q}(2)$ & 0.5 & 1 & $\mathbf{0.51}$ & $0.55$ & $1.38$ & $0.55$ \\
$\mathbf{Q}(2)$ & 0.9 & 1 & $\mathbf{0.21}$ & $0.50$ & $0.81$ & $0.59$ \\
$\mathbf{Q}_\text{lat}$ & 0.5 & 1 & $0.54$ & $0.58$ & $1.26$ & $\mathbf{0.51}$ \\
$\mathbf{Q}_\text{lat}$ & 0.9 & 1 & $\mathbf{0.37}$ & $0.59$ & $0.89$ & $0.52$ \\
$\mathbf{Q}_\text{con}$ & 0.5 & 1 & $\mathbf{0.29}$ & $6.32$ & $0.77$ & $0.78$ \\
$\mathbf{Q}_\text{con}$ & 0.9 & 1 & $\mathbf{0.00}$ & $10.22$ & $0.13$ & $2.82$ \\
$\mathbf{Q}(2)$ & 0.5 & 10 & $0.75$ & $0.72$ & $1.68$ & $\mathbf{0.60}$ \\
$\mathbf{Q}(2)$ & 0.9 & 10 & $\mathbf{0.84}$ & $1.41$ & $1.72$ & $1.40$ \\
$\mathbf{Q}_\text{lat}$ & 0.5 & 10 & $0.64$ & $0.66$ & $1.68$ & $\mathbf{0.55}$ \\
$\mathbf{Q}_\text{lat}$ & 0.9 & 10 & $\mathbf{0.78}$ & $1.07$ & $1.41$ & $0.83$ \\
$\mathbf{Q}_\text{con}$ & 0.5 & 10 & $\mathbf{0.74}$ & $13.44$ & $0.92$ & $2.21$ \\
$\mathbf{Q}_\text{con}$ & 0.9 & 10 & $0.20$ & $27.31$ & $\mathbf{0.18}$ & $8.43$ \\
$\mathbf{Q}(2)$ & 0.5 & 100 & $\mathbf{0.73}$ & $0.77$ & $3.07$ & $1.11$ \\
$\mathbf{Q}(2)$ & 0.9 & 100 & $\mathbf{0.73}$ & $1.71$ & $3.80$ & $2.12$ \\
$\mathbf{Q}_\text{lat}$ & 0.5 & 100 & $\mathbf{0.77}$ & $0.79$ & $3.02$ & $1.11$ \\
$\mathbf{Q}_\text{lat}$ & 0.9 & 100 & $\mathbf{1.05}$ & $2.98$ & $3.71$ & $1.92$ \\
$\mathbf{Q}_\text{con}$ & 0.5 & 100 & $\mathbf{2.63}$ & $65.10$ & $5.89$ & $16.20$ \\
$\mathbf{Q}_\text{con}$ & 0.9 & 100 & $\mathbf{8.38}$ & $113.87$ & $18.99$ & $38.35$ \\
\bottomrule
\end{tabular}
\caption{RMSE of changepoint estimates for $p = 100$, $n = 200$, $\tau = 140$, $\vartheta = 3$, and $\bmu_{(\bSigma)}$ changes. The smallest value is given in bold. 1000 random datasets were used for each RMSE estimate.}
\label{tab:mse_p100_vartheta3_shape6_rho0509}
\end{table}

We also compare the RMSE of estimated changepoints for the four methods.
For these comparisons, to avoid conflation due to methods having different powers, we assume the existence of a single changepoint to be known \textit{a priori} (as recommended by \citet{fearnhead_relating_2020}), and let all methods output their estimate of the changepoint location.
For the \mvcptcor\ methods, this is not unproblematic because testing for the existence of a changepoint is built into the method through the penalty function.
As a consequence, estimates of unsignificant changepoints (when $\tilde{S} < 0$) tend to be placed at either end of the data set.
For a fairer comparison, we have therefore chosen to set $\vartheta = 3$, such that almost all changes are significant.
A subset of the results for $\bmu^{(\bJ)} \sim N(\mathbf{0}, \bSigma_{\bJ, \bJ})$ are given in Table \ref{tab:mse_p100_vartheta3_shape6_rho0509}.
We see that \mvcapacor($\hat{\bQ}(4)$) also performs well in terms of RMSE, but that inspect is more competitive, especially for the medium sparse changes of $J = 10$.
For the other change classes, the same trends as seen in the power simulations can be observed (see Section \ref{sec:Bsim_cpt}), but with a stronger performance of inspect for $J = 10$;
\mvcapacor($\hat{\bQ}(4)$) is almost uniformly better for $\bmu_{(0)}$, while for $\bmu_{(1)}$, MVCAPA is better when $J = 100$, and either inspect($\bI$) or inspect($\bQ$) is best for $J = 10$.

\section{Additional simulation results} \label{sec:appendixB}
\subsection{Approximation vs. MLE} \label{sec:Bapprox_performance}

\begin{figure}
  \centering
  \includegraphics[width=0.85\textwidth] {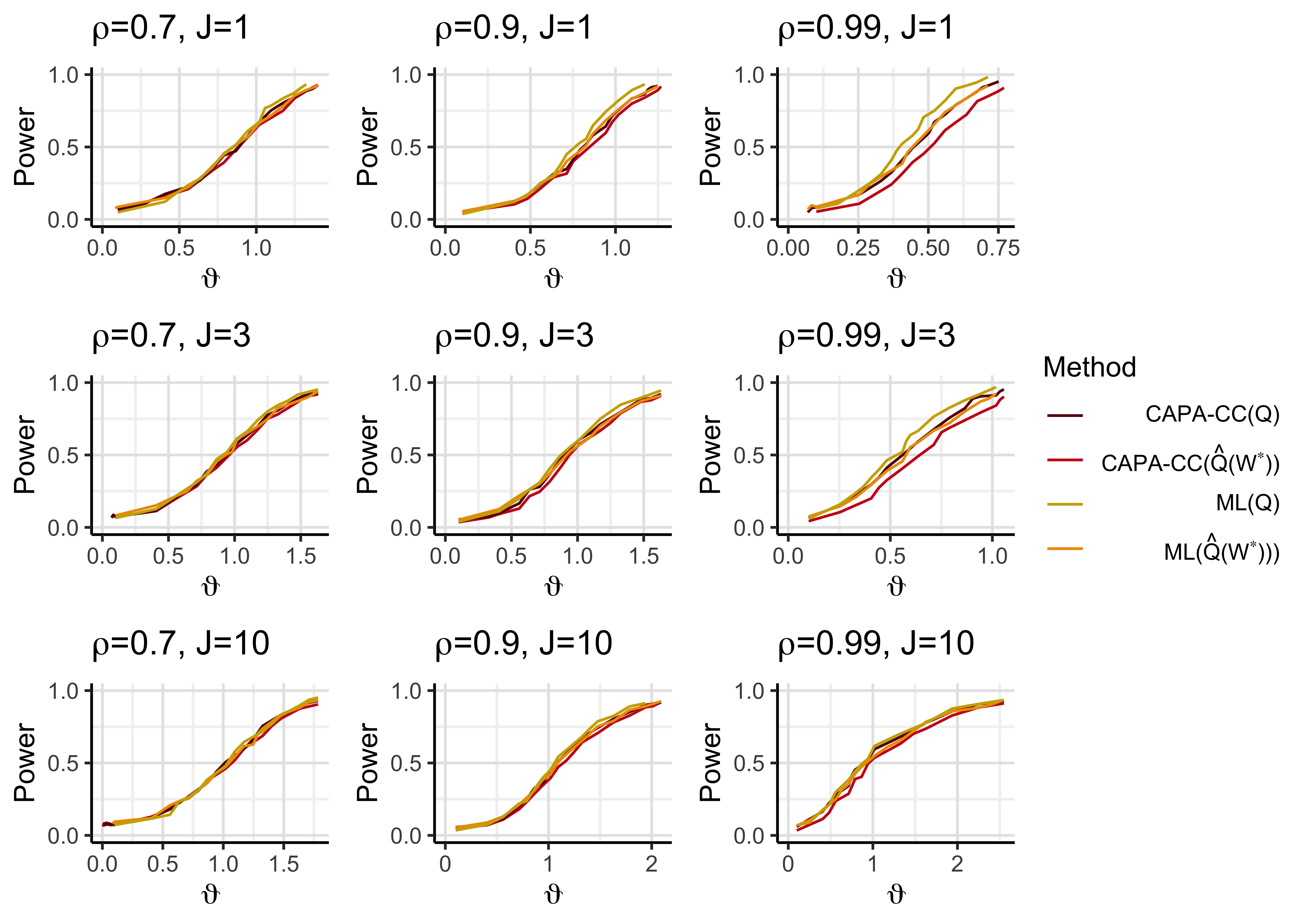}
  \caption{Power curves of our approximation and the exact ML method with the true precision matrix (dark red and gold lines) and an estimate using the true adjacency matrix (red and orange lines). Plot-wise from left to right, the correlation grows, and the number of anomalous variables grows from top to bottom. Other parameters: $n = 100$, $p = 10$, $\bQ = \bQ(2)$, $s = 50$, $e = 60$, change class $\bmu_{(\bSigma)}$, and 1000 repetitions were used during tuning and power estimation.}
  \label{fig:compare_aMLE}
\end{figure}

In this section, we compare the power of our approximation in \mvcapacor\ with the exact ML method in a data scenario with $n = 100$ observations from a $N(\bmu_t, \bQ(\rho, 2)^{-1})$ distribution with a single collective anomaly at $(s, e) = (50, 60)$ when $p = 10$.
As in Section \ref{sec:single_anom_sim} in the main text, we assume that the location of the anomaly is known.
Within this setup, we focus on varying the change class, $\rho$, $p$ and $J = |\bJ|$.
The penalty function for a given precision matrix was tuned for \mvcapacor\ and reused in the ML method for computational reasons.
Proposition \ref{prop:aMLE_bound} guarantees that \mvcapacor\ is in a disadvantage, if anything, under this choice.

\begin{figure}
  \centering
  \includegraphics[width=0.85\textwidth] {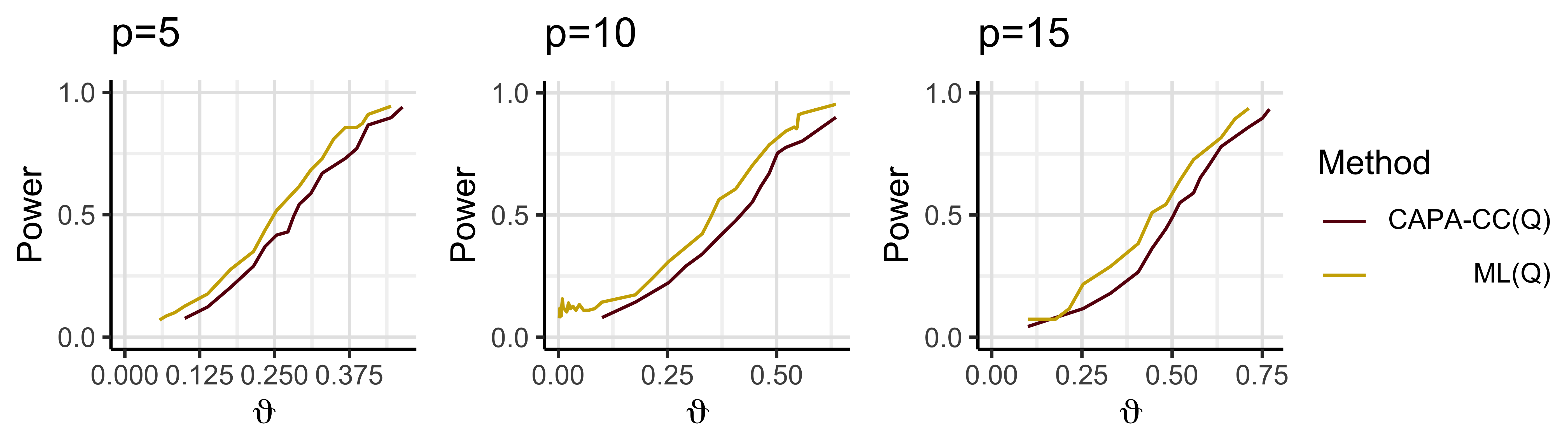}
  \caption{Power comparison of the approximation and the MLE with the true precision matrix for $p = 5, 10, 15$ in the worst-case scenario of a single changing variable in highly correlated data (the top right scenario in Figure \ref{fig:compare_aMLE}). Other parameters: $n = 100$, $s = 50$, $e = 60$, and 1000 repetitions were used during tuning and power estimation.}
  \label{fig:compare_aMLE_sparse_highcor}
\end{figure}

As can be seen from Figure \ref{fig:compare_aMLE}, almost no power is lost in the low dimensional setting by using our approximation rather than the exact ML method, both when using the true precision matrix and when the precision matrix is estimated from the true adjacency matrix.
It is only in the scenarios with a very high correlation of $0.99$ and a relatively sparse change of $J = 1, 3$ that there is a notable difference between the two methods for each precision matrix $\bQ$ and $\hat{\bQ}(\bW^*)$.
We should point out that this difference may become bigger as $p$ grows.
For $p = 5, 10 \text{ and} 15$, however, the results are very similar (Figure \ref{fig:compare_aMLE_sparse_highcor}).
All the results for $\bmu_{(0)}$ (i.i.d.) and $\bmu_{(1)}$ (equal) changes were qualitatively similar to the results for $\bmu_{(\bSigma)}$ shown here.

\subsection{Single anomaly detection} \label{sec:Banomaly_detection}
\begin{figure}
  \centering
  \includegraphics[width=0.85\textwidth] {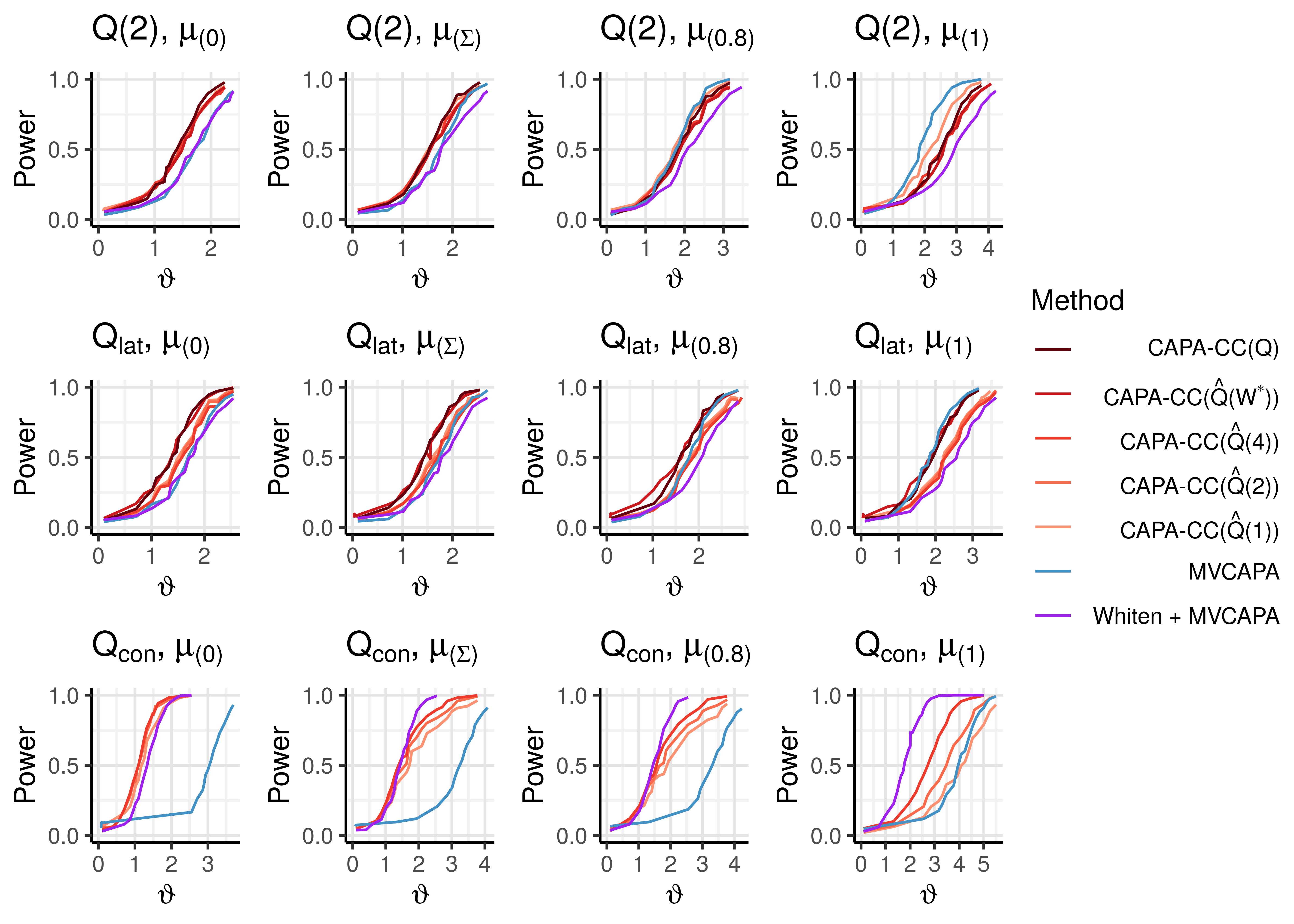}
  \caption{Power curves for $J = 10$, $p = 100$, $\rho = 0.7$, $n = 200$, $(s, e) = (100, 110)$, $\alpha = 0.05$.}
  \label{fig:known_anomJ10_rho07}
\end{figure}

\begin{figure}
  \centering
  \includegraphics[width=0.85\textwidth] {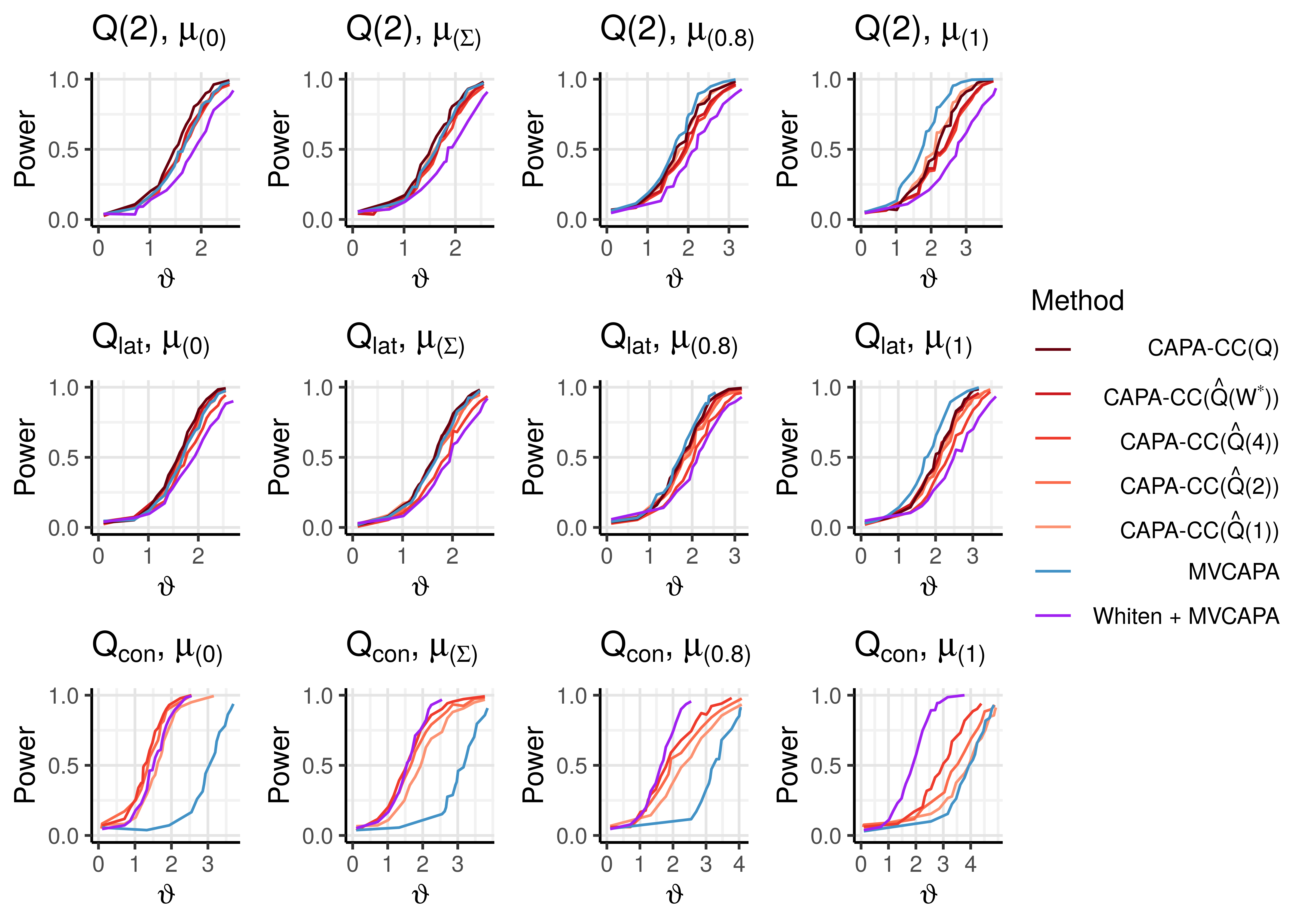}
  \caption{Power curves for $J = 10$, $p = 100$, $\rho = 0.5$, $n = 200$, $(s, e) = (100, 110)$, $\alpha = 0.05$.}
  \label{fig:known_anomJ10_rho05}
\end{figure}

\begin{figure}
  \centering
  \includegraphics[width=0.85\textwidth] {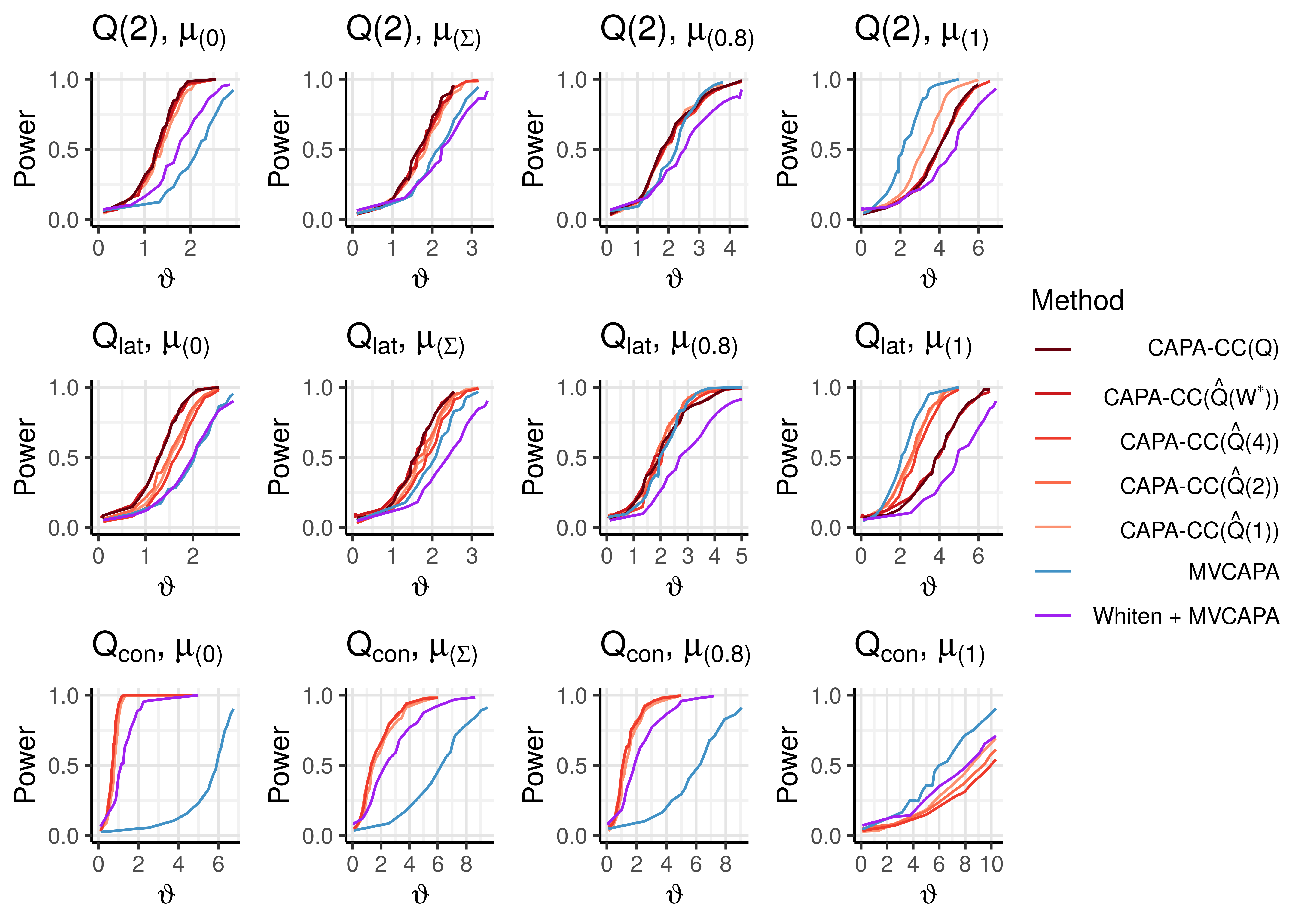}
  \caption{Power curves for $J = 100$, $p = 100$, $\rho = 0.9$, $n = 200$, $(s, e) = (100, 110)$, $\alpha = 0.05$.}
  \label{fig:known_anomJ100_rho09}
\end{figure}

\begin{figure}
  \centering
  \includegraphics[width=0.85\textwidth] {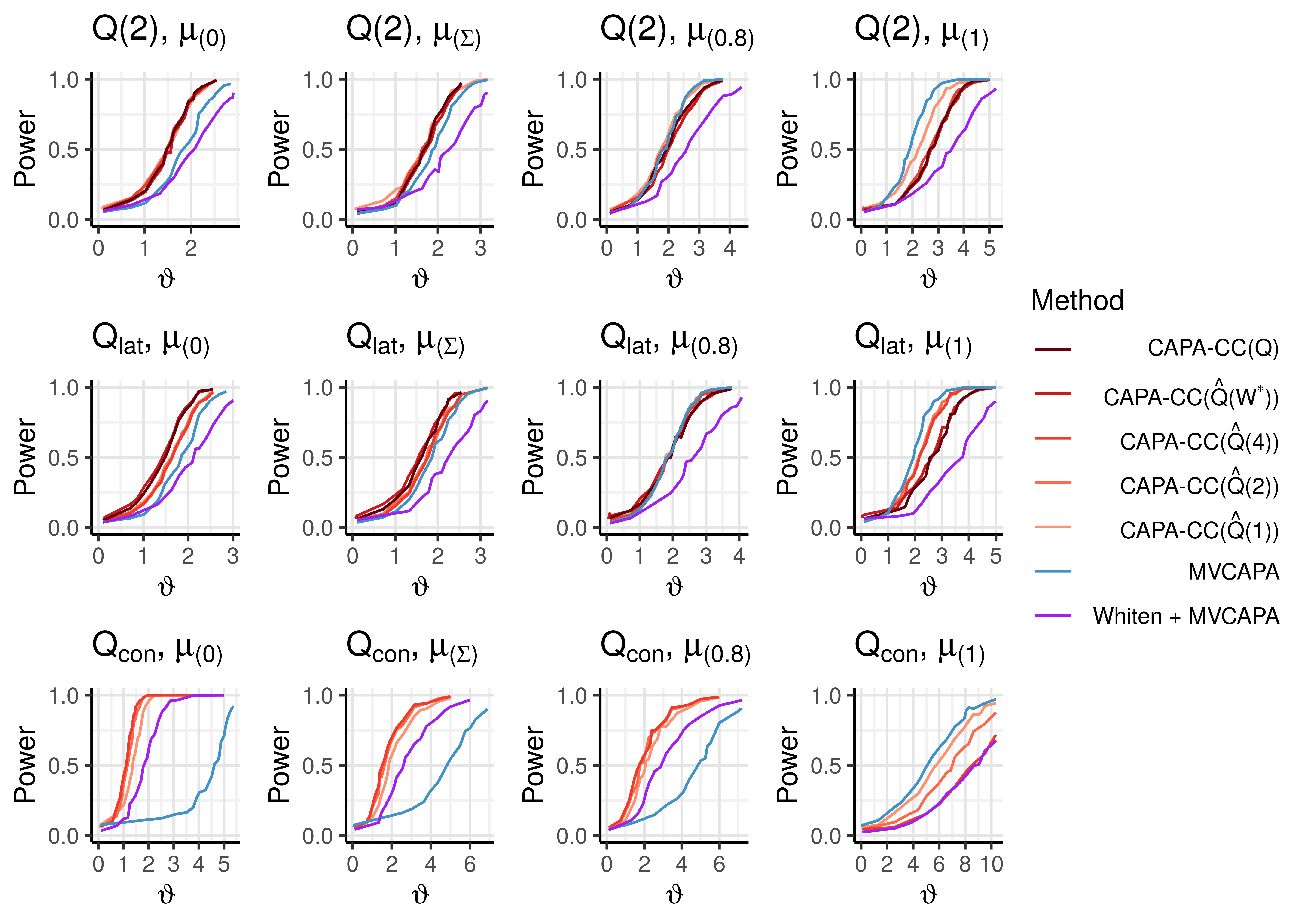}
  \caption{Power curves for $J = 100$, $p = 100$, $\rho = 0.7$, $n = 200$, $(s, e) = (100, 110)$, $\alpha = 0.05$.}
  \label{fig:known_anomJ100_rho07}
\end{figure}

\begin{figure}
  \centering
  \includegraphics[width=0.85\textwidth] {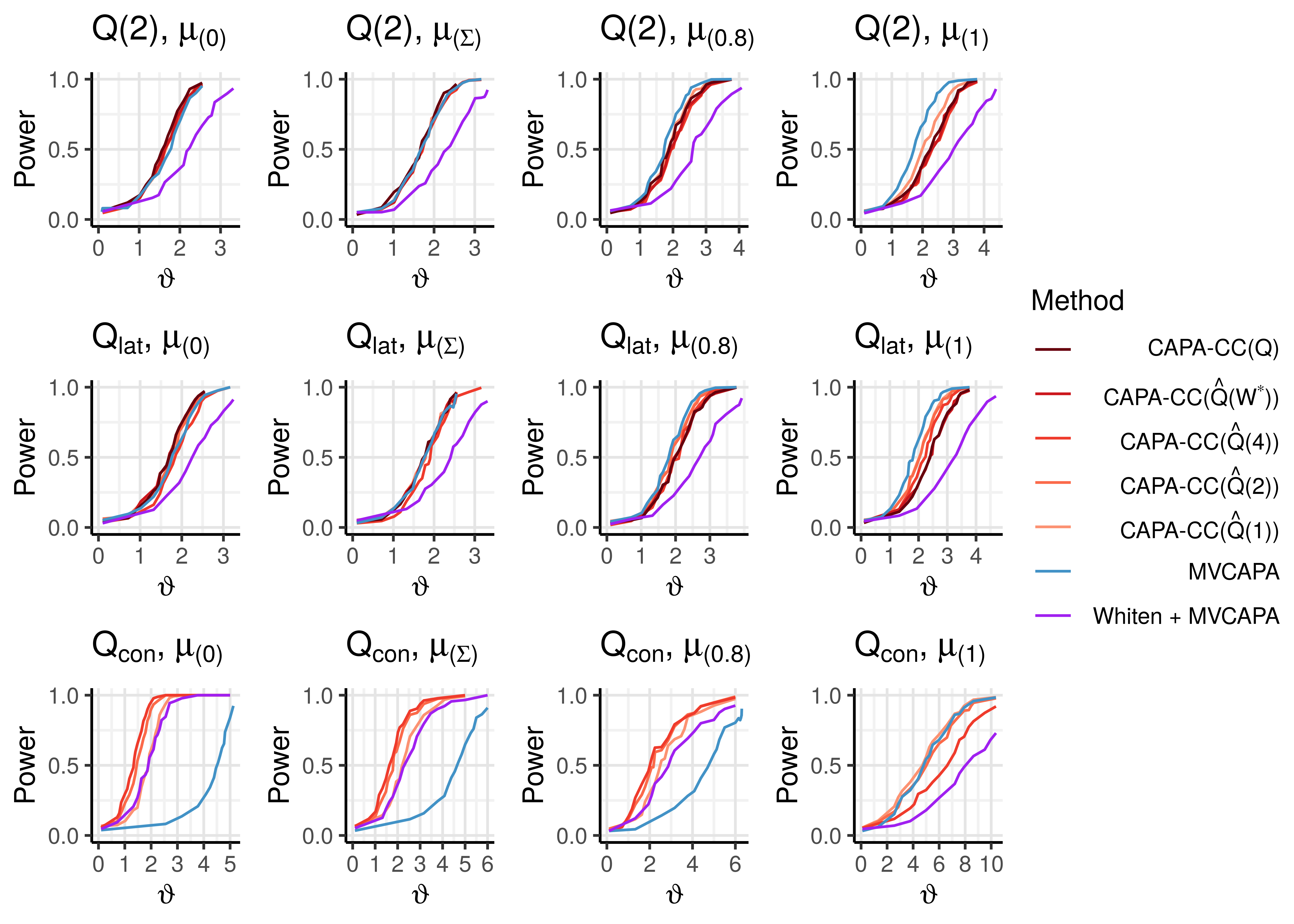}
  \caption{Power curves for $J = 100$, $p = 100$, $\rho = 0.5$, $n = 200$, $(s, e) = (100, 110)$, $\alpha = 0.05$.}
  \label{fig:known_anomJ100_rho05}
\end{figure}

\begin{figure}
  \centering
  \includegraphics[width=0.85\textwidth] {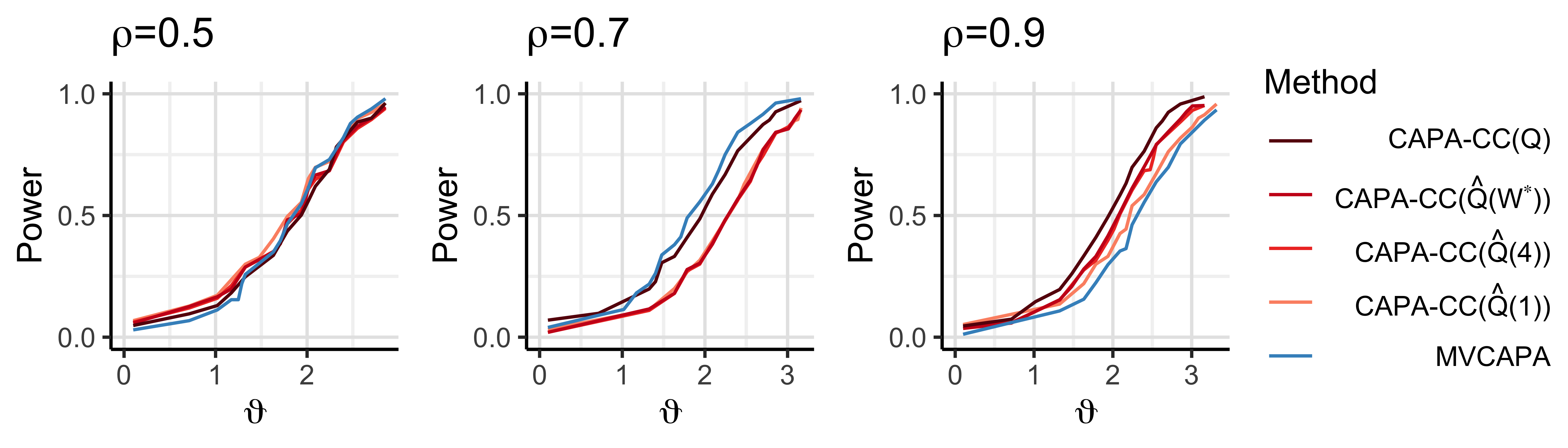}
  \caption{Power curves for $\bJ = \{1, 2, 3, 4, 50, 51, 52, 98, 99, 100 \}$, $p = 100$, $n = 200$, $\bQ = \bQ(2)$ $(s, e) = (100, 110)$, $\bmu_1^{(\bJ)} \sim \bmu_{(1)}$, $\alpha = 0.05$.}
  \label{fig:known_anomJ10_block_scattered}
\end{figure}

\begin{figure}
  \centering
  \includegraphics[width=0.85\textwidth] {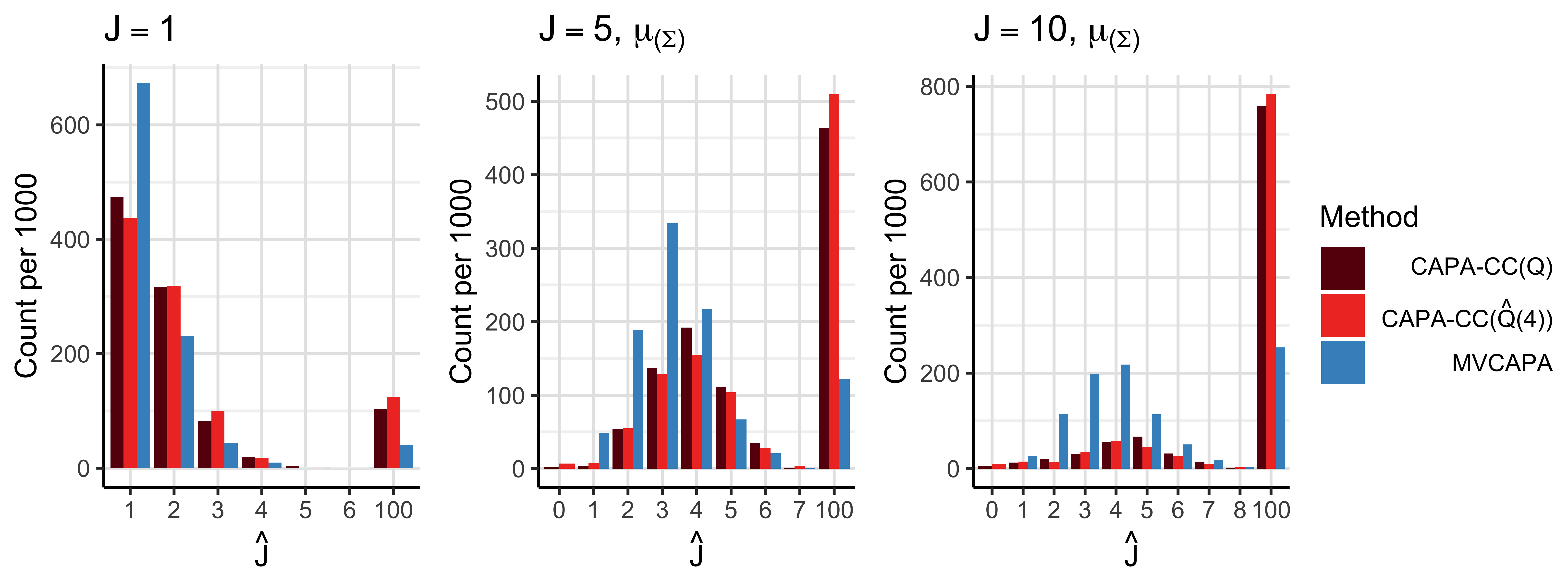}
  \caption{Estimated sizes of $\bJ$ for $\bJ = \{1\}$ (left) $\bJ = \{1, \ldots, 5\}$ (middle) and $\bJ = \{1, \ldots, 10\}$ when $p = 100$.
  Other parameters: $n = 200$, $\bQ = \bQ(2, 0.9)$, $s = 10$, $e = 20$, $\vartheta = 3$, $\bmu_{(\bSigma)}$ , $\alpha = 0.005$.}
  \label{fig:variable_selection_p100}
\end{figure}

Figures \ref{fig:known_anomJ10_rho07}-\ref{fig:known_anomJ10_block_scattered} display additional simulation results for comparing power when incorporating dependence in the method versus ignoring it in the single anomaly setting of Section \ref{sec:single_anom_sim} in the main text.
For more results on variable selection in the single anomaly setting, see Figure \ref{fig:variable_selection_p100} and Tables \ref{tab:subset_est_p10} and \ref{tab:subset_est_p100}.

\begin{table}
\caption{Average precision, recall and $\hat{J}$ over 1000 repetitions for $p = 10$ and  $n = 100$. Other parameters: $\bQ = \bQ(2)$, $s = n / 10$ and $e = s + 10$, $\bmu_{(\Sigma)}$, $\alpha = 0.005$.}
\label{tab:subset_est_p10}
\centering
\begin{tabular}{ccccccc}
\toprule
$J$ & $\vartheta$ & $\rho$ & Method & $\hat{J}$ & Precision & Recall \\
\midrule
1 & 2 & 0.5 & MVCAPA & 1.66 & 0.73 & 1.00 \\
1 & 2 & 0.5 & CAPA-CC($\bQ$) & 1.97 & 0.66 & 1.00 \\
1 & 2 & 0.5 & ML($\bQ$) & 1.94 & 0.65 & 1.00 \\
1 & 2 & 0.5 & CAPA-CC($\hat{\bQ}(4)$) & 1.77 & 0.70 & 1.00 \\
1 & 2 & 0.5 & ML($\hat{\bQ}(4)$) & 1.77 & 0.70 & 1.00 \\
1 & 2 & 0.9 & MVCAPA & 1.79 & 0.78 & 1.00 \\
1 & 2 & 0.9 & CAPA-CC($\bQ$) & 1.87 & 0.73 & 1.00 \\
1 & 2 & 0.9 & ML($\bQ$) & 1.77 & 0.71 & 1.00 \\
1 & 2 & 0.9 & CAPA-CC($\hat{\bQ}(4)$) & 1.89 & 0.72 & 1.00 \\
1 & 2 & 0.9 & ML($\hat{\bQ}(4)$) & 1.79 & 0.70 & 1.00 \\
1 & 5 & 0.5 & MVCAPA & 1.66 & 0.74 & 1.00 \\
1 & 5 & 0.5 & CAPA-CC($\bQ$) & 1.92 & 0.68 & 1.00 \\
1 & 5 & 0.5 & ML($\bQ$) & 1.90 & 0.68 & 1.00 \\
1 & 5 & 0.5 & CAPA-CC($\hat{\bQ}(4)$) & 1.77 & 0.71 & 1.00 \\
1 & 5 & 0.5 & ML($\hat{\bQ}(4)$) & 1.73 & 0.71 & 1.00 \\
1 & 5 & 0.9 & MVCAPA & 1.66 & 0.81 & 1.00 \\
1 & 5 & 0.9 & CAPA-CC($\bQ$) & 1.84 & 0.72 & 1.00 \\
1 & 5 & 0.9 & ML($\bQ$) & 1.75 & 0.71 & 1.00 \\
1 & 5 & 0.9 & CAPA-CC($\hat{\bQ}(4)$) & 1.85 & 0.73 & 1.00 \\
1 & 5 & 0.9 & ML($\hat{\bQ}(4)$) & 1.81 & 0.69 & 1.00 \\
3 & 2 & 0.5 & MVCAPA & 2.80 & 0.83 & 0.70 \\
3 & 2 & 0.5 & CAPA-CC($\bQ$) & 3.25 & 0.78 & 0.74 \\
3 & 2 & 0.5 & ML($\bQ$) & 3.15 & 0.78 & 0.73 \\
3 & 2 & 0.5 & CAPA-CC($\hat{\bQ}(4)$) & 2.97 & 0.81 & 0.72 \\
3 & 2 & 0.5 & ML($\hat{\bQ}(4)$) & 2.88 & 0.81 & 0.70 \\
3 & 2 & 0.9 & MVCAPA & 2.86 & 0.87 & 0.72 \\
3 & 2 & 0.9 & CAPA-CC($\bQ$) & 3.47 & 0.82 & 0.81 \\
3 & 2 & 0.9 & ML($\bQ$) & 3.05 & 0.83 & 0.77 \\
3 & 2 & 0.9 & CAPA-CC($\hat{\bQ}(4)$) & 3.55 & 0.80 & 0.81 \\
3 & 2 & 0.9 & ML($\hat{\bQ}(4)$) & 3.10 & 0.81 & 0.77 \\
3 & 5 & 0.5 & MVCAPA & 3.42 & 0.85 & 0.88 \\
3 & 5 & 0.5 & CAPA-CC($\bQ$) & 3.85 & 0.80 & 0.90 \\
3 & 5 & 0.5 & ML($\bQ$) & 3.80 & 0.80 & 0.90 \\
3 & 5 & 0.5 & CAPA-CC($\hat{\bQ}(4)$) & 3.56 & 0.83 & 0.89 \\
3 & 5 & 0.5 & ML($\hat{\bQ}(4)$) & 3.47 & 0.84 & 0.89 \\
3 & 5 & 0.9 & MVCAPA & 3.53 & 0.88 & 0.90 \\
3 & 5 & 0.9 & CAPA-CC($\bQ$) & 4.12 & 0.81 & 0.93 \\
3 & 5 & 0.9 & ML($\bQ$) & 3.63 & 0.83 & 0.92 \\
3 & 5 & 0.9 & CAPA-CC($\hat{\bQ}(4)$) & 4.16 & 0.80 & 0.93 \\
3 & 5 & 0.9 & ML($\hat{\bQ}(4)$) & 3.65 & 0.82 & 0.92 \\
\bottomrule
\end{tabular}
\end{table}

\begin{table}
\caption{Average precision, recall and $\hat{J}$ over 1000 repetitions for $p = 100$ and  $n = 200$. Other parameters: $\bQ = \bQ(2)$, $s = n / 10$ and $e = s + 10$, $\bmu_{(\Sigma)}$, $\alpha = 0.005$.}
\label{tab:subset_est_p100}
\centering
\begin{tabular}{ccccccc}
\toprule
$J$ & $\vartheta$ & $\rho$ & Method & $\hat{J}$ & Precision & Recall \\
\midrule
1 & 2 & 0.5 & MVCAPA & 10.25 & 0.63 & 1.00 \\
1 & 2 & 0.5 & CAPA-CC($\bQ$) & 10.40 & 0.64 & 1.00 \\
1 & 2 & 0.5 & CAPA-CC($\hat{\bQ}(4)$) & 11.62 & 0.62 & 1.00 \\
1 & 2 & 0.9 & MVCAPA & 4.45 & 0.80 & 1.00 \\
1 & 2 & 0.9 & CAPA-CC($\bQ$) & 10.74 & 0.69 & 1.00 \\
1 & 2 & 0.9 & CAPA-CC($\hat{\bQ}(4)$) & 13.73 & 0.66 & 1.00 \\
1 & 5 & 0.5 & MVCAPA & 9.65 & 0.63 & 1.00 \\
1 & 5 & 0.5 & CAPA-CC($\bQ$) & 10.15 & 0.62 & 1.00 \\
1 & 5 & 0.5 & CAPA-CC($\hat{\bQ}(4)$) & 9.79 & 0.62 & 1.00 \\
1 & 5 & 0.9 & MVCAPA & 5.21 & 0.81 & 1.00 \\
1 & 5 & 0.9 & CAPA-CC($\bQ$) & 11.44 & 0.68 & 1.00 \\
1 & 5 & 0.9 & CAPA-CC($\hat{\bQ}(4)$) & 11.84 & 0.67 & 1.00 \\
5 & 2 & 0.5 & MVCAPA & 27.83 & 0.59 & 0.55 \\
5 & 2 & 0.5 & CAPA-CC($\bQ$) & 29.32 & 0.59 & 0.57 \\
5 & 2 & 0.5 & CAPA-CC($\hat{\bQ}(4)$) & 30.00 & 0.56 & 0.56 \\
5 & 2 & 0.9 & MVCAPA & 11.35 & 0.79 & 0.42 \\
5 & 2 & 0.9 & CAPA-CC($\bQ$) & 34.55 & 0.56 & 0.63 \\
5 & 2 & 0.9 & CAPA-CC($\hat{\bQ}(4)$) & 38.06 & 0.50 & 0.63 \\
5 & 5 & 0.5 & MVCAPA & 44.22 & 0.53 & 0.84 \\
5 & 5 & 0.5 & CAPA-CC($\bQ$) & 47.75 & 0.51 & 0.85 \\
5 & 5 & 0.5 & CAPA-CC($\hat{\bQ}(4)$) & 51.34 & 0.47 & 0.86 \\
5 & 5 & 0.9 & MVCAPA & 21.58 & 0.78 & 0.79 \\
5 & 5 & 0.9 & CAPA-CC($\bQ$) & 55.21 & 0.46 & 0.91 \\
5 & 5 & 0.9 & CAPA-CC($\hat{\bQ}(4)$) & 58.26 & 0.42 & 0.91 \\
10 & 2 & 0.5 & MVCAPA & 38.26 & 0.55 & 0.50 \\
10 & 2 & 0.5 & CAPA-CC($\bQ$) & 42.77 & 0.51 & 0.54 \\
10 & 2 & 0.5 & CAPA-CC($\hat{\bQ}(4)$) & 44.63 & 0.48 & 0.54 \\
10 & 2 & 0.9 & MVCAPA & 14.15 & 0.76 & 0.28 \\
10 & 2 & 0.9 & CAPA-CC($\bQ$) & 49.74 & 0.43 & 0.60 \\
10 & 2 & 0.9 & CAPA-CC($\hat{\bQ}(4)$) & 52.24 & 0.40 & 0.60 \\
10 & 5 & 0.5 & MVCAPA & 85.29 & 0.23 & 0.93 \\
10 & 5 & 0.5 & CAPA-CC($\bQ$) & 88.40 & 0.20 & 0.94 \\
10 & 5 & 0.5 & CAPA-CC($\hat{\bQ}(4)$) & 89.42 & 0.19 & 0.95 \\
10 & 5 & 0.9 & MVCAPA & 51.88 & 0.55 & 0.78 \\
10 & 5 & 0.9 & CAPA-CC($\bQ$) & 94.77 & 0.15 & 0.98 \\
10 & 5 & 0.9 & CAPA-CC($\hat{\bQ}(4)$) & 95.97 & 0.14 & 0.98 \\
\bottomrule
\end{tabular}
\end{table}

\subsection{Single changepoint detection and estimation} \label{sec:Bsim_cpt}
\begin{figure}
  \centering
  \includegraphics[width=0.85\textwidth] {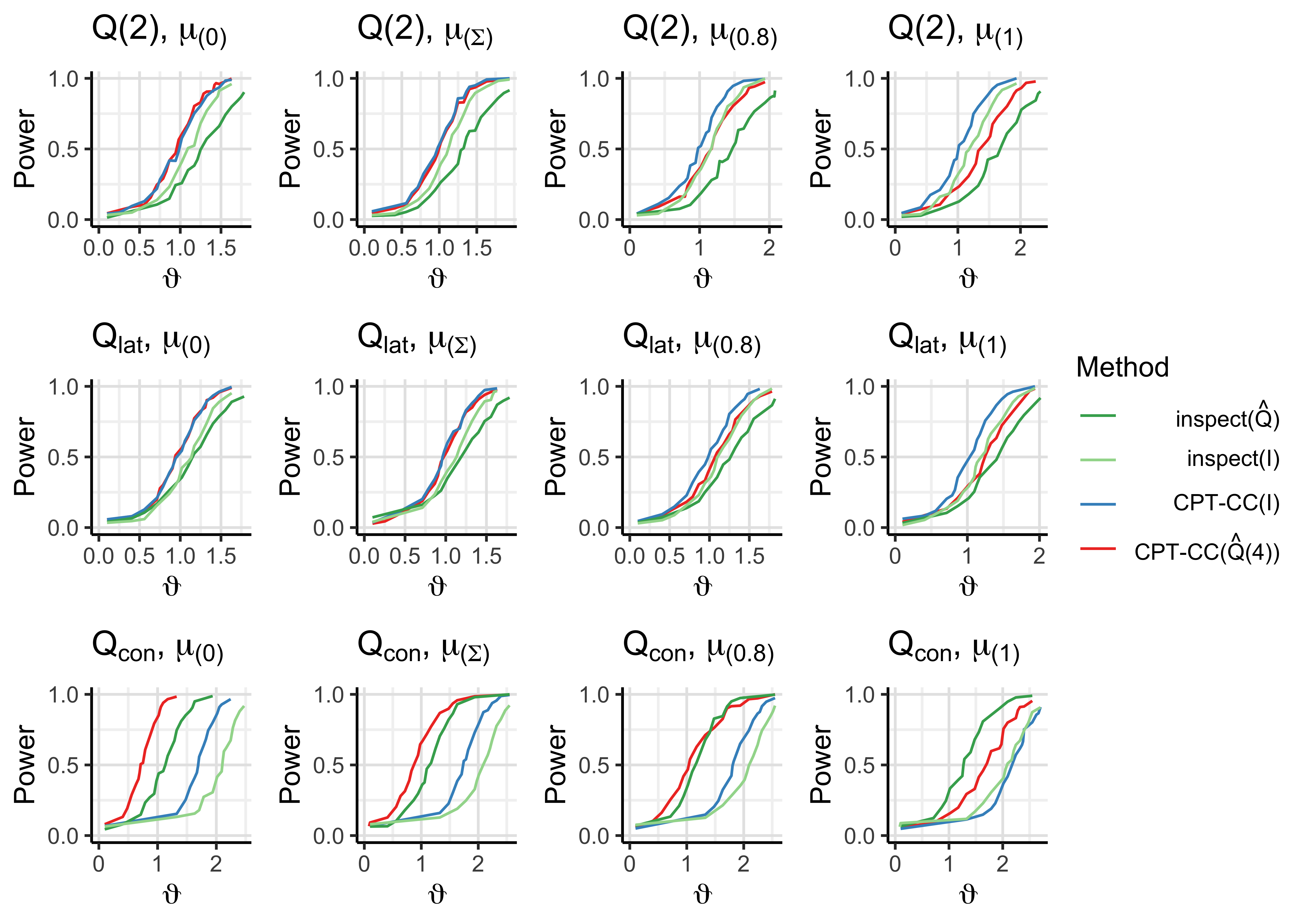}
  \caption{Power curves for a single known changepoint at $\tau = 170$ when $J = 10$, $p = 100$ and $\rho = 0.5$.
  Other parameters: $n = 200$, $\rho = 0.5$, $\alpha = 0.05$.}
  \label{fig:known_cpt_J10_rho05}
\end{figure}

\begin{figure}
  \centering
  \includegraphics[width=0.85\textwidth] {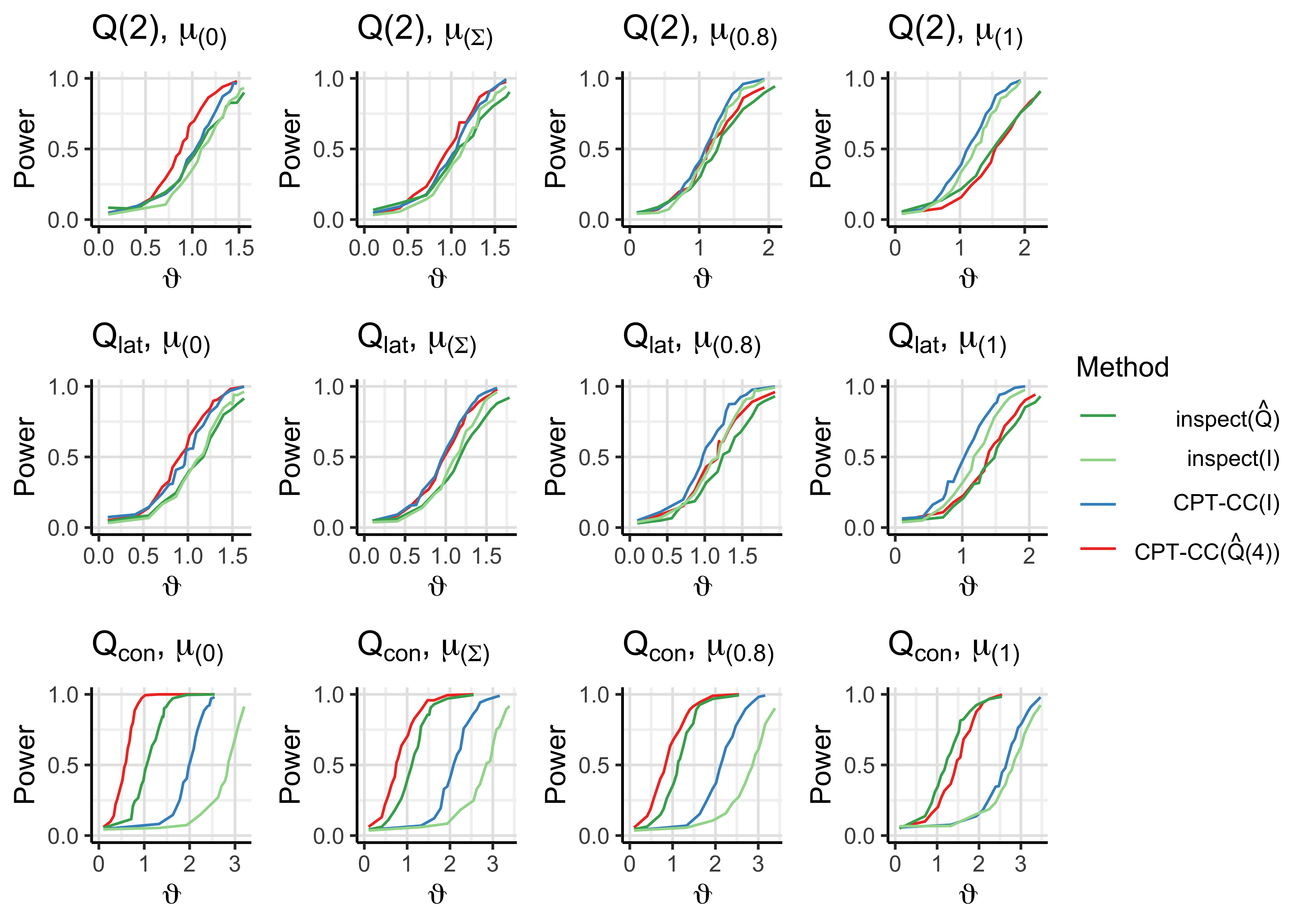}
  \caption{Power curves for a single known changepoint at $\tau = 170$ when $J = 10$, $p = 100$ and $\rho = 0.7$.
  Other parameters: $n = 200$, $\rho = 0.7$, $\alpha = 0.05$.}
  \label{fig:known_cpt_J10_rho07}
\end{figure}

\begin{figure}
  \centering
  \includegraphics[width=0.85\textwidth] {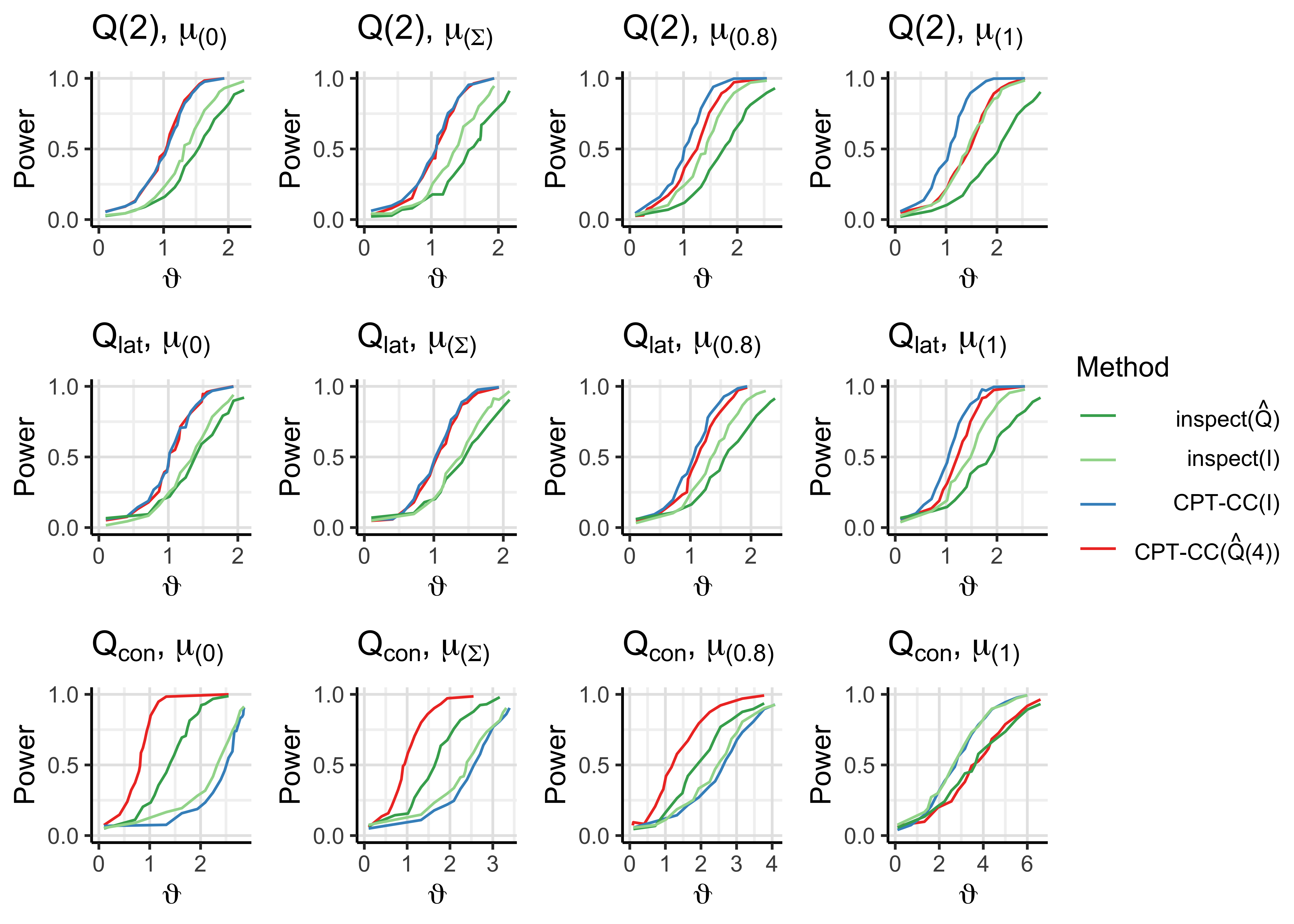}
  \caption{Power curves for a single known changepoint at $\tau = 170$ when $J = 100$, $p = 100$ and $\rho = 0.5$.
  Other parameters: $n = 200$, $\rho = 0.5$, $\alpha = 0.05$.}
  \label{fig:known_cpt_J100_rho05}
\end{figure}

\begin{figure}
  \centering
  \includegraphics[width=0.85\textwidth] {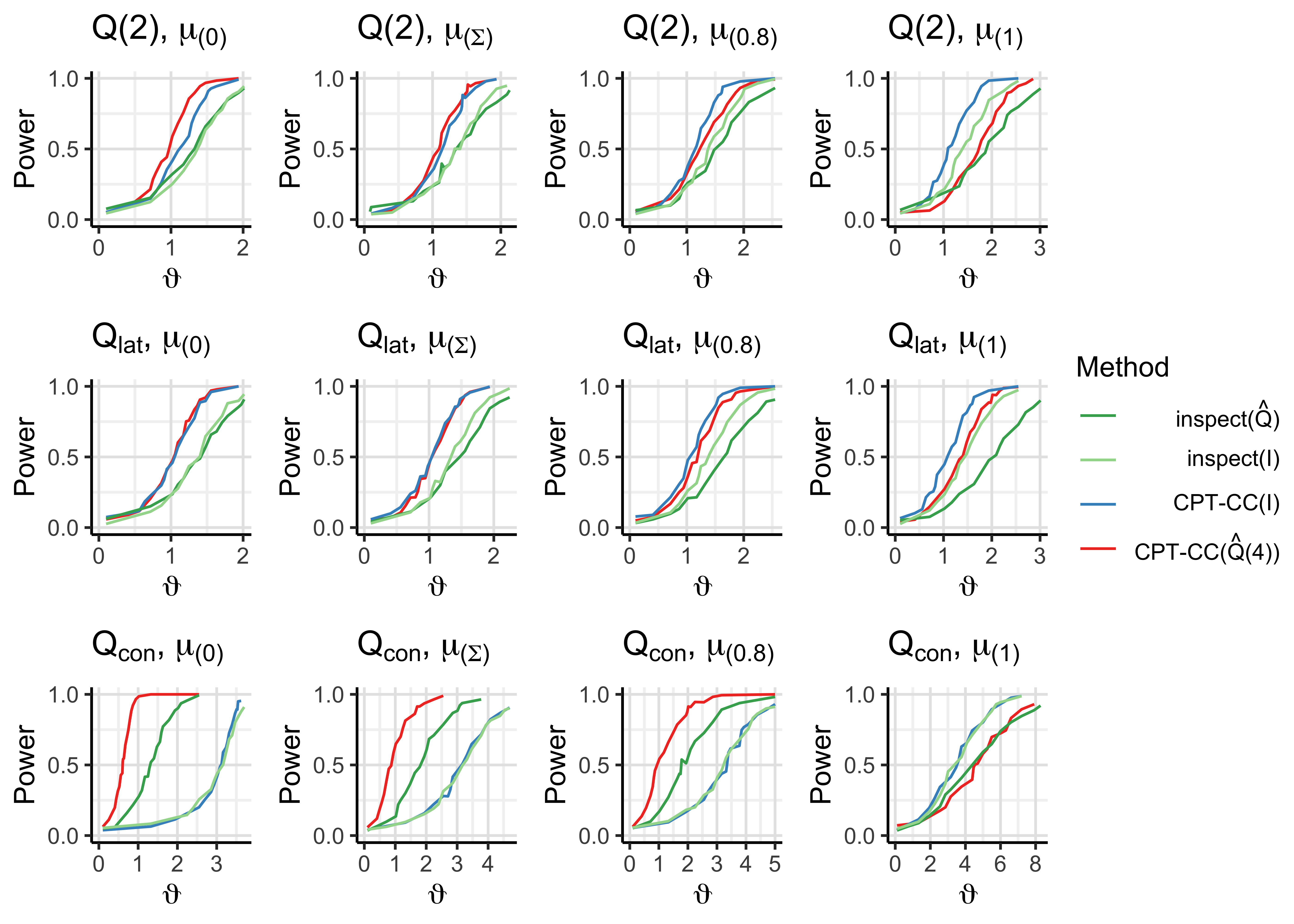}
  \caption{Power curves for a single known changepoint at $\tau = 170$ when $J = 100$, $p = 100$ and $\rho = 0.7$.
  Other parameters: $n = 200$, $\rho = 0.7$, $\alpha = 0.05$.}
  \label{fig:known_cpt_J100_rho07}
\end{figure}

\begin{figure}
  \centering
  \includegraphics[width=0.85\textwidth] {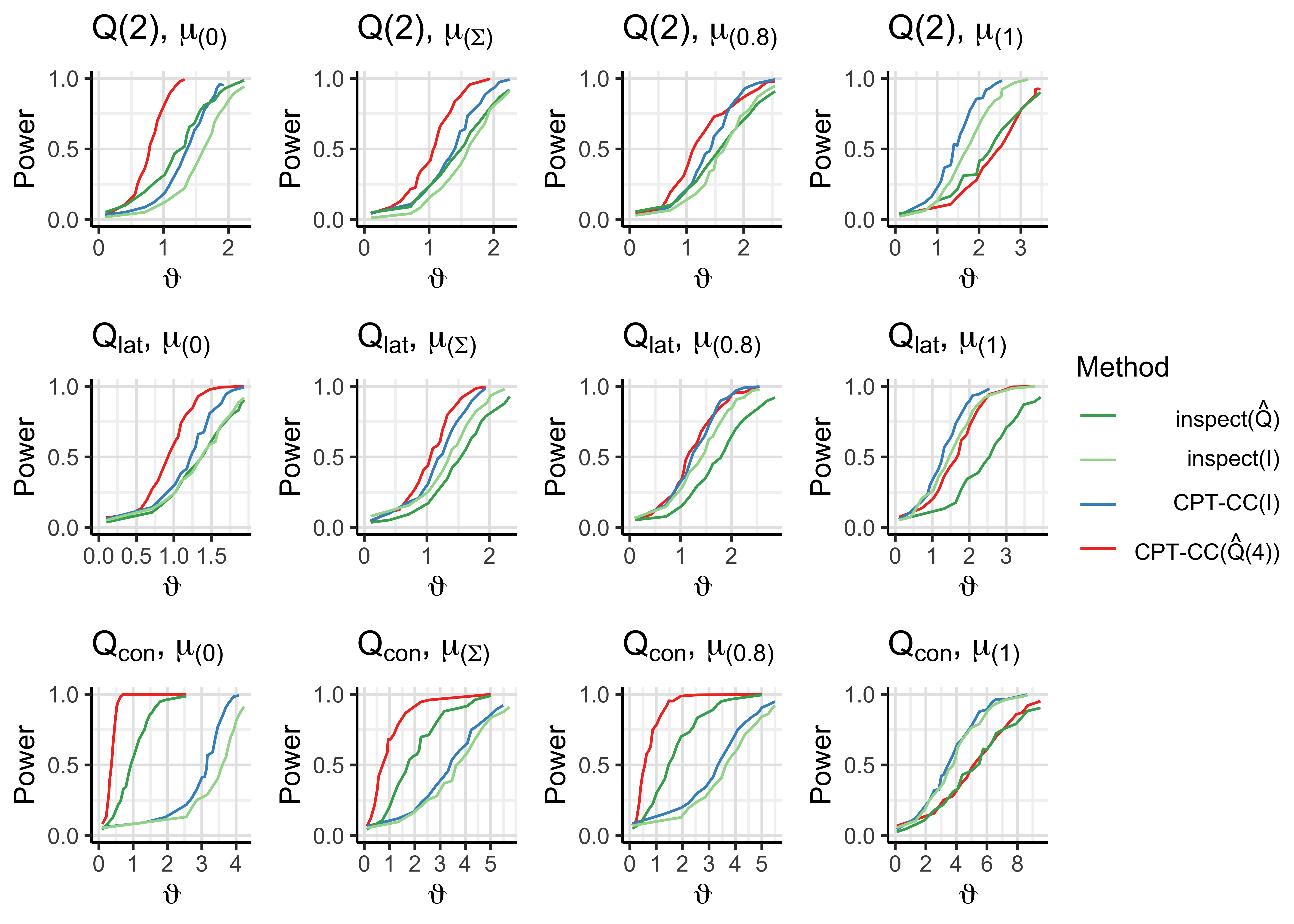}
  \caption{Power curves for a single known changepoint at $\tau = 170$ when $J = 100$, $p = 100$ and $\rho = 0.9$.
  Other parameters: $n = 200$, $\alpha = 0.05$.}
  \label{fig:known_cpt_J100_rho09}
\end{figure}

Further simulation results on power in the single changepoint setting is given in Figure \ref{fig:known_cpt_J10_rho05}-\ref{fig:known_cpt_J100_rho09}.
Tables \ref{tab:mse_p100_vartheta2_shape5}-\ref{tab:mse_p100_vartheta3_shape0} give additional results on the RMSE of changepoint estimates.

\begin{table}
\caption{RMSE for $p = 100$, $n = 200$, $\tau = 140$, $\vartheta = 2$ and $\bmu_{(0)}$ changes. The smallest value is given in bold. 1000 random samples were used for each RMSE estimate.}
\label{tab:mse_p100_vartheta2_shape5}
\centering
\begin{tabular}{ccccccc}
\toprule
$\bQ$ & $\rho$ & $J$ & CPT-CC($\hat{\bQ}(\bW(4))$) & inspect($\hat{\bQ}$) & CPT-CC($\bI$) & inspect($\bI$) \\
\midrule
$\mathbf{Q}(2)$ & 0.5 & 1 & $\mathbf{1.3}$ & $3.4$ & $1.4$ & $1.4$ \\
$\mathbf{Q}(2)$ & 0.5 & 10 & $1.9$ & $4.4$ & $1.9$ & $\mathbf{1.7}$ \\
$\mathbf{Q}(2)$ & 0.5 & 100 & $\mathbf{2.3}$ & $9.5$ & $2.6$ & $5.4$ \\
$\mathbf{Q}(2)$ & 0.7 & 1 & $\mathbf{1.2}$ & $3.1$ & $1.4$ & $1.3$ \\
$\mathbf{Q}(2)$ & 0.7 & 10 & $\mathbf{1.7}$ & $3.7$ & $2.2$ & $1.8$ \\
$\mathbf{Q}(2)$ & 0.7 & 100 & $\mathbf{1.8}$ & $7.4$ & $3.0$ & $5.7$ \\
$\mathbf{Q}(2)$ & 0.9 & 1 & $\mathbf{0.7}$ & $2.1$ & $1.5$ & $1.3$ \\
$\mathbf{Q}(2)$ & 0.9 & 10 & $\mathbf{1.0}$ & $2.3$ & $3.6$ & $1.7$ \\
$\mathbf{Q}(2)$ & 0.9 & 100 & $\mathbf{1.1}$ & $5.5$ & $4.9$ & $6.8$ \\
$\mathbf{Q}_\text{lat}$ & 0.5 & 1 & $\mathbf{1.3}$ & $3.3$ & $1.4$ & $1.4$ \\
$\mathbf{Q}_\text{lat}$ & 0.5 & 10 & $2.1$ & $4.4$ & $2.3$ & $\mathbf{1.6}$ \\
$\mathbf{Q}_\text{lat}$ & 0.5 & 100 & $2.7$ & $8.6$ & $\mathbf{2.7}$ & $4.6$ \\
$\mathbf{Q}_\text{lat}$ & 0.7 & 1 & $1.4$ & $2.8$ & $1.4$ & $\mathbf{1.3}$ \\
$\mathbf{Q}_\text{lat}$ & 0.7 & 10 & $\mathbf{1.7}$ & $4.0$ & $2.0$ & $1.9$ \\
$\mathbf{Q}_\text{lat}$ & 0.7 & 100 & $\mathbf{2.1}$ & $8.8$ & $2.5$ & $5.7$ \\
$\mathbf{Q}_\text{lat}$ & 0.9 & 1 & $\mathbf{1.0}$ & $2.2$ & $1.4$ & $1.5$ \\
$\mathbf{Q}_\text{lat}$ & 0.9 & 10 & $\mathbf{1.4}$ & $3.3$ & $2.4$ & $1.8$ \\
$\mathbf{Q}_\text{lat}$ & 0.9 & 100 & $\mathbf{1.8}$ & $6.4$ & $4.5$ & $6.9$ \\
$\mathbf{Q}_\text{con}$ & 0.5 & 1 & $\mathbf{0.8}$ & $1.8$ & $13.3$ & $2.4$ \\
$\mathbf{Q}_\text{con}$ & 0.5 & 10 & $\mathbf{1.1}$ & $2.0$ & $21.2$ & $5.1$ \\
$\mathbf{Q}_\text{con}$ & 0.5 & 100 & $\mathbf{1.1}$ & $7.7$ & $115.9$ & $23.4$ \\
$\mathbf{Q}_\text{con}$ & 0.7 & 1 & $\mathbf{0.5}$ & $1.1$ & $17.1$ & $6.4$ \\
$\mathbf{Q}_\text{con}$ & 0.7 & 10 & $\mathbf{0.6}$ & $1.7$ & $32.0$ & $9.8$ \\
$\mathbf{Q}_\text{con}$ & 0.7 & 100 & $\mathbf{0.5}$ & $5.5$ & $126.1$ & $34.9$ \\
$\mathbf{Q}_\text{con}$ & 0.9 & 1 & $\mathbf{0.1}$ & $0.4$ & $20.5$ & $11.7$ \\
$\mathbf{Q}_\text{con}$ & 0.9 & 10 & $\mathbf{0.1}$ & $3.7$ & $53.9$ & $17.7$ \\
$\mathbf{Q}_\text{con}$ & 0.9 & 100 & $\mathbf{0.1}$ & $4.0$ & $131.2$ & $34.5$ \\
\bottomrule
\end{tabular}
\end{table}

\begin{table}
\caption{RMSE for $p = 100$, $n = 200$, $\tau = 140$, $\vartheta = 2$ and $\bmu_{(\bSigma)}$ changes. The smallest value is given in bold. 1000 random samples were used for each RMSE estimate.}
\label{tab:mse_p100_vartheta2_shape6}
\centering
\begin{tabular}{ccccccc}
\toprule
$\bQ$ & $\rho$ & $J$ & CPT-CC($\hat{\bQ}(\bW(4))$) & inspect($\hat{\bQ}$) & CPT-CC($\bI$) & inspect($\bI$) \\
\midrule
$\mathbf{Q}(2)$ & 0.5 & 1 & $\mathbf{1.3}$ & $3.4$ & $1.4$ & $1.4$ \\
$\mathbf{Q}(2)$ & 0.5 & 10 & $2.3$ & $4.8$ & $2.4$ & $\mathbf{1.9}$ \\
$\mathbf{Q}(2)$ & 0.5 & 100 & $\mathbf{2.4}$ & $9.7$ & $2.7$ & $5.0$ \\
$\mathbf{Q}(2)$ & 0.7 & 1 & $\mathbf{1.2}$ & $3.1$ & $1.4$ & $1.3$ \\
$\mathbf{Q}(2)$ & 0.7 & 10 & $2.7$ & $5.6$ & $\mathbf{2.6}$ & $2.6$ \\
$\mathbf{Q}(2)$ & 0.7 & 100 & $\mathbf{2.4}$ & $10.9$ & $3.4$ & $6.0$ \\
$\mathbf{Q}(2)$ & 0.9 & 1 & $\mathbf{0.7}$ & $2.1$ & $1.5$ & $1.3$ \\
$\mathbf{Q}(2)$ & 0.9 & 10 & $4.5$ & $5.3$ & $5.2$ & $\mathbf{3.1}$ \\
$\mathbf{Q}(2)$ & 0.9 & 100 & $\mathbf{3.2}$ & $13.4$ & $9.5$ & $10.5$ \\
$\mathbf{Q}_\text{lat}$ & 0.5 & 1 & $\mathbf{1.3}$ & $3.3$ & $1.4$ & $1.4$ \\
$\mathbf{Q}_\text{lat}$ & 0.5 & 10 & $2.4$ & $6.1$ & $2.3$ & $\mathbf{1.8}$ \\
$\mathbf{Q}_\text{lat}$ & 0.5 & 100 & $\mathbf{2.5}$ & $9.9$ & $2.6$ & $5.1$ \\
$\mathbf{Q}_\text{lat}$ & 0.7 & 1 & $1.4$ & $2.8$ & $1.4$ & $\mathbf{1.3}$ \\
$\mathbf{Q}_\text{lat}$ & 0.7 & 10 & $2.2$ & $4.4$ & $2.4$ & $\mathbf{1.9}$ \\
$\mathbf{Q}_\text{lat}$ & 0.7 & 100 & $\mathbf{2.5}$ & $10.1$ & $3.3$ & $5.2$ \\
$\mathbf{Q}_\text{lat}$ & 0.9 & 1 & $\mathbf{1.0}$ & $2.2$ & $1.4$ & $1.5$ \\
$\mathbf{Q}_\text{lat}$ & 0.9 & 10 & $3.0$ & $4.0$ & $4.5$ & $\mathbf{2.3}$ \\
$\mathbf{Q}_\text{lat}$ & 0.9 & 100 & $\mathbf{3.1}$ & $12.2$ & $8.3$ & $9.1$ \\
$\mathbf{Q}_\text{con}$ & 0.5 & 1 & $\mathbf{0.8}$ & $1.8$ & $13.3$ & $2.4$ \\
$\mathbf{Q}_\text{con}$ & 0.5 & 10 & $5.0$ & $\mathbf{3.0}$ & $23.9$ & $7.8$ \\
$\mathbf{Q}_\text{con}$ & 0.5 & 100 & $\mathbf{12.3}$ & $22.4$ & $117.2$ & $35.2$ \\
$\mathbf{Q}_\text{con}$ & 0.7 & 1 & $\mathbf{0.5}$ & $1.1$ & $17.1$ & $6.4$ \\
$\mathbf{Q}_\text{con}$ & 0.7 & 10 & $6.6$ & $\mathbf{3.0}$ & $49.2$ & $13.8$ \\
$\mathbf{Q}_\text{con}$ & 0.7 & 100 & $\mathbf{17.3}$ & $29.9$ & $123.3$ & $45.6$ \\
$\mathbf{Q}_\text{con}$ & 0.9 & 1 & $\mathbf{0.1}$ & $0.4$ & $20.5$ & $11.7$ \\
$\mathbf{Q}_\text{con}$ & 0.9 & 10 & $0.8$ & $\mathbf{0.5}$ & $92.9$ & $21.0$ \\
$\mathbf{Q}_\text{con}$ & 0.9 & 100 & $\mathbf{23.6}$ & $42.3$ & $127.2$ & $58.8$ \\
\bottomrule
\end{tabular}
\end{table}

\begin{table}
\caption{RMSE for $p = 100$, $n = 200$, $\tau = 140$, $\vartheta = 2$ and $\bmu_{(1)}$ changes. The smallest value is given in bold. 1000 random samples were used for each RMSE estimate.}
\label{tab:mse_p100_vartheta2_shape0}
\centering
\begin{tabular}{ccccccc}
\toprule
$\bQ$ & $\rho$ & $J$ & CPT-CC($\hat{\bQ}(\bW(4))$) & inspect($\hat{\bQ}$) & CPT-CC($\bI$) & inspect($\bI$) \\
\midrule
$\mathbf{Q}(2)$ & 0.5 & 1 & $\mathbf{1.3}$ & $3.4$ & $1.4$ & $1.4$ \\
$\mathbf{Q}(2)$ & 0.5 & 10 & $11.5$ & $8.9$ & $3.9$ & $\mathbf{3.3}$ \\
$\mathbf{Q}(2)$ & 0.5 & 100 & $14.1$ & $18.1$ & $\mathbf{3.5}$ & $9.0$ \\
$\mathbf{Q}(2)$ & 0.7 & 1 & $\mathbf{1.2}$ & $3.1$ & $1.4$ & $1.3$ \\
$\mathbf{Q}(2)$ & 0.7 & 10 & $27.9$ & $14.4$ & $5.3$ & $\mathbf{5.3}$ \\
$\mathbf{Q}(2)$ & 0.7 & 100 & $36.8$ & $21.6$ & $\mathbf{5.6}$ & $12.4$ \\
$\mathbf{Q}(2)$ & 0.9 & 1 & $\mathbf{0.7}$ & $2.1$ & $1.5$ & $1.3$ \\
$\mathbf{Q}(2)$ & 0.9 & 10 & $43.9$ & $17.7$ & $11.9$ & $\mathbf{8.1}$ \\
$\mathbf{Q}(2)$ & 0.9 & 100 & $68.7$ & $29.7$ & $18.7$ & $\mathbf{18.3}$ \\
$\mathbf{Q}_\text{lat}$ & 0.5 & 1 & $\mathbf{1.3}$ & $3.3$ & $1.4$ & $1.4$ \\
$\mathbf{Q}_\text{lat}$ & 0.5 & 10 & $7.1$ & $6.5$ & $3.2$ & $\mathbf{2.8}$ \\
$\mathbf{Q}_\text{lat}$ & 0.5 & 100 & $5.3$ & $20.2$ & $\mathbf{3.5}$ & $9.5$ \\
$\mathbf{Q}_\text{lat}$ & 0.7 & 1 & $1.4$ & $2.8$ & $1.4$ & $\mathbf{1.3}$ \\
$\mathbf{Q}_\text{lat}$ & 0.7 & 10 & $11.4$ & $8.9$ & $3.8$ & $\mathbf{3.0}$ \\
$\mathbf{Q}_\text{lat}$ & 0.7 & 100 & $13.8$ & $24.4$ & $\mathbf{7.0}$ & $13.0$ \\
$\mathbf{Q}_\text{lat}$ & 0.9 & 1 & $\mathbf{1.0}$ & $2.2$ & $1.4$ & $1.5$ \\
$\mathbf{Q}_\text{lat}$ & 0.9 & 10 & $40.9$ & $8.4$ & $9.1$ & $\mathbf{4.5}$ \\
$\mathbf{Q}_\text{lat}$ & 0.9 & 100 & $37.6$ & $28.1$ & $19.1$ & $\mathbf{17.6}$ \\
$\mathbf{Q}_\text{con}$ & 0.5 & 1 & $\mathbf{0.8}$ & $1.8$ & $13.3$ & $2.4$ \\
$\mathbf{Q}_\text{con}$ & 0.5 & 10 & $42.8$ & $\mathbf{5.2}$ & $79.8$ & $15.4$ \\
$\mathbf{Q}_\text{con}$ & 0.5 & 100 & $79.7$ & $\mathbf{50.7}$ & $116.5$ & $52.6$ \\
$\mathbf{Q}_\text{con}$ & 0.7 & 1 & $\mathbf{0.5}$ & $1.1$ & $17.1$ & $6.4$ \\
$\mathbf{Q}_\text{con}$ & 0.7 & 10 & $23.1$ & $\mathbf{6.9}$ & $114.2$ & $23.9$ \\
$\mathbf{Q}_\text{con}$ & 0.7 & 100 & $82.9$ & $\mathbf{57.8}$ & $118.2$ & $62.7$ \\
$\mathbf{Q}_\text{con}$ & 0.9 & 1 & $\mathbf{0.1}$ & $0.4$ & $20.5$ & $11.7$ \\
$\mathbf{Q}_\text{con}$ & 0.9 & 10 & $\mathbf{1.4}$ & $4.2$ & $128.4$ & $29.1$ \\
$\mathbf{Q}_\text{con}$ & 0.9 & 100 & $84.1$ & $\mathbf{67.5}$ & $124.7$ & $71.2$ \\
\bottomrule
\end{tabular}
\end{table}

\begin{table}
\caption{RMSE for $p = 100$, $n = 200$, $\tau = 140$, $\vartheta = 3$ and $\bmu_{(0)}$ changes. The smallest value is given in bold. 1000 random samples were used for each RMSE estimate.}
\label{tab:mse_p100_vartheta3_shape5}
\centering
\begin{tabular}{ccccccc}
\toprule
$\bQ$ & $\rho$ & $J$ & CPT-CC($\hat{\bQ}(\bW(4))$) & CPT-CC($\bI$) & inspect($\hat{\bQ}$) & inspect($\bI$) \\
\midrule
$\mathbf{Q}(2)$ & 0.5 & 1 & $\mathbf{0.5}$ & $0.6$ & $1.4$ & $0.5$ \\
$\mathbf{Q}(2)$ & 0.7 & 1 & $\mathbf{0.4}$ & $0.6$ & $1.2$ & $0.5$ \\
$\mathbf{Q}(2)$ & 0.9 & 1 & $\mathbf{0.2}$ & $0.5$ & $0.8$ & $0.6$ \\
$\mathbf{Q}_\text{lat}$ & 0.5 & 1 & $0.5$ & $0.6$ & $1.3$ & $\mathbf{0.5}$ \\
$\mathbf{Q}_\text{lat}$ & 0.7 & 1 & $\mathbf{0.4}$ & $0.5$ & $1.2$ & $0.5$ \\
$\mathbf{Q}_\text{lat}$ & 0.9 & 1 & $\mathbf{0.4}$ & $0.6$ & $0.9$ & $0.5$ \\
$\mathbf{Q}_\text{con}$ & 0.5 & 1 & $\mathbf{0.3}$ & $6.3$ & $0.8$ & $0.8$ \\
$\mathbf{Q}_\text{con}$ & 0.7 & 1 & $\mathbf{0.1}$ & $7.2$ & $0.4$ & $2.1$ \\
$\mathbf{Q}_\text{con}$ & 0.9 & 1 & $\mathbf{0.0}$ & $10.2$ & $0.1$ & $2.8$ \\
$\mathbf{Q}(2)$ & 0.5 & 10 & $0.6$ & $0.7$ & $1.4$ & $\mathbf{0.6}$ \\
$\mathbf{Q}(2)$ & 0.7 & 10 & $\mathbf{0.6}$ & $0.8$ & $1.4$ & $0.6$ \\
$\mathbf{Q}(2)$ & 0.9 & 10 & $\mathbf{0.4}$ & $0.7$ & $1.0$ & $0.6$ \\
$\mathbf{Q}_\text{lat}$ & 0.5 & 10 & $0.7$ & $0.7$ & $1.5$ & $\mathbf{0.6}$ \\
$\mathbf{Q}_\text{lat}$ & 0.7 & 10 & $\mathbf{0.6}$ & $0.8$ & $1.4$ & $0.6$ \\
$\mathbf{Q}_\text{lat}$ & 0.9 & 10 & $\mathbf{0.5}$ & $0.8$ & $1.0$ & $0.6$ \\
$\mathbf{Q}_\text{con}$ & 0.5 & 10 & $\mathbf{0.3}$ & $12.3$ & $0.9$ & $1.2$ \\
$\mathbf{Q}_\text{con}$ & 0.7 & 10 & $\mathbf{0.2}$ & $15.5$ & $0.5$ & $2.0$ \\
$\mathbf{Q}_\text{con}$ & 0.9 & 10 & $\mathbf{0.0}$ & $16.4$ & $0.2$ & $6.2$ \\
$\mathbf{Q}(2)$ & 0.5 & 100 & $\mathbf{0.7}$ & $0.7$ & $2.9$ & $1.0$ \\
$\mathbf{Q}(2)$ & 0.7 & 100 & $\mathbf{0.5}$ & $0.7$ & $2.2$ & $1.1$ \\
$\mathbf{Q}(2)$ & 0.9 & 100 & $\mathbf{0.3}$ & $1.1$ & $1.5$ & $1.4$ \\
$\mathbf{Q}_\text{lat}$ & 0.5 & 100 & $\mathbf{0.7}$ & $0.7$ & $2.8$ & $1.1$ \\
$\mathbf{Q}_\text{lat}$ & 0.7 & 100 & $\mathbf{0.7}$ & $0.8$ & $2.2$ & $1.2$ \\
$\mathbf{Q}_\text{lat}$ & 0.9 & 100 & $\mathbf{0.5}$ & $0.8$ & $2.0$ & $1.4$ \\
$\mathbf{Q}_\text{con}$ & 0.5 & 100 & $\mathbf{0.4}$ & $40.2$ & $1.3$ & $8.1$ \\
$\mathbf{Q}_\text{con}$ & 0.7 & 100 & $\mathbf{0.1}$ & $87.0$ & $0.8$ & $12.4$ \\
$\mathbf{Q}_\text{con}$ & 0.9 & 100 & $\mathbf{0.0}$ & $117.2$ & $0.3$ & $18.6$ \\
\bottomrule
\end{tabular}
\end{table}

\begin{table}
\caption{RMSE for $p = 100$, $n = 200$, $\tau = 140$, $\vartheta = 3$ and $\bmu_{(\bSigma)}$ changes. The smallest value is given in bold. 1000 random samples were used for each RMSE estimate.}
\label{tab:mse_p100_vartheta3_shape6}
\centering
\begin{tabular}{ccccccc}
\toprule
$\bQ$ & $\rho$ & $J$ & CPT-CC($\hat{\bQ}(\bW(4))$) & CPT-CC($\bI$) & inspect($\hat{\bQ}$) & inspect($\bI$) \\
\midrule
$\mathbf{Q}(2)$ & 0.5 & 1 & $\mathbf{0.5}$ & $0.6$ & $1.4$ & $0.5$ \\
$\mathbf{Q}(2)$ & 0.7 & 1 & $\mathbf{0.4}$ & $0.6$ & $1.2$ & $0.5$ \\
$\mathbf{Q}(2)$ & 0.9 & 1 & $\mathbf{0.2}$ & $0.5$ & $0.8$ & $0.6$ \\
$\mathbf{Q}_\text{lat}$ & 0.5 & 1 & $0.5$ & $0.6$ & $1.3$ & $\mathbf{0.5}$ \\
$\mathbf{Q}_\text{lat}$ & 0.7 & 1 & $\mathbf{0.4}$ & $0.5$ & $1.2$ & $0.5$ \\
$\mathbf{Q}_\text{lat}$ & 0.9 & 1 & $\mathbf{0.4}$ & $0.6$ & $0.9$ & $0.5$ \\
$\mathbf{Q}_\text{con}$ & 0.5 & 1 & $\mathbf{0.3}$ & $6.3$ & $0.8$ & $0.8$ \\
$\mathbf{Q}_\text{con}$ & 0.7 & 1 & $\mathbf{0.1}$ & $7.2$ & $0.4$ & $2.1$ \\
$\mathbf{Q}_\text{con}$ & 0.9 & 1 & $\mathbf{0.0}$ & $10.2$ & $0.1$ & $2.8$ \\
$\mathbf{Q}(2)$ & 0.5 & 10 & $0.8$ & $0.7$ & $1.7$ & $\mathbf{0.6}$ \\
$\mathbf{Q}(2)$ & 0.7 & 10 & $\mathbf{0.8}$ & $1.0$ & $1.8$ & $0.8$ \\
$\mathbf{Q}(2)$ & 0.9 & 10 & $\mathbf{0.8}$ & $1.4$ & $1.7$ & $1.4$ \\
$\mathbf{Q}_\text{lat}$ & 0.5 & 10 & $0.6$ & $0.7$ & $1.7$ & $\mathbf{0.6}$ \\
$\mathbf{Q}_\text{lat}$ & 0.7 & 10 & $0.7$ & $0.8$ & $2.2$ & $\mathbf{0.7}$ \\
$\mathbf{Q}_\text{lat}$ & 0.9 & 10 & $\mathbf{0.8}$ & $1.1$ & $1.4$ & $0.8$ \\
$\mathbf{Q}_\text{con}$ & 0.5 & 10 & $\mathbf{0.7}$ & $13.4$ & $0.9$ & $2.2$ \\
$\mathbf{Q}_\text{con}$ & 0.7 & 10 & $0.7$ & $15.2$ & $\mathbf{0.5}$ & $4.1$ \\
$\mathbf{Q}_\text{con}$ & 0.9 & 10 & $0.2$ & $27.3$ & $\mathbf{0.2}$ & $8.4$ \\
$\mathbf{Q}(2)$ & 0.5 & 100 & $\mathbf{0.7}$ & $0.8$ & $3.1$ & $1.1$ \\
$\mathbf{Q}(2)$ & 0.7 & 100 & $\mathbf{0.8}$ & $1.0$ & $3.2$ & $1.5$ \\
$\mathbf{Q}(2)$ & 0.9 & 100 & $\mathbf{0.7}$ & $1.7$ & $3.8$ & $2.1$ \\
$\mathbf{Q}_\text{lat}$ & 0.5 & 100 & $\mathbf{0.8}$ & $0.8$ & $3.0$ & $1.1$ \\
$\mathbf{Q}_\text{lat}$ & 0.7 & 100 & $0.9$ & $\mathbf{0.8}$ & $3.6$ & $1.2$ \\
$\mathbf{Q}_\text{lat}$ & 0.9 & 100 & $\mathbf{1.0}$ & $3.0$ & $3.7$ & $1.9$ \\
$\mathbf{Q}_\text{con}$ & 0.5 & 100 & $\mathbf{2.6}$ & $65.1$ & $5.9$ & $16.2$ \\
$\mathbf{Q}_\text{con}$ & 0.7 & 100 & $\mathbf{4.4}$ & $99.5$ & $10.8$ & $27.4$ \\
$\mathbf{Q}_\text{con}$ & 0.9 & 100 & $\mathbf{8.4}$ & $113.9$ & $19.0$ & $38.3$ \\
\bottomrule
\end{tabular}
\end{table}

\begin{table}
\caption{RMSE for $p = 100$, $n = 200$, $\tau = 140$, $\vartheta = 3$ and $\bmu_{(1)}$ changes. The smallest value is given in bold. 1000 random samples were used for each RMSE estimate.}
\label{tab:mse_p100_vartheta3_shape0}
\centering
\begin{tabular}{ccccccc}
\toprule
$\bQ$ & $\rho$ & $J$ & CPT-CC($\hat{\bQ}(\bW(4))$) & CPT-CC($\bI$) & inspect($\hat{\bQ}$) & inspect($\bI$) \\
\midrule
$\mathbf{Q}(2)$ & 0.5 & 1 & $\mathbf{0.5}$ & $0.6$ & $1.4$ & $0.5$ \\
$\mathbf{Q}(2)$ & 0.7 & 1 & $\mathbf{0.4}$ & $0.6$ & $1.2$ & $0.5$ \\
$\mathbf{Q}(2)$ & 0.9 & 1 & $\mathbf{0.2}$ & $0.5$ & $0.8$ & $0.6$ \\
$\mathbf{Q}_\text{lat}$ & 0.5 & 1 & $0.5$ & $0.6$ & $1.3$ & $\mathbf{0.5}$ \\
$\mathbf{Q}_\text{lat}$ & 0.7 & 1 & $\mathbf{0.4}$ & $0.5$ & $1.2$ & $0.5$ \\
$\mathbf{Q}_\text{lat}$ & 0.9 & 1 & $\mathbf{0.4}$ & $0.6$ & $0.9$ & $0.5$ \\
$\mathbf{Q}_\text{con}$ & 0.5 & 1 & $\mathbf{0.3}$ & $6.3$ & $0.8$ & $0.8$ \\
$\mathbf{Q}_\text{con}$ & 0.7 & 1 & $\mathbf{0.1}$ & $7.2$ & $0.4$ & $2.1$ \\
$\mathbf{Q}_\text{con}$ & 0.9 & 1 & $\mathbf{0.0}$ & $10.2$ & $0.1$ & $2.8$ \\
$\mathbf{Q}(2)$ & 0.5 & 10 & $1.6$ & $1.2$ & $3.2$ & $\mathbf{1.2}$ \\
$\mathbf{Q}(2)$ & 0.7 & 10 & $2.5$ & $1.6$ & $5.4$ & $\mathbf{1.5}$ \\
$\mathbf{Q}(2)$ & 0.9 & 10 & $5.7$ & $3.0$ & $6.5$ & $\mathbf{2.7}$ \\
$\mathbf{Q}_\text{lat}$ & 0.5 & 10 & $1.4$ & $1.1$ & $2.3$ & $\mathbf{0.8}$ \\
$\mathbf{Q}_\text{lat}$ & 0.7 & 10 & $1.7$ & $1.2$ & $2.4$ & $\mathbf{1.0}$ \\
$\mathbf{Q}_\text{lat}$ & 0.9 & 10 & $3.2$ & $1.7$ & $2.5$ & $\mathbf{1.4}$ \\
$\mathbf{Q}_\text{con}$ & 0.5 & 10 & $3.7$ & $17.6$ & $\mathbf{0.9}$ & $4.2$ \\
$\mathbf{Q}_\text{con}$ & 0.7 & 10 & $1.7$ & $39.9$ & $\mathbf{0.6}$ & $8.2$ \\
$\mathbf{Q}_\text{con}$ & 0.9 & 10 & $0.3$ & $81.9$ & $\mathbf{0.2}$ & $11.3$ \\
$\mathbf{Q}(2)$ & 0.5 & 100 & $1.8$ & $\mathbf{1.3}$ & $8.7$ & $2.7$ \\
$\mathbf{Q}(2)$ & 0.7 & 100 & $3.4$ & $\mathbf{2.0}$ & $12.1$ & $3.5$ \\
$\mathbf{Q}(2)$ & 0.9 & 100 & $30.2$ & $\mathbf{4.0}$ & $18.5$ & $6.8$ \\
$\mathbf{Q}_\text{lat}$ & 0.5 & 100 & $1.4$ & $\mathbf{1.2}$ & $8.3$ & $2.6$ \\
$\mathbf{Q}_\text{lat}$ & 0.7 & 100 & $2.3$ & $\mathbf{1.8}$ & $12.6$ & $3.0$ \\
$\mathbf{Q}_\text{lat}$ & 0.9 & 100 & $6.7$ & $\mathbf{5.5}$ & $18.5$ & $5.6$ \\
$\mathbf{Q}_\text{con}$ & 0.5 & 100 & $71.0$ & $88.5$ & $36.6$ & $\mathbf{36.2}$ \\
$\mathbf{Q}_\text{con}$ & 0.7 & 100 & $80.8$ & $99.8$ & $\mathbf{42.5}$ & $46.7$ \\
$\mathbf{Q}_\text{con}$ & 0.9 & 100 & $83.8$ & $111.7$ & $\mathbf{49.7}$ & $55.2$ \\
\bottomrule
\end{tabular}
\end{table}

\subsection{Multiple anomaly detection} \label{sec:Bmultiple_anom_sim}
Supplementary results on multiple anomaly detection are given in Table \ref{tab:ari_p100_vartheta1} and \ref{tab:ari_p100_vartheta2}.
The setup is precisely the same as in Table \ref{tab:ari_p100_vartheta1_shape6} in the main text, but with different signal strengths $\vartheta$ and change classes included.
Note that the results for change class $\bmu_{(0.8)}$ are very similar to $\bmu_{(\bSigma)}$, we have therefor omitted them to allow the tables to fit on one page each.

\begin{table}[htb]
\caption{ARI for $p = 100$, $\vartheta = 1$. The largest value for each data setting is given in bold.}
\label{tab:ari_p100_vartheta1}
\centering
\begin{tabular}{cccccccc}
\toprule
$\mathbf{Q}$ & $\rho$ & $\bmu_{(\cdot)}$ & Pt. anoms & CAPA-CC($\hat{\mathbf{Q}}(4)$) & W + MVCAPA & MVCAPA & inspect($\hat{\mathbf{Q}}$) \\
\midrule
$\mathbf{Q}(2)$ & 0.5 & $0$ & -- & $\mathbf{0.27}$ & $0.13$ & $0.18$ & $0.04$ \\
$\mathbf{Q}(2)$ & 0.5 & $0$ & \checkmark & $\mathbf{0.39}$ & $0.28$ & $0.37$ & $0.01$ \\
$\mathbf{Q}(2)$ & 0.7 & $0$ & -- & $\mathbf{0.32}$ & $0.14$ & $0.14$ & $0.08$ \\
$\mathbf{Q}(2)$ & 0.7 & $0$ & \checkmark & $\mathbf{0.43}$ & $0.28$ & $0.30$ & $0.02$ \\
$\mathbf{Q}(2)$ & 0.9 & $0$ & -- & $\mathbf{0.60}$ & $0.43$ & $0.03$ & $0.16$ \\
$\mathbf{Q}(2)$ & 0.9 & $0$ & \checkmark & $\mathbf{0.68}$ & $0.48$ & $0.26$ & $0.02$ \\
$\mathbf{Q}(2)$ & 0.5 & $\Sigma$ & -- & $\mathbf{0.23}$ & $0.09$ & $0.20$ & $0.05$ \\
$\mathbf{Q}(2)$ & 0.5 & $\Sigma$ & \checkmark & $\mathbf{0.40}$ & $0.25$ & $0.37$ & $0.01$ \\
$\mathbf{Q}(2)$ & 0.7 & $\Sigma$ & -- & $\mathbf{0.34}$ & $0.19$ & $0.12$ & $0.06$ \\
$\mathbf{Q}(2)$ & 0.7 & $\Sigma$ & \checkmark & $\mathbf{0.43}$ & $0.30$ & $0.31$ & $0.00$ \\
$\mathbf{Q}(2)$ & 0.9 & $\Sigma$ & -- & $\mathbf{0.53}$ & $0.43$ & $0.05$ & $0.13$ \\
$\mathbf{Q}(2)$ & 0.9 & $\Sigma$ & \checkmark & $\mathbf{0.61}$ & $0.46$ & $0.26$ & $0.03$ \\
$\mathbf{Q}_\text{lat}$ & 0.5 & $0$ & -- & $\mathbf{0.21}$ & $0.10$ & $0.12$ & $0.06$ \\
$\mathbf{Q}_\text{lat}$ & 0.5 & $0$ & \checkmark & $\mathbf{0.31}$ & $0.28$ & $0.23$ & $0.04$ \\
$\mathbf{Q}_\text{lat}$ & 0.7 & $0$ & -- & $\mathbf{0.27}$ & $0.18$ & $0.16$ & $0.05$ \\
$\mathbf{Q}_\text{lat}$ & 0.7 & $0$ & \checkmark & $\mathbf{0.33}$ & $0.31$ & $0.21$ & $0.06$ \\
$\mathbf{Q}_\text{lat}$ & 0.9 & $0$ & -- & $0.34$ & $\mathbf{0.37}$ & $0.08$ & $0.12$ \\
$\mathbf{Q}_\text{lat}$ & 0.9 & $0$ & \checkmark & $0.40$ & $\mathbf{0.41}$ & $0.20$ & $0.07$ \\
$\mathbf{Q}_\text{lat}$ & 0.5 & $\Sigma$ & -- & $\mathbf{0.21}$ & $0.08$ & $0.12$ & $0.05$ \\
$\mathbf{Q}_\text{lat}$ & 0.5 & $\Sigma$ & \checkmark & $\mathbf{0.29}$ & $0.26$ & $0.25$ & $0.08$ \\
$\mathbf{Q}_\text{lat}$ & 0.7 & $\Sigma$ & -- & $\mathbf{0.27}$ & $0.21$ & $0.13$ & $0.05$ \\
$\mathbf{Q}_\text{lat}$ & 0.7 & $\Sigma$ & \checkmark & $\mathbf{0.35}$ & $0.31$ & $0.25$ & $0.10$ \\
$\mathbf{Q}_\text{lat}$ & 0.9 & $\Sigma$ & -- & $\mathbf{0.34}$ & $0.28$ & $0.09$ & $0.08$ \\
$\mathbf{Q}_\text{lat}$ & 0.9 & $\Sigma$ & \checkmark & $0.33$ & $\mathbf{0.42}$ & $0.18$ & $0.14$ \\
$\mathbf{Q}_\text{con}$ & 0.5 & $0$ & -- & $0.47$ & $\mathbf{0.50}$ & $0.00$ & $0.07$ \\
$\mathbf{Q}_\text{con}$ & 0.5 & $0$ & \checkmark & $\mathbf{0.53}$ & $0.49$ & $0.16$ & $0.02$ \\
$\mathbf{Q}_\text{con}$ & 0.7 & $0$ & -- & $0.63$ & $\mathbf{0.66}$ & $0.00$ & $0.10$ \\
$\mathbf{Q}_\text{con}$ & 0.7 & $0$ & \checkmark & $\mathbf{0.68}$ & $0.66$ & $0.13$ & $0.04$ \\
$\mathbf{Q}_\text{con}$ & 0.9 & $0$ & -- & $0.83$ & $\mathbf{0.89}$ & $0.00$ & $0.28$ \\
$\mathbf{Q}_\text{con}$ & 0.9 & $0$ & \checkmark & $0.83$ & $\mathbf{0.87}$ & $0.10$ & $0.08$ \\
$\mathbf{Q}_\text{con}$ & 0.5 & $\Sigma$ & -- & $0.44$ & $\mathbf{0.52}$ & $0.00$ & $0.06$ \\
$\mathbf{Q}_\text{con}$ & 0.5 & $\Sigma$ & \checkmark & $\mathbf{0.50}$ & $0.49$ & $0.11$ & $0.03$ \\
$\mathbf{Q}_\text{con}$ & 0.7 & $\Sigma$ & -- & $0.60$ & $\mathbf{0.65}$ & $0.00$ & $0.08$ \\
$\mathbf{Q}_\text{con}$ & 0.7 & $\Sigma$ & \checkmark & $\mathbf{0.66}$ & $0.64$ & $0.10$ & $0.04$ \\
$\mathbf{Q}_\text{con}$ & 0.9 & $\Sigma$ & -- & $0.66$ & $\mathbf{0.82}$ & $0.00$ & $0.26$ \\
$\mathbf{Q}_\text{con}$ & 0.9 & $\Sigma$ & \checkmark & $0.71$ & $\mathbf{0.82}$ & $0.09$ & $0.10$ \\
\bottomrule
\end{tabular}
\end{table}

\begin{table}[htb]
\caption{ARI for $p = 100$, $\vartheta = 1.5$. The largest value for each data setting is given in bold.}
\label{tab:ari_p100_vartheta1.5}
\centering
\begin{tabular}{cccccccc}
\toprule
$\mathbf{Q}$ & $\rho$ & $\bmu_{(\cdot)}$ & Pt. anoms & CAPA-CC($\hat{\mathbf{Q}}(4)$) & W + MVCAPA & MVCAPA & inspect($\hat{\mathbf{Q}}$) \\
\midrule
$\mathbf{Q}(2)$ & 0.5 & $0$ & -- & $\mathbf{0.66}$ & $0.59$ & $0.62$ & $0.34$ \\
$\mathbf{Q}(2)$ & 0.5 & $0$ & \checkmark & $\mathbf{0.70}$ & $0.63$ & $0.68$ & $0.11$ \\
$\mathbf{Q}(2)$ & 0.7 & $0$ & -- & $\mathbf{0.68}$ & $0.62$ & $0.61$ & $0.41$ \\
$\mathbf{Q}(2)$ & 0.7 & $0$ & \checkmark & $\mathbf{0.73}$ & $0.66$ & $0.68$ & $0.13$ \\
$\mathbf{Q}(2)$ & 0.9 & $0$ & -- & $\mathbf{0.82}$ & $0.74$ & $0.57$ & $0.57$ \\
$\mathbf{Q}(2)$ & 0.9 & $0$ & \checkmark & $\mathbf{0.82}$ & $0.74$ & $0.63$ & $0.16$ \\
$\mathbf{Q}(2)$ & 0.5 & $\Sigma$ & -- & $\mathbf{0.64}$ & $0.60$ & $0.63$ & $0.34$ \\
$\mathbf{Q}(2)$ & 0.5 & $\Sigma$ & \checkmark & $\mathbf{0.69}$ & $0.62$ & $\mathbf{0.69}$ & $0.12$ \\
$\mathbf{Q}(2)$ & 0.7 & $\Sigma$ & -- & $\mathbf{0.67}$ & $0.63$ & $0.62$ & $0.39$ \\
$\mathbf{Q}(2)$ & 0.7 & $\Sigma$ & \checkmark & $\mathbf{0.71}$ & $0.64$ & $0.69$ & $0.10$ \\
$\mathbf{Q}(2)$ & 0.9 & $\Sigma$ & -- & $\mathbf{0.68}$ & $\mathbf{0.68}$ & $0.56$ & $0.52$ \\
$\mathbf{Q}(2)$ & 0.9 & $\Sigma$ & \checkmark & $\mathbf{0.74}$ & $0.68$ & $0.60$ & $0.17$ \\
$\mathbf{Q}_\text{lat}$ & 0.5 & $0$ & -- & $\mathbf{0.65}$ & $0.61$ & $0.59$ & $0.33$ \\
$\mathbf{Q}_\text{lat}$ & 0.5 & $0$ & \checkmark & $0.64$ & $\mathbf{0.65}$ & $0.63$ & $0.18$ \\
$\mathbf{Q}_\text{lat}$ & 0.7 & $0$ & -- & $\mathbf{0.68}$ & $0.60$ & $0.58$ & $0.30$ \\
$\mathbf{Q}_\text{lat}$ & 0.7 & $0$ & \checkmark & $\mathbf{0.67}$ & $\mathbf{0.67}$ & $0.61$ & $0.20$ \\
$\mathbf{Q}_\text{lat}$ & 0.9 & $0$ & -- & $0.69$ & $\mathbf{0.70}$ & $0.57$ & $0.50$ \\
$\mathbf{Q}_\text{lat}$ & 0.9 & $0$ & \checkmark & $0.71$ & $\mathbf{0.72}$ & $0.57$ & $0.25$ \\
$\mathbf{Q}_\text{lat}$ & 0.5 & $\Sigma$ & -- & $\mathbf{0.62}$ & $0.59$ & $0.60$ & $0.35$ \\
$\mathbf{Q}_\text{lat}$ & 0.5 & $\Sigma$ & \checkmark & $\mathbf{0.65}$ & $\mathbf{0.65}$ & $0.61$ & $0.23$ \\
$\mathbf{Q}_\text{lat}$ & 0.7 & $\Sigma$ & -- & $\mathbf{0.64}$ & $0.62$ & $0.60$ & $0.36$ \\
$\mathbf{Q}_\text{lat}$ & 0.7 & $\Sigma$ & \checkmark & $0.64$ & $\mathbf{0.65}$ & $0.60$ & $0.23$ \\
$\mathbf{Q}_\text{lat}$ & 0.9 & $\Sigma$ & -- & $\mathbf{0.66}$ & $0.65$ & $0.58$ & $0.50$ \\
$\mathbf{Q}_\text{lat}$ & 0.9 & $\Sigma$ & \checkmark & $0.63$ & $\mathbf{0.67}$ & $0.58$ & $0.29$ \\
$\mathbf{Q}_\text{con}$ & 0.5 & $0$ & -- & $\mathbf{0.76}$ & $\mathbf{0.76}$ & $0.01$ & $0.21$ \\
$\mathbf{Q}_\text{con}$ & 0.5 & $0$ & \checkmark & $\mathbf{0.81}$ & $0.72$ & $0.16$ & $0.07$ \\
$\mathbf{Q}_\text{con}$ & 0.7 & $0$ & -- & $0.86$ & $\mathbf{0.88}$ & $0.00$ & $0.31$ \\
$\mathbf{Q}_\text{con}$ & 0.7 & $0$ & \checkmark & $\mathbf{0.85}$ & $\mathbf{0.85}$ & $0.13$ & $0.15$ \\
$\mathbf{Q}_\text{con}$ & 0.9 & $0$ & -- & $0.93$ & $\mathbf{0.96}$ & $0.00$ & $0.30$ \\
$\mathbf{Q}_\text{con}$ & 0.9 & $0$ & \checkmark & $0.94$ & $\mathbf{0.95}$ & $0.10$ & $0.15$ \\
$\mathbf{Q}_\text{con}$ & 0.5 & $\Sigma$ & -- & $0.70$ & $\mathbf{0.74}$ & $0.00$ & $0.19$ \\
$\mathbf{Q}_\text{con}$ & 0.5 & $\Sigma$ & \checkmark & $\mathbf{0.74}$ & $0.72$ & $0.11$ & $0.09$ \\
$\mathbf{Q}_\text{con}$ & 0.7 & $\Sigma$ & -- & $0.74$ & $\mathbf{0.85}$ & $0.00$ & $0.30$ \\
$\mathbf{Q}_\text{con}$ & 0.7 & $\Sigma$ & \checkmark & $0.75$ & $\mathbf{0.83}$ & $0.10$ & $0.12$ \\
$\mathbf{Q}_\text{con}$ & 0.9 & $\Sigma$ & -- & $0.74$ & $\mathbf{0.91}$ & $0.00$ & $0.39$ \\
$\mathbf{Q}_\text{con}$ & 0.9 & $\Sigma$ & \checkmark & $0.77$ & $\mathbf{0.89}$ & $0.09$ & $0.16$ \\
\bottomrule
\end{tabular}
\end{table}

\begin{table}[htb]
\caption{ARI for $p = 100$, $\vartheta = 2$. The largest value for each data setting is given in bold.}
\label{tab:ari_p100_vartheta2}
\centering
\begin{tabular}{cccccccc}
\toprule
$\mathbf{Q}$ & $\rho$ & $\bmu_{(\cdot)}$ & Pt. anoms & CAPA-CC($\hat{\mathbf{Q}}(4)$) & W + MVCAPA & MVCAPA & inspect($\hat{\mathbf{Q}}$) \\
\midrule
$\mathbf{Q}(2)$ & 0.5 & $0$ & -- & $\mathbf{0.82}$ & $0.77$ & $0.80$ & $0.52$ \\
$\mathbf{Q}(2)$ & 0.5 & $0$ & \checkmark & $\mathbf{0.84}$ & $0.75$ & $0.83$ & $0.21$ \\
$\mathbf{Q}(2)$ & 0.7 & $0$ & -- & $\mathbf{0.84}$ & $0.80$ & $0.78$ & $0.60$ \\
$\mathbf{Q}(2)$ & 0.7 & $0$ & \checkmark & $\mathbf{0.87}$ & $0.80$ & $0.82$ & $0.20$ \\
$\mathbf{Q}(2)$ & 0.9 & $0$ & -- & $\mathbf{0.90}$ & $0.88$ & $0.72$ & $0.69$ \\
$\mathbf{Q}(2)$ & 0.9 & $0$ & \checkmark & $\mathbf{0.91}$ & $0.86$ & $0.78$ & $0.23$ \\
$\mathbf{Q}(2)$ & 0.5 & $\Sigma$ & -- & $0.79$ & $0.73$ & $\mathbf{0.80}$ & $0.52$ \\
$\mathbf{Q}(2)$ & 0.5 & $\Sigma$ & \checkmark & $0.82$ & $0.77$ & $\mathbf{0.84}$ & $0.18$ \\
$\mathbf{Q}(2)$ & 0.7 & $\Sigma$ & -- & $\mathbf{0.80}$ & $0.75$ & $0.76$ & $0.55$ \\
$\mathbf{Q}(2)$ & 0.7 & $\Sigma$ & \checkmark & $0.81$ & $0.76$ & $\mathbf{0.82}$ & $0.19$ \\
$\mathbf{Q}(2)$ & 0.9 & $\Sigma$ & -- & $\mathbf{0.80}$ & $0.78$ & $0.70$ & $0.64$ \\
$\mathbf{Q}(2)$ & 0.9 & $\Sigma$ & \checkmark & $\mathbf{0.82}$ & $0.78$ & $0.75$ & $0.21$ \\
$\mathbf{Q}_\text{lat}$ & 0.5 & $0$ & -- & $\mathbf{0.82}$ & $0.75$ & $0.79$ & $0.56$ \\
$\mathbf{Q}_\text{lat}$ & 0.5 & $0$ & \checkmark & $\mathbf{0.77}$ & $0.76$ & $0.75$ & $0.25$ \\
$\mathbf{Q}_\text{lat}$ & 0.7 & $0$ & -- & $\mathbf{0.85}$ & $0.79$ & $0.81$ & $0.67$ \\
$\mathbf{Q}_\text{lat}$ & 0.7 & $0$ & \checkmark & $0.80$ & $\mathbf{0.81}$ & $0.76$ & $0.36$ \\
$\mathbf{Q}_\text{lat}$ & 0.9 & $0$ & -- & $\mathbf{0.85}$ & $\mathbf{0.85}$ & $0.73$ & $0.60$ \\
$\mathbf{Q}_\text{lat}$ & 0.9 & $0$ & \checkmark & $0.83$ & $\mathbf{0.86}$ & $0.71$ & $0.30$ \\
$\mathbf{Q}_\text{lat}$ & 0.5 & $\Sigma$ & -- & $\mathbf{0.79}$ & $0.73$ & $\mathbf{0.79}$ & $0.48$ \\
$\mathbf{Q}_\text{lat}$ & 0.5 & $\Sigma$ & \checkmark & $\mathbf{0.77}$ & $\mathbf{0.77}$ & $0.76$ & $0.29$ \\
$\mathbf{Q}_\text{lat}$ & 0.7 & $\Sigma$ & -- & $\mathbf{0.79}$ & $0.76$ & $0.77$ & $0.64$ \\
$\mathbf{Q}_\text{lat}$ & 0.7 & $\Sigma$ & \checkmark & $\mathbf{0.78}$ & $0.76$ & $0.76$ & $0.38$ \\
$\mathbf{Q}_\text{lat}$ & 0.9 & $\Sigma$ & -- & $\mathbf{0.79}$ & $\mathbf{0.79}$ & $0.73$ & $0.65$ \\
$\mathbf{Q}_\text{lat}$ & 0.9 & $\Sigma$ & \checkmark & $0.76$ & $\mathbf{0.80}$ & $0.72$ & $0.31$ \\
$\mathbf{Q}_\text{con}$ & 0.5 & $0$ & -- & $\mathbf{0.88}$ & $\mathbf{0.88}$ & $0.01$ & $0.31$ \\
$\mathbf{Q}_\text{con}$ & 0.5 & $0$ & \checkmark & $\mathbf{0.90}$ & $0.85$ & $0.16$ & $0.11$ \\
$\mathbf{Q}_\text{con}$ & 0.7 & $0$ & -- & $0.93$ & $\mathbf{0.94}$ & $0.00$ & $0.37$ \\
$\mathbf{Q}_\text{con}$ & 0.7 & $0$ & \checkmark & $\mathbf{0.93}$ & $0.89$ & $0.13$ & $0.16$ \\
$\mathbf{Q}_\text{con}$ & 0.9 & $0$ & -- & $\mathbf{1.00}$ & $\mathbf{1.00}$ & $0.00$ & $0.39$ \\
$\mathbf{Q}_\text{con}$ & 0.9 & $0$ & \checkmark & $\mathbf{1.00}$ & $0.97$ & $0.10$ & $0.14$ \\
$\mathbf{Q}_\text{con}$ & 0.5 & $\Sigma$ & -- & $0.81$ & $\mathbf{0.88}$ & $0.00$ & $0.24$ \\
$\mathbf{Q}_\text{con}$ & 0.5 & $\Sigma$ & \checkmark & $0.81$ & $\mathbf{0.84}$ & $0.12$ & $0.14$ \\
$\mathbf{Q}_\text{con}$ & 0.7 & $\Sigma$ & -- & $0.79$ & $\mathbf{0.90}$ & $0.00$ & $0.40$ \\
$\mathbf{Q}_\text{con}$ & 0.7 & $\Sigma$ & \checkmark & $0.82$ & $\mathbf{0.87}$ & $0.14$ & $0.16$ \\
$\mathbf{Q}_\text{con}$ & 0.9 & $\Sigma$ & -- & $0.82$ & $\mathbf{0.98}$ & $0.00$ & $0.43$ \\
$\mathbf{Q}_\text{con}$ & 0.9 & $\Sigma$ & \checkmark & $0.84$ & $\mathbf{0.95}$ & $0.10$ & $0.13$ \\
\bottomrule
\end{tabular}
\end{table}

\end{document}